\documentclass[preprint2]{aastex631}

\usepackage{amsmath}	
\usepackage{amssymb}	
\usepackage{graphicx}	
\usepackage{subfigure}

\received{\today}
\revised{\today}
\accepted{\today}

\submitjournal{AAS Publishing}

\shorttitle{Lensed Radio Sources in VLASS}
\shortauthors{Martinez et al.}

\begin{document}

\title{Finding Lensed Radio Sources with the VLA Sky Survey}

\correspondingauthor{Michael N. Martinez}
\email{mnmartinez@wisc.edu}

\author[0000-0002-8397-8412]{Michael N. Martinez}
\affiliation{Department of Physics, University of Wisconsin, Madison, 1150 University Avenue, Madison, WI 53706, USA}

\author[0000-0003-1432-253X]{Yjan~A. Gordon}
\affil{Department of Physics, University of Wisconsin, Madison, 1150 University Avenue, Madison, WI 53706, USA}

\author[0000-0001-8156-0429]{Keith Bechtol}
\affil{Department of Physics, University of Wisconsin, Madison, 1150 University Avenue, Madison, WI 53706, USA}

\author[0009-0005-7030-9948]{Gillian Cartwright}
\affil{Department of Physics, University of Wisconsin, Madison, 1150 University Avenue, Madison, WI 53706, USA}

\author[0000-0001-6957-1627]{Peter S. Ferguson}
\affil{Department of Physics, University of Wisconsin, Madison, 1150 University Avenue, Madison, WI 53706, USA}


\author[0000-0002-3135-3824]{Miranda Gorsuch}
\affil{Department of Physics, University of Wisconsin, Madison, 1150 University Avenue, Madison, WI 53706, USA}



\begin{abstract}
Radio observations of strongly lensed objects are valuable as cosmological probes.
Lensed radio sources have proven difficult to identify in large part due to the limited depth and angular resolution of the previous generation of radio sky surveys, and as such, only a few dozen lensed radio sources are known.
In this work we present the results of a pilot study using the Very Large Array Sky Survey (VLASS) in combination with optical data to more efficiently identify lensed radio sources.
We obtain high-resolution ($0.2''$) VLA follow-up observations for 11 targets that we identify using three different techniques: i) a search for compact radio sources offset from galaxies with high lensing potential, ii) VLASS detections of known lensed galaxies, iii) VLASS detections of known lensed quasars.
5 of our targets show radio emission from the lensed images, including $100\,\%$ of the lensed optical quasar systems.
This work demonstrates the efficacy of combining deep and high-resolution wide-area radio and optical survey data to efficiently find lensed radio sources, and we discuss the potential impact of such an approach using next-generation surveys with the Vera C. Rubin Observatory, Euclid, and Nancy Grace Roman Space Telescope.
\end{abstract}


\section{Introduction}  \label{sec:intro}

Strong gravitational lensing, the phenomenon by which multiple images of a background source are created by a foreground lens, has been an active and growing field of study since the discovery of the first lensed object by \citet{walsh79}.
Since then, the advent of high-resolution space-based optical imaging from the Hubble Space Telescope and large ground-based optical surveys have increased the number of known lenses today to many hundreds \citep[e.g.,][]{Jacobs2019, Huang2020, zaborowski23, Lemon2024}.

Gravitational lensing is achromatic, and strong lenses can be observed in any wavelength of light, though relative abundances vary across the electromagnetic spectrum.
At radio frequencies, under 100 lensed sources are known, as opposed to the thousands of optical ones.
This is due in part to the relative scarcity of radio sources. For example, the Faint Images of the Radio Sky at Twenty-centimeters \citep[FIRST,][]{FIRST, FIRSTfinal} and the imaging portion of the optical Sloan Digital Sky Survey \citep[SDSS][]{SDSS, SDSSDR7}, which covered the same area and were roughly contemporaneous, had source densities of $\approx90\,\text{deg}^{-2}$ and $\approx30,000\,\text{deg}^{-2}$, respectively.
Furthermore, the angular resolution needed to identify strong lensing, typically on the scale of $\sim1\,$arcsecond for galaxy-galaxy lenses \citep{McKean2015, collett15}, also presents a large barrier to finding lensed radio sources as the angular resolution of wide area surveys has historically been on the order of a few tens of arcseconds \citep[e.g,][]{Condon1998, Bock1999, Intema2017}.
This has historically resulted in samples of rare candidate lensed radio sources being overwhelmingly contaminated by non-lensed objects \citep[e.g.][]{JB07}.
Successful radio searches for lensing, such as the Jodrell Bank Astrometric Survey \citep[JVAS, ][]{1999MNRAS.307..225K} and the Cosmic Lens All-Sky Survey \citep[CLASS, ][]{Myers2003, 2003MNRAS.341...13B}, began with a flux-limited sample to limit the amount of necessary high-resolution follow-up to confirm lensing.
More recently, radio lens searches have taken advantage of the abundance of optical lensed quasars by conducting deep observations of these to try to detect radio emission \citep{2015MNRAS.454..287J, dobie23}.
In the future, facilities such as the Square Kilometer Array \citep[SKA, ][]{skaperformance} and next generation Very Large Array \citep[ngVLA,][]{Carilli2015} will provide depth and sub-arcsecond resolution in combination with high survey-speeds, making them efficient lens-finding tools.
Currently however, only limited sky areas (of order a few square degrees) have been observed with the requisite combination of depth and angular resolution to readily identify strong lensing at radio wavelengths \citep{Morabito2022}. 

The Very Large Array Sky Survey \citep[VLASS,][]{VLASS} provides $\approx 2''.5$ angular resolution across $\approx 34,000\,\text{deg}^2$ of sky at 3 GHz.
By 2025 VLASS will have observed its entire footprint over three distinct epochs, and at the time of writing, VLASS has already completed two of these epochs with the third epoch already underway.
While VLASS does not posses the resolution necessary to separate the images of most lensed quasars \citep{Lemon19}, the $2''.5$ beam of survey allows for high confidence associations with optical sources and is less subject to contamination from interloping sources than other near-all-sky radio surveys.

The scientific applications of radio lenses are numerous, and range from probing the structure of AGN jets at high redshift \citep{spignola2019} to studying the magnetic fields of lens galaxies \citep{mao17}.
One particularly exciting possibility lies in the characterization of low-mass dark matter halos to constrain the microphysics of dark matter.
Due to the sensitivity of image magnifications and deflections to all mass along the line of sight between source and observer, lensing observations are sensitive to the lower end of the dark matter halo mass function, especially the ``completely dark halos" not massive enough to form stars \citep{vegetti23, bechtol22}.
The milliarcsecond-scale astrometric perturbations caused by these halos \citep{Metcalf2001} currently can only be accessed using the resolution of radio Very Long baseline Interferometry (VLBI).
Such gravitational imaging analyses can potentially differentiate between different models of dark matter phenomenology \citep[e.g.,][]{2019MNRAS.483.2125S, powell23}.
Next-generation telescopes such as SKA and ngVLA will be able to perform observations of lens systems quickly and robustly -- larger samples of candidate systems are important to inform both the theory and technical development of those dark matter analyses.

In this paper, we present the results of a VLASS-based search for strong lensed radio sources, and report the detection of radio emission from five previously known optically lensed quasars.
In Section \ref{sec:search} we describe our candidate selection process, 
Section \ref{sec:obs} provides a summary of our observations, and Section \ref{sec:results} presents the results of each candidate observed in detail.
In Section \ref{sec:discussion} we discuss the population of known lensed radio sources and the potential for future survey-based radio lens searches.
We summarize this work in Section \ref{sec:summary}.

\section{Candidate Identification} \label{sec:search}


In selecting sources for the VLA observations, we took a two-pronged approach based on both known lens systems and catalog-based optical-radio cross-matching.
We identify radio sources using the VLASS epoch 1 quick-look catalog from \citet{Gordon2021}, which contains $\sim 1.8\times10^{6}$ reliable detections with $S_{3\text{GHz}}\gtrsim 1\,$mJy at $\delta > -40^{\circ}$.
To account for the known $\sim 0''.25$ astrometric errors in the quick-look data, we have corrected the source positions based on the method of \citet{Bruzewski2021}.
\footnote{Since the identification of these targets in 2022, a version of the epoch 1 VLASS Quick Look catalog with corrected astrometry has been made available (B. Sebastian et al., in prep.)}

\subsection{Known Lensed Optical Sources} \label{existing}

As lensed radio sources are rare, knowing a priori that a system is a gravitational lens maximises the efficiency in searching for these objects.
To this end, we cross match the VLASS catalog with two catalogs of known optical lenses using data from Gaia \citep{Gaia2016} and the Dark Energy Survey \citep[DES,][]{DES2016}. 

We first used the catalog of lensed quasars in Gaia \citep{Lemon2017, Lemon2018, Lemon19}, finding 43 matches with VLASS.
Of these, 31 were previously known lensed radio sources, and a further 7 had existing archival observations at sufficient resolution and depth to confirm or reject the radio lensing hypothesis without the need for additional telescope time.
An additional candidate was also rejected after visual inspection of the VLASS data showed the lens galaxy to be an FR I radio galaxy, implying the radio emission in the system came from the lens rather than the lensed source.
After these cuts we were left with 5 candidate new radio lenses.

We also cross-matched VLASS with strongly lensed systems in DES \citep{Jacobs2019}.
Here we found 17 matches, all of which were neither previously known strong lenses nor had archival high-resolution VLA data.
Visually inspecting these 17 objects showed that in most of these cases, the radio emission was more likely due to the lens galaxy being a radio galaxy. While in theory it is possible to observe radio emission from both the lens and source, we did not prioritize these targets.
In two cases we found the radio emission to be consistent with being from the \textit{lensed images} and require higher resolution follow up to confirm their nature.
However, due to limited observing time we only observed one of these with the VLA for this paper.

\subsection{Blind Search for Lensed Sources From Optical/Radio Cross Associations} \label{catalog}

\begin{figure*}
    \centering  
    \subfigure{\includegraphics[trim={10mm 10mm 0 0}, clip, width=0.23\textwidth]{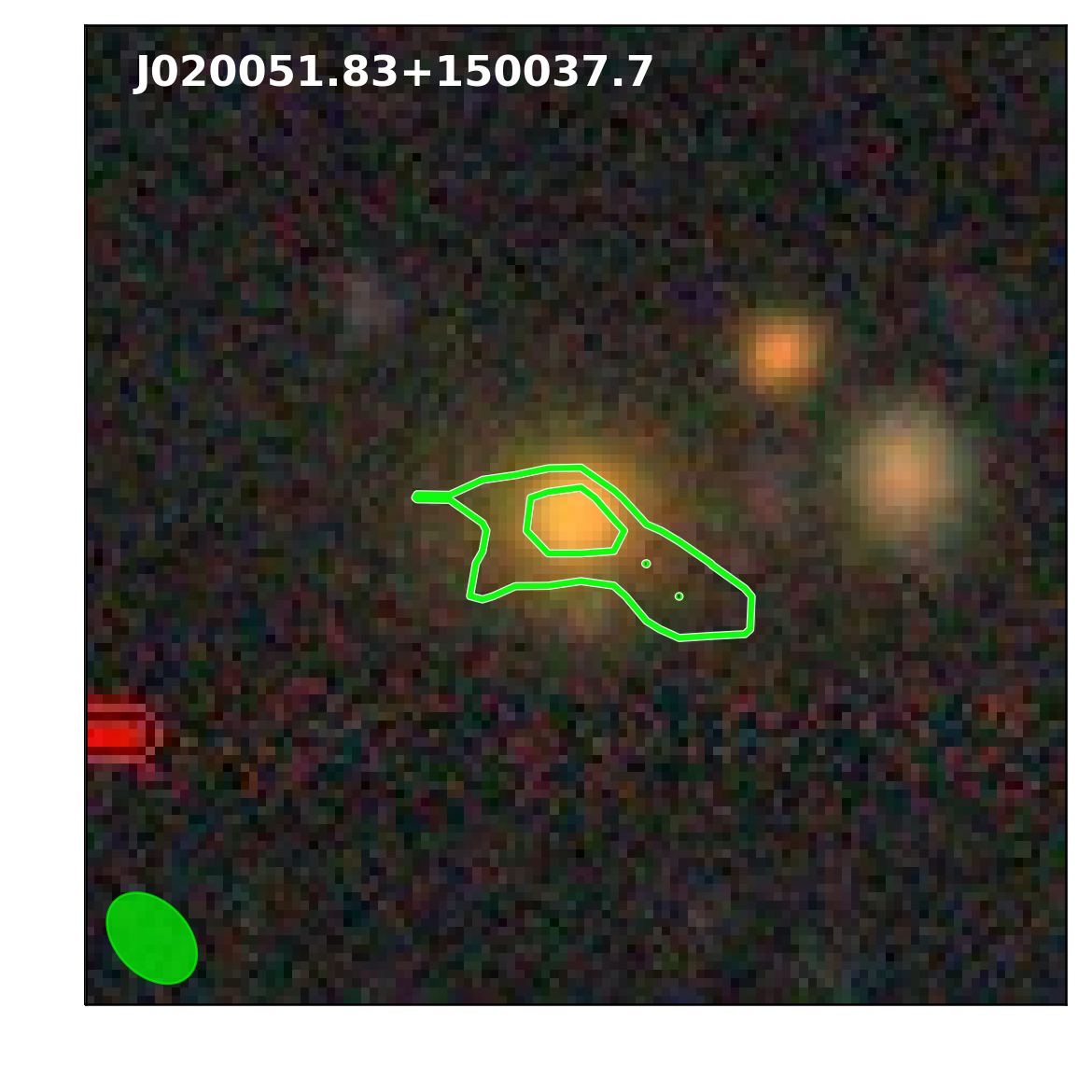}}
    \hspace{1mm}
    \subfigure{\includegraphics[trim={10mm 10mm 0 0}, clip, width=0.23\textwidth]{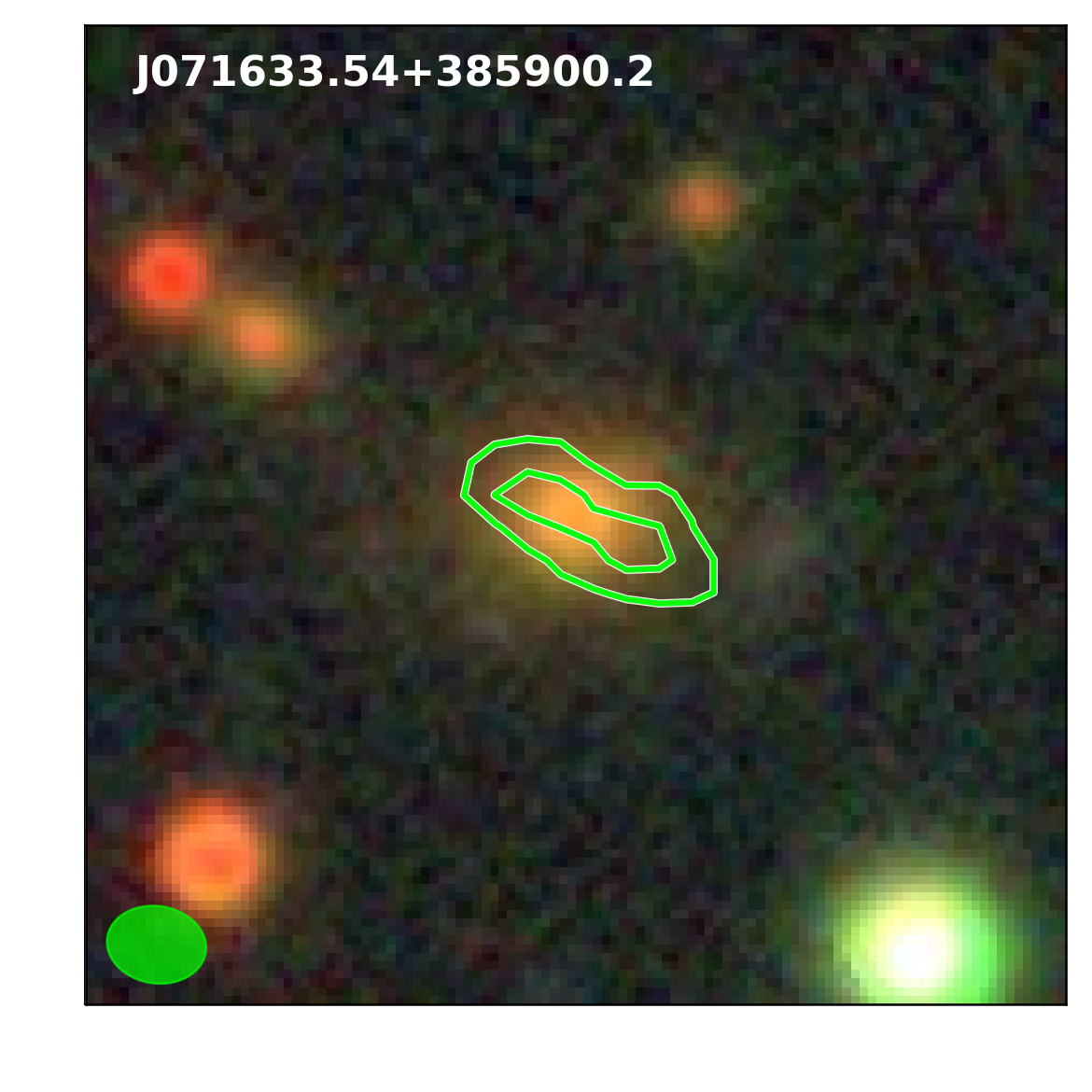}}
    \hspace{1mm}
    \subfigure{\includegraphics[trim={10mm 10mm 0 0}, clip, width=0.23\textwidth]{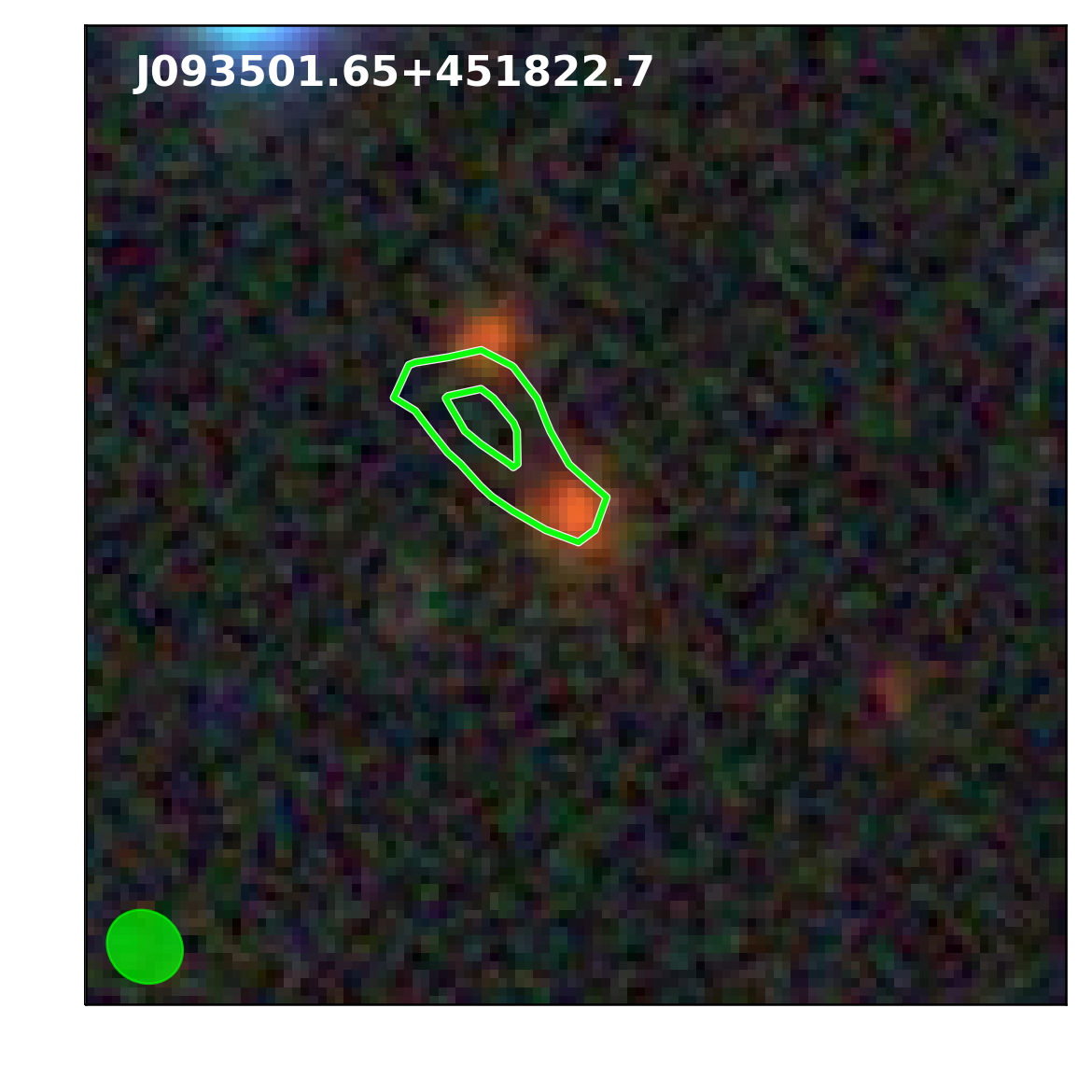}}
    \hspace{1mm}
    \subfigure{\includegraphics[trim={10mm 10mm 0 0}, clip, width=0.23\textwidth]{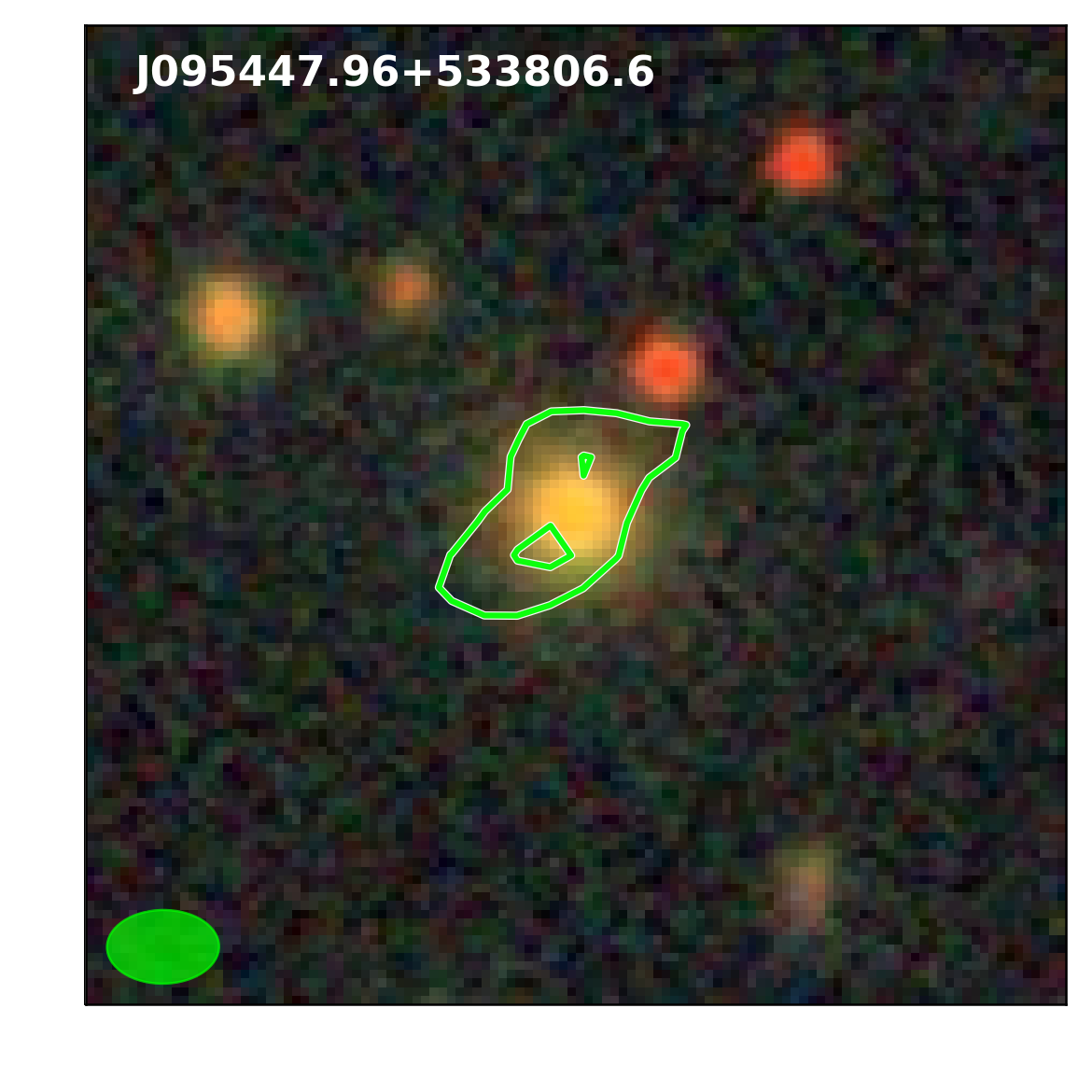}}
    
    \caption{Postage stamp optical cutouts ($30''\times30''$) of sources where the radio emission is attributed to the lens galaxy on visual inspection and thus rejected as candidate lensed radio sources.}
    \label{fig:eg_rejects}
\end{figure*}

In addition to combining VLASS with catalogs of known optical lenses, we adopted the approach of \citet{JB07} (hereafter JB07) to conduct a blind search for lensed systems in the radio catalog data.
The JB07 method assumes the lensed source flux is blended together into a single detection at the survey resolution, and predicts an offset from the lens galaxy due to the unequal magnifications inherent in lensed images.
Additionally, these blended components should have position angles either close to or perpendicular to that of the lens galaxy's optical position angle for 2 and 4-image systems, respectively.
JB07 matched the SDSS and FIRST surveys, identifying $\sim 70$ candidates, none of which were lenses.
However, the wealth of additional candidates afforded by increased depth and sky coverage since JB07 has led us to use their method with VLASS and DES to attempt to identify further candidate lensed radio systems.

We begin by narrowing our optical selection to luminous red galaxies (LRGs), which due to their high mass are the most common type of lens galaxy, and are often embedded in larger structures which can increase lensing probability.
We used the Dark Energy Spectroscopic Instrument Legacy Survey 9th data release \citep[LS-DR9,][]{Dey2019} as the optical survey. 
LS-DR9 covers nearly the entire sky at $|b| > 18^{\circ}$ in the $g,r,z$ bands down to a point source depth of $r\lesssim23\,$mag in the Legacy Survey northern fields ($\delta > +32^{\circ}, b>+18^{\circ}$) and $r\lesssim 23.5\,$mag in the southern sky.
Additionally LS-DR9 provides mid-infrared forced photometry from the unblended Wide-field Infrared Survey Experiment \cite[unWISE,][]{WISE, unWISE} bands. 
We follow the selection criteria of \citet{Zhou2020},  to select LRGs with high purity.
We then cross matched these with VLASS sources that were marginally resolved ($0 < \Psi < 0''.5$; where $\Psi$ is the major axis of the source after deconvolution from the beam), selecting only those matches that satisfy the angular separation and misalignment criteria used in JB07.

We are interested in radio emission from the background lensed source rather than extended radio lobes from the lens galaxy, the latter of which may mimic the configuration of lensed radio sources in catalog space and as such are a likely contaminant for this selection technique.
As radio lobes are expected to have steep radio spectra, a spectral index cut identifying only flat spectrum radio sources is a straightforward way to minimise such contamination of our sample.
We achieved this by estimating the $1.4\,\text{GHz} - 3\,\text{GHz}$ spectral index\footnote{throughout this work we adopt the convention relating flux density, $S$, and frequency, $\nu$, by $S\propto \nu^{\alpha}$}, $\alpha$,
using data from the NRAO VLA Sky Survey \citep[NVSS,][]{Condon1998}, and selecting only those sources with $-0.5 < \alpha < +0.5$.
In determining the spectral indices, VLASS flux measurements are scaled by $1/0.87$, as per the recommendation of \citet{Gordon2021}, to correct for the known underestimation of flux densities in the VLASS quick look catalog.
Furthermore, given the large difference in angular resolution between VLASS ($2''.5$) and NVSS ($45''$), spectral indices are only considered reliable when single VLASS sources are matched to an NVSS source.

\begin{figure*}
    \centering  
    \subfigure{\includegraphics[trim={10mm 10mm 0 0}, clip, width=0.23\textwidth]{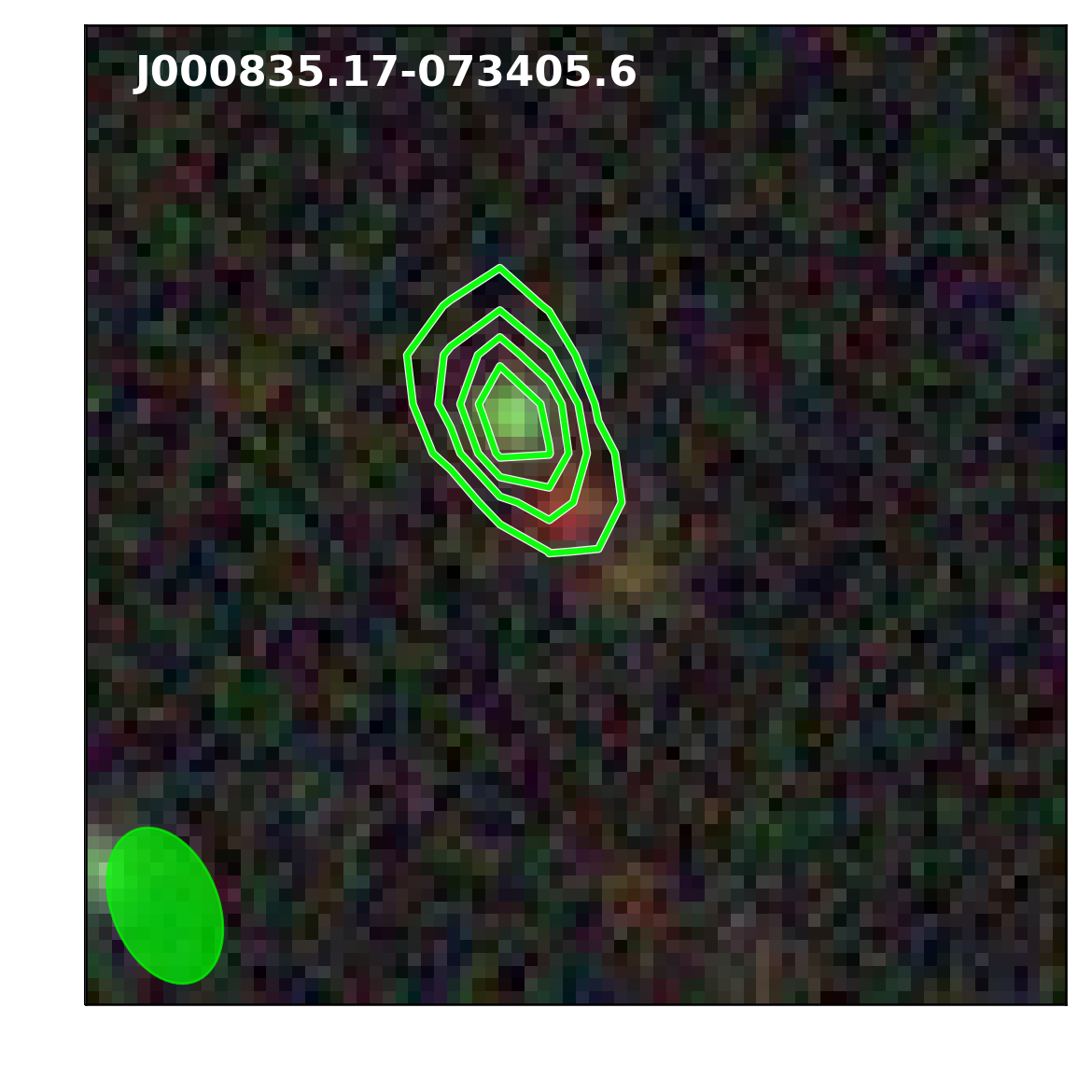}}
    \hspace{1mm}
    \subfigure{\includegraphics[trim={10mm 10mm 0 0}, clip, width=0.23\textwidth]{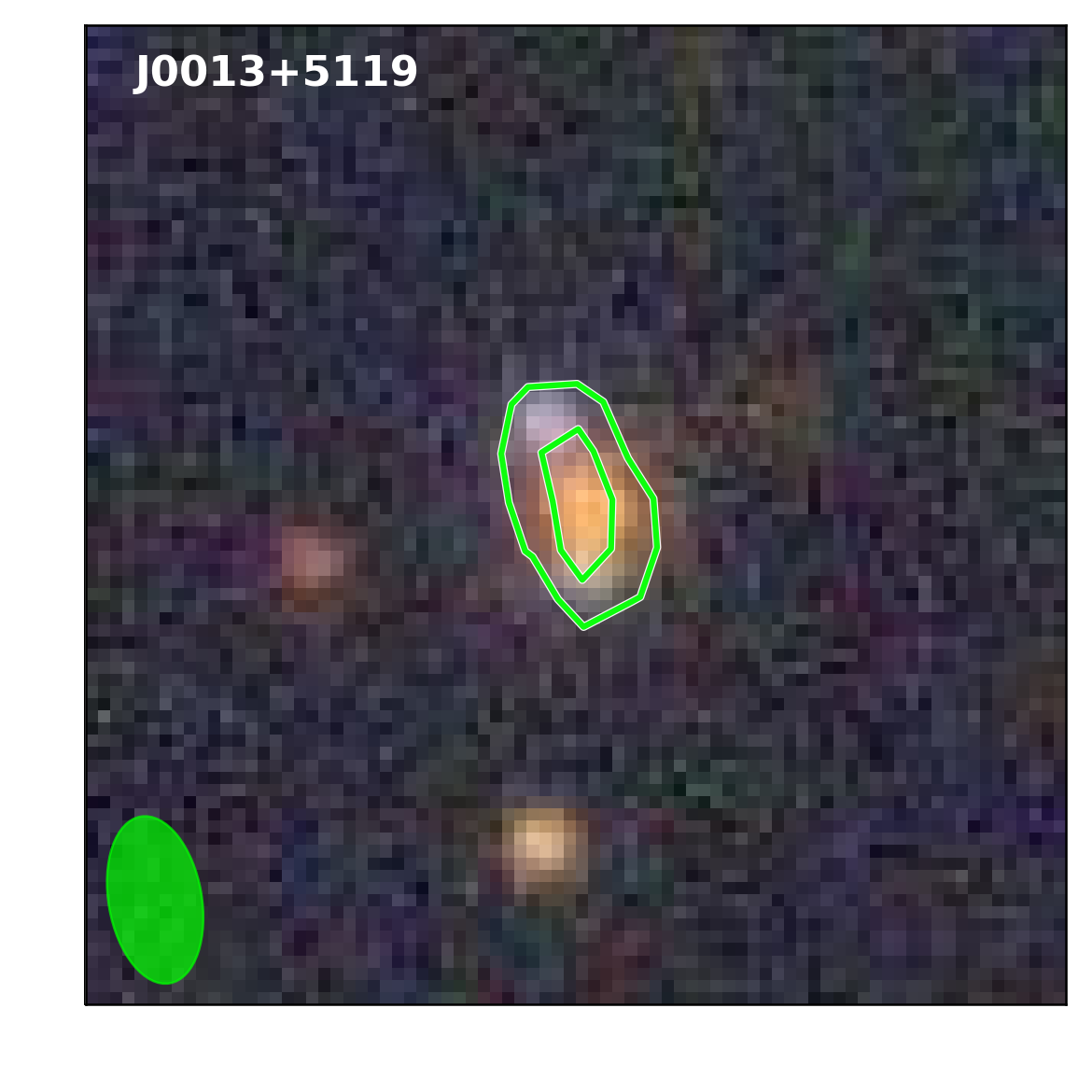}}
    \hspace{1mm}
    \subfigure{\includegraphics[trim={10mm 10mm 0 0}, clip, width=0.23\textwidth]{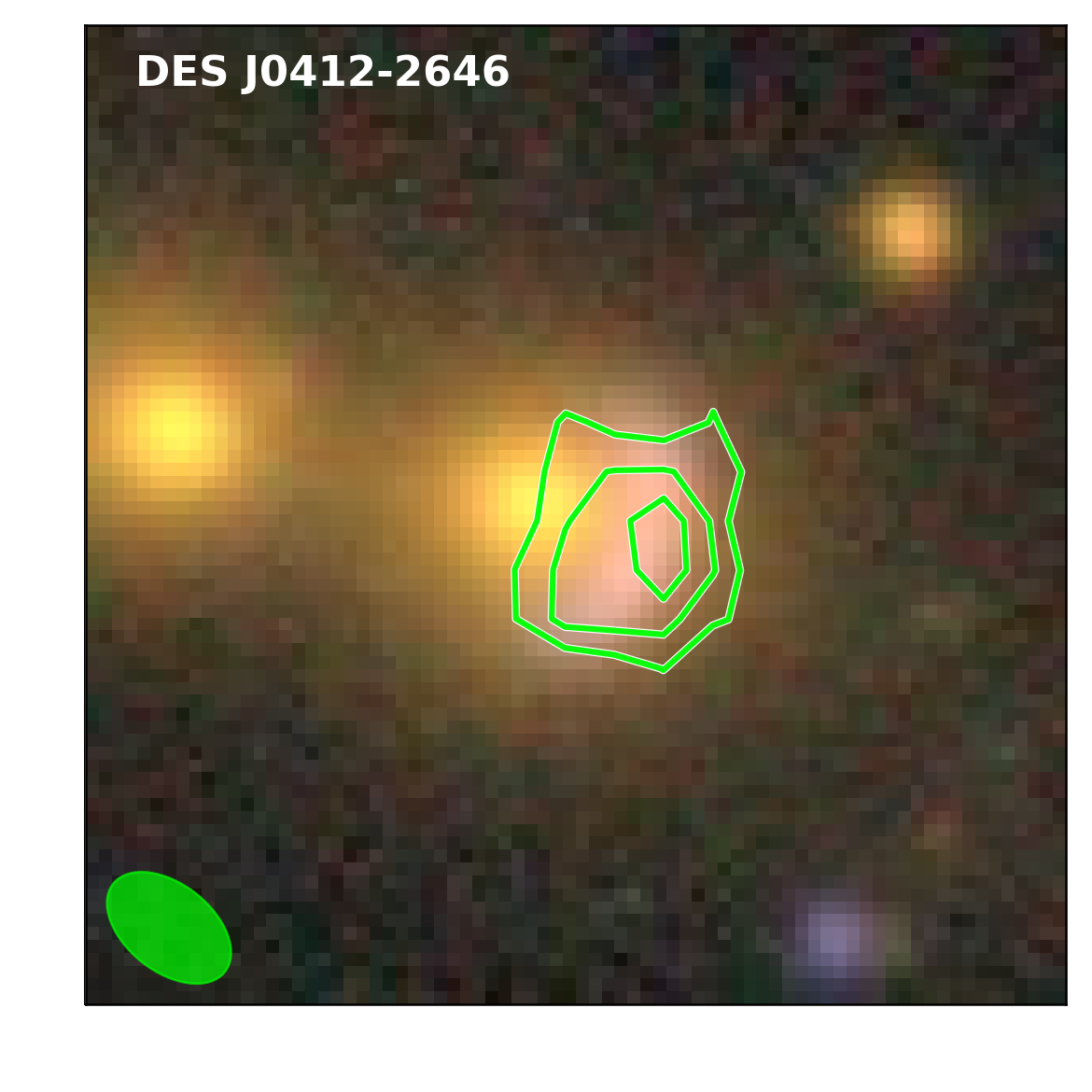}}
    \hspace{1mm}
    \subfigure{\includegraphics[trim={10mm 10mm 0 0}, clip, width=0.23\textwidth]{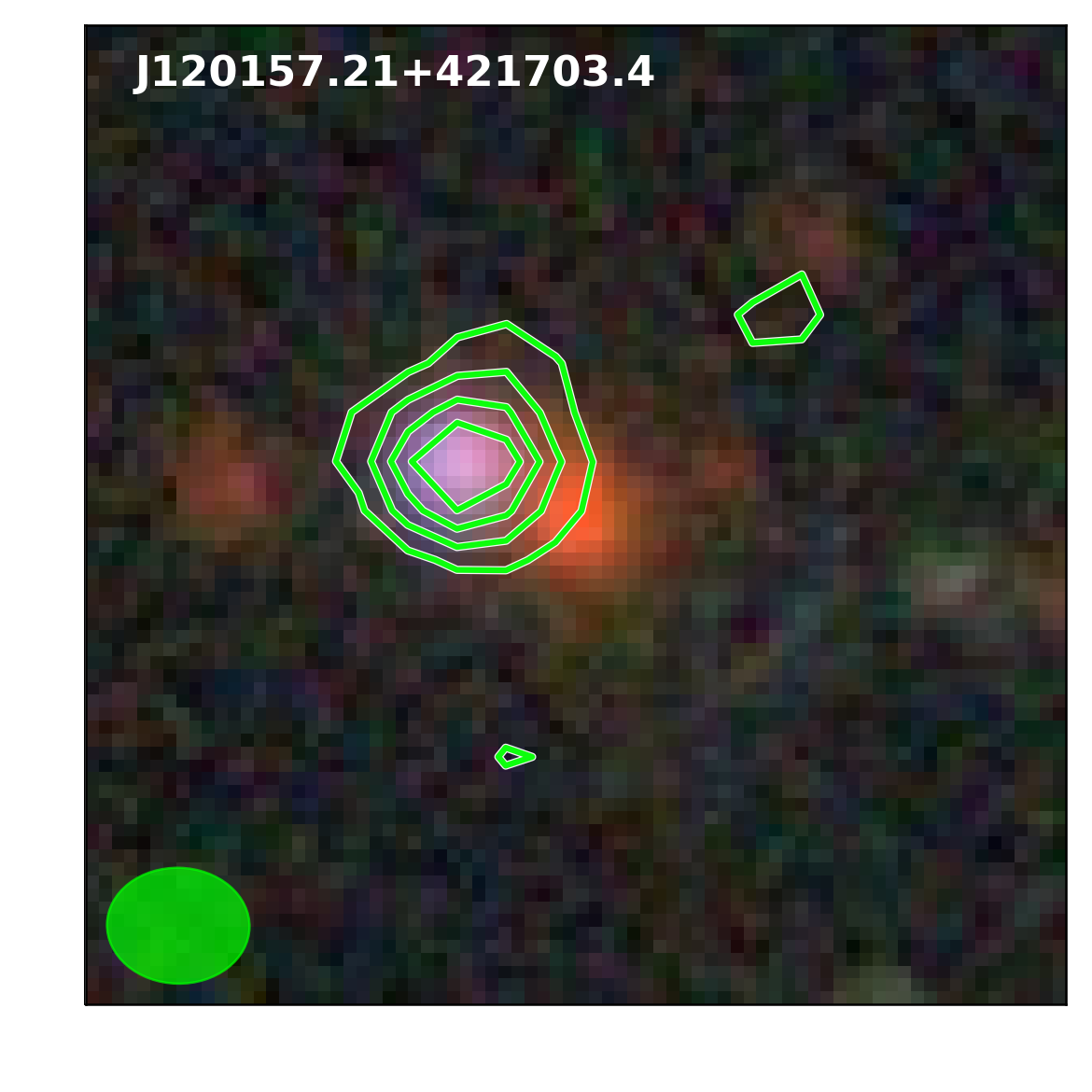}}
    \subfigure{\includegraphics[trim={10mm 10mm 0 0}, clip, width=0.23\textwidth]{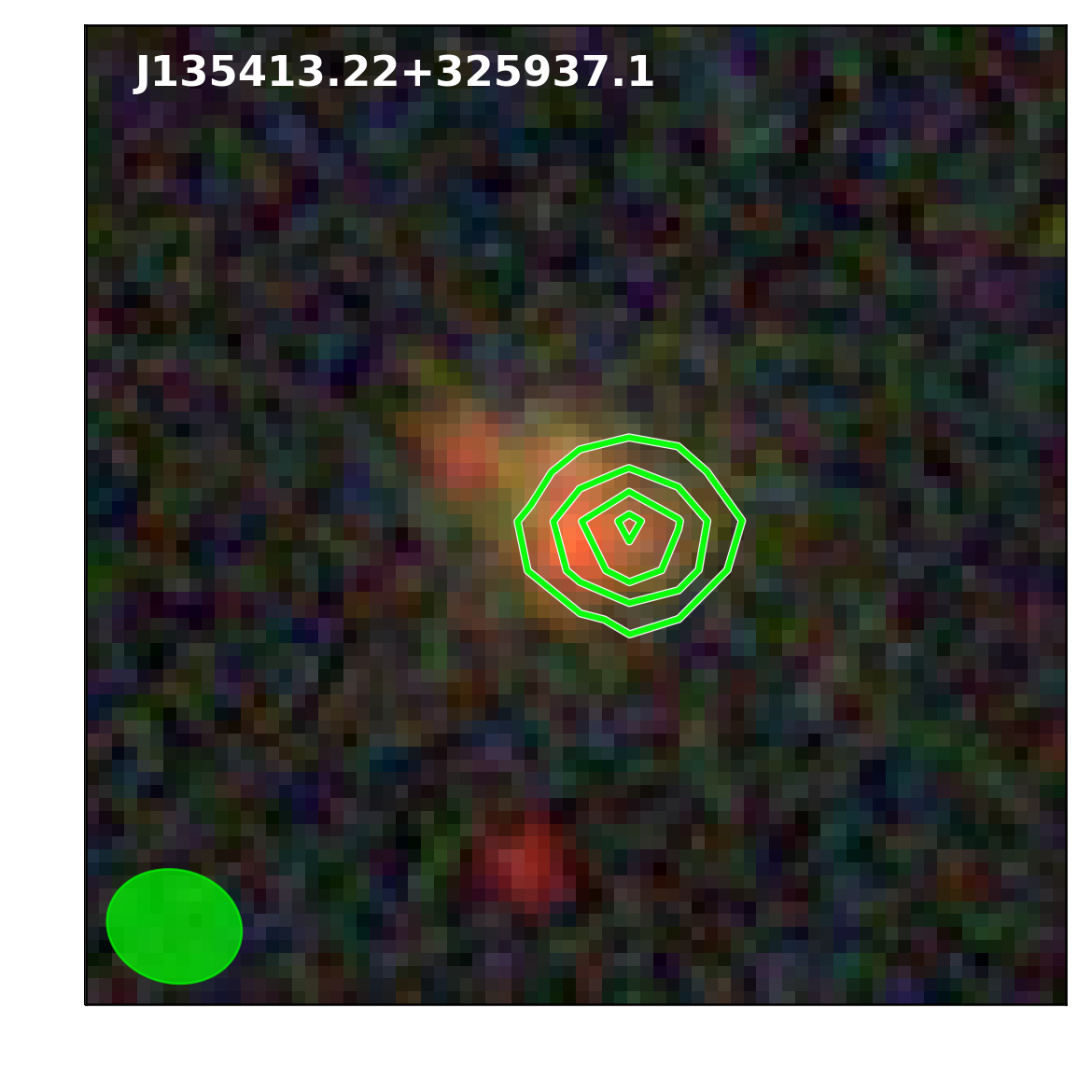}}
    \hspace{1mm}
    \subfigure{\includegraphics[trim={10mm 10mm 0 0}, clip, width=0.23\textwidth]{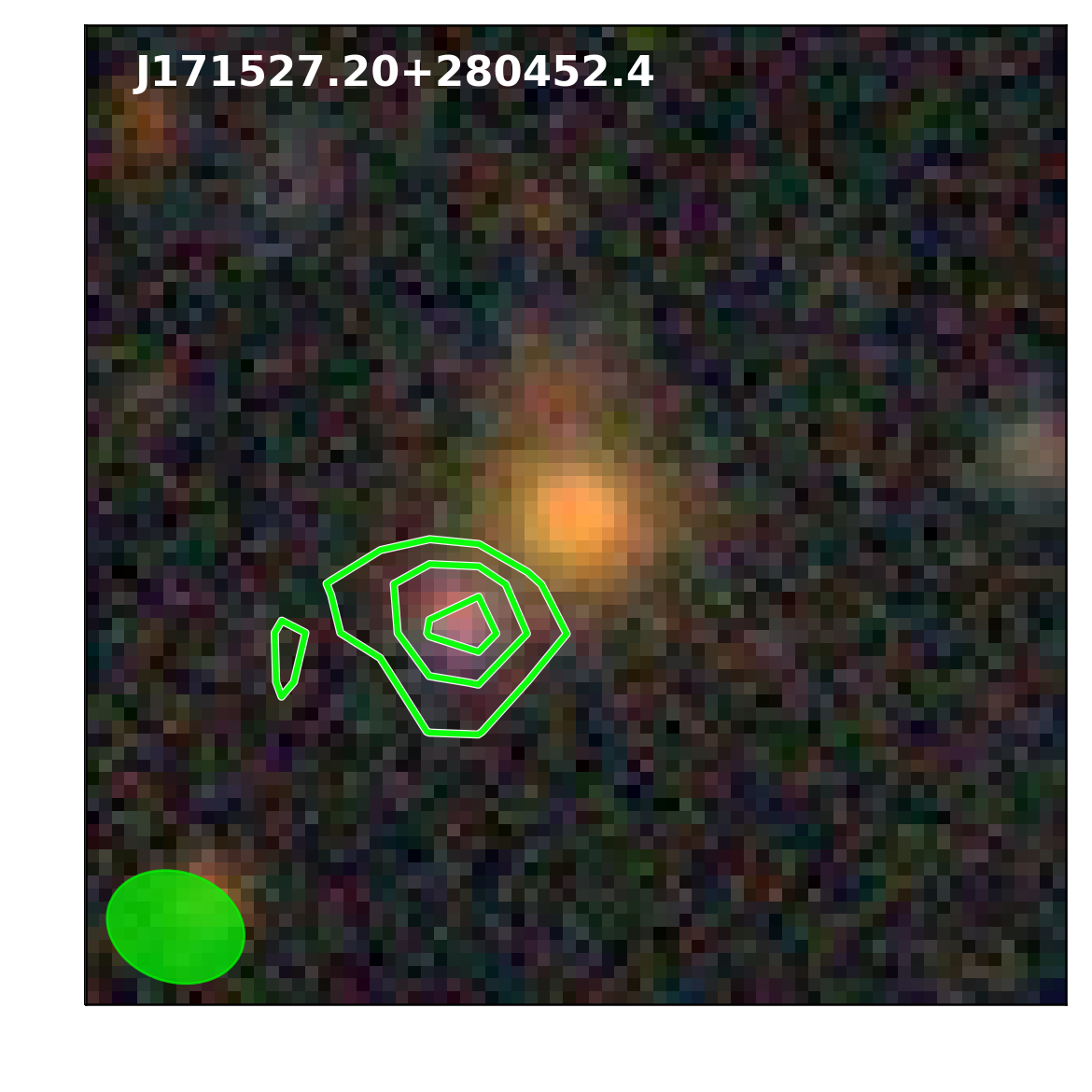}}
    \hspace{1mm}
    \subfigure{\includegraphics[trim={10mm 10mm 0 0}, clip, width=0.23\textwidth]{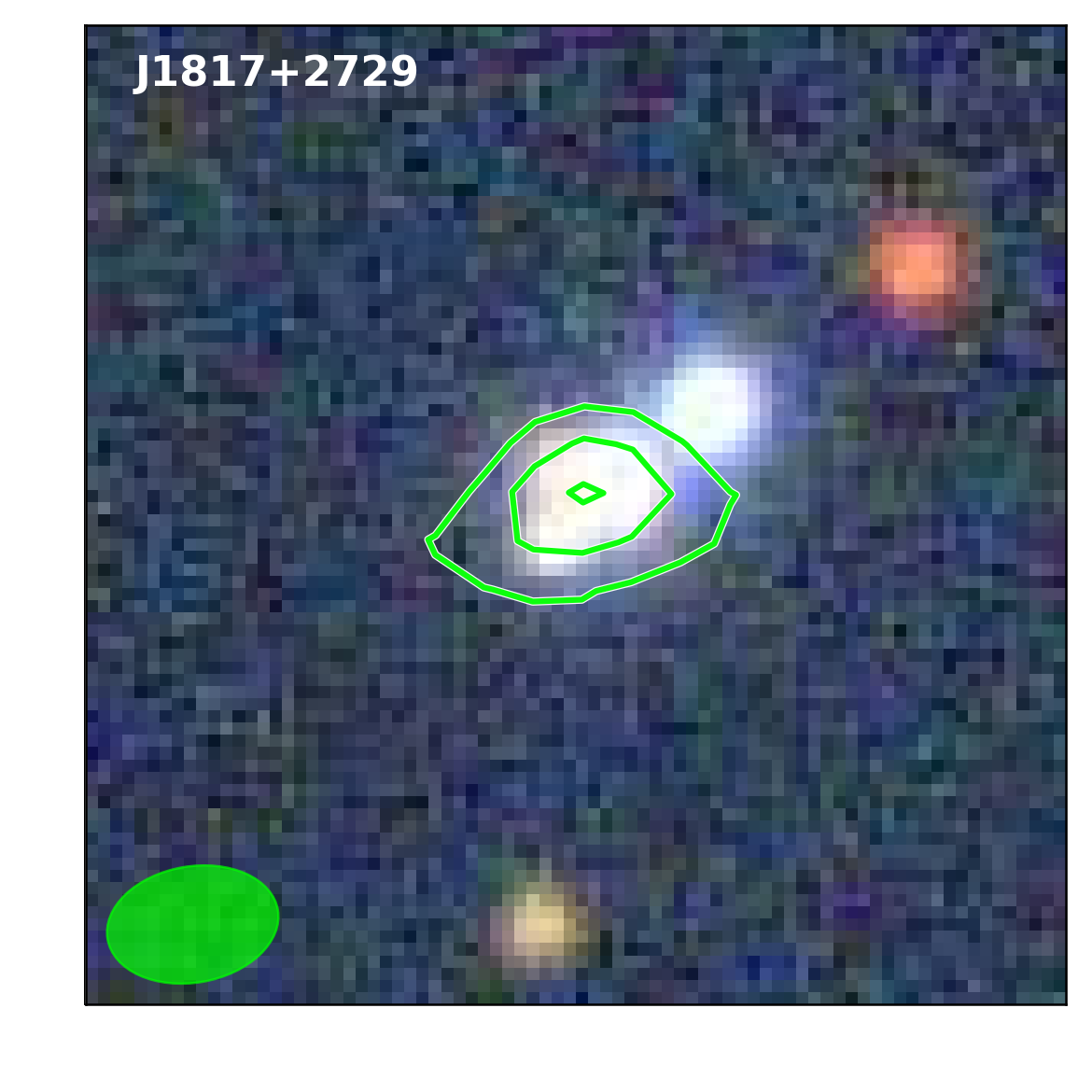}}
    \hspace{1mm}
    \subfigure{\includegraphics[trim={10mm 10mm 0 0}, clip, width=0.23\textwidth]{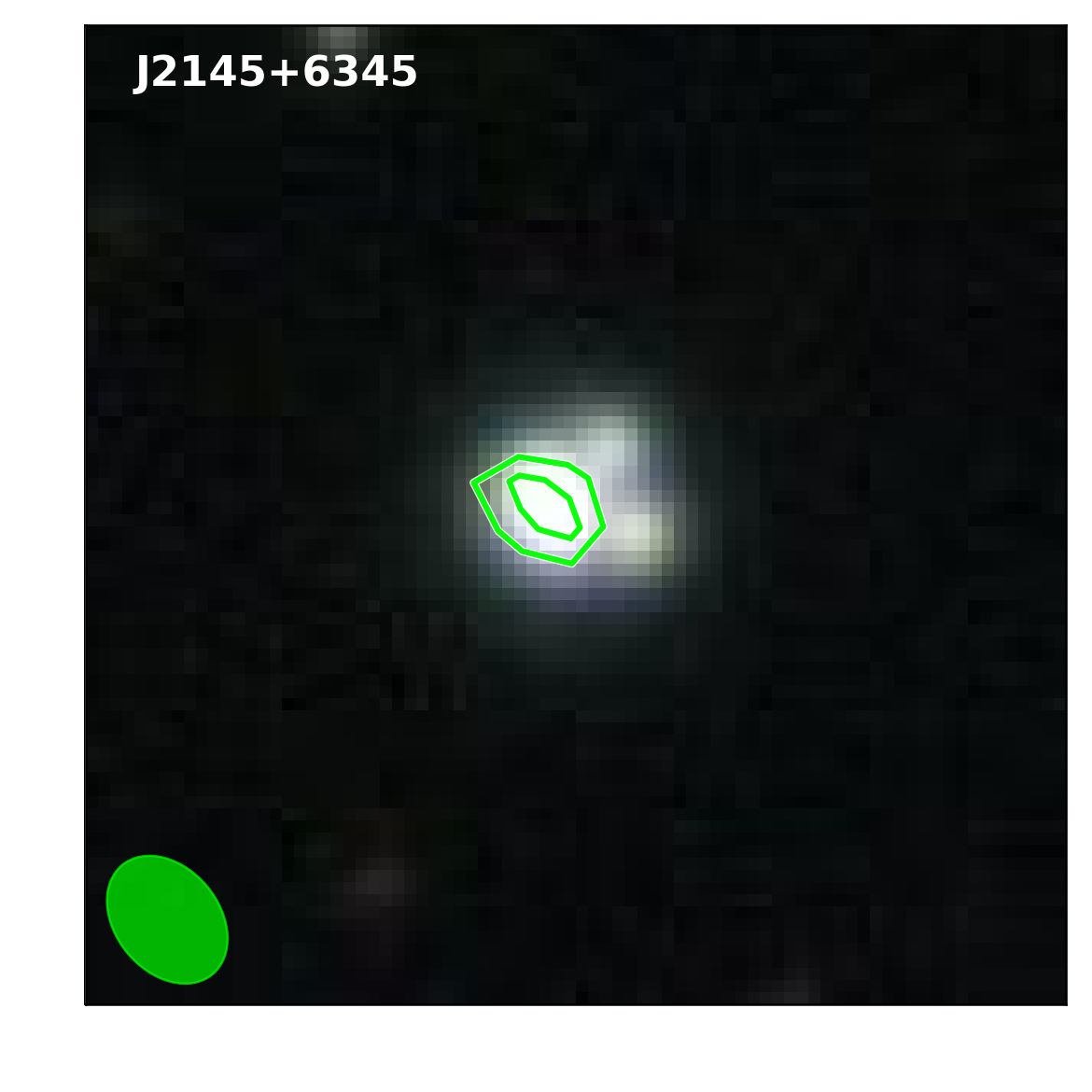}}
    \subfigure{\includegraphics[trim={10mm 10mm 0 0}, clip, width=0.23\textwidth]{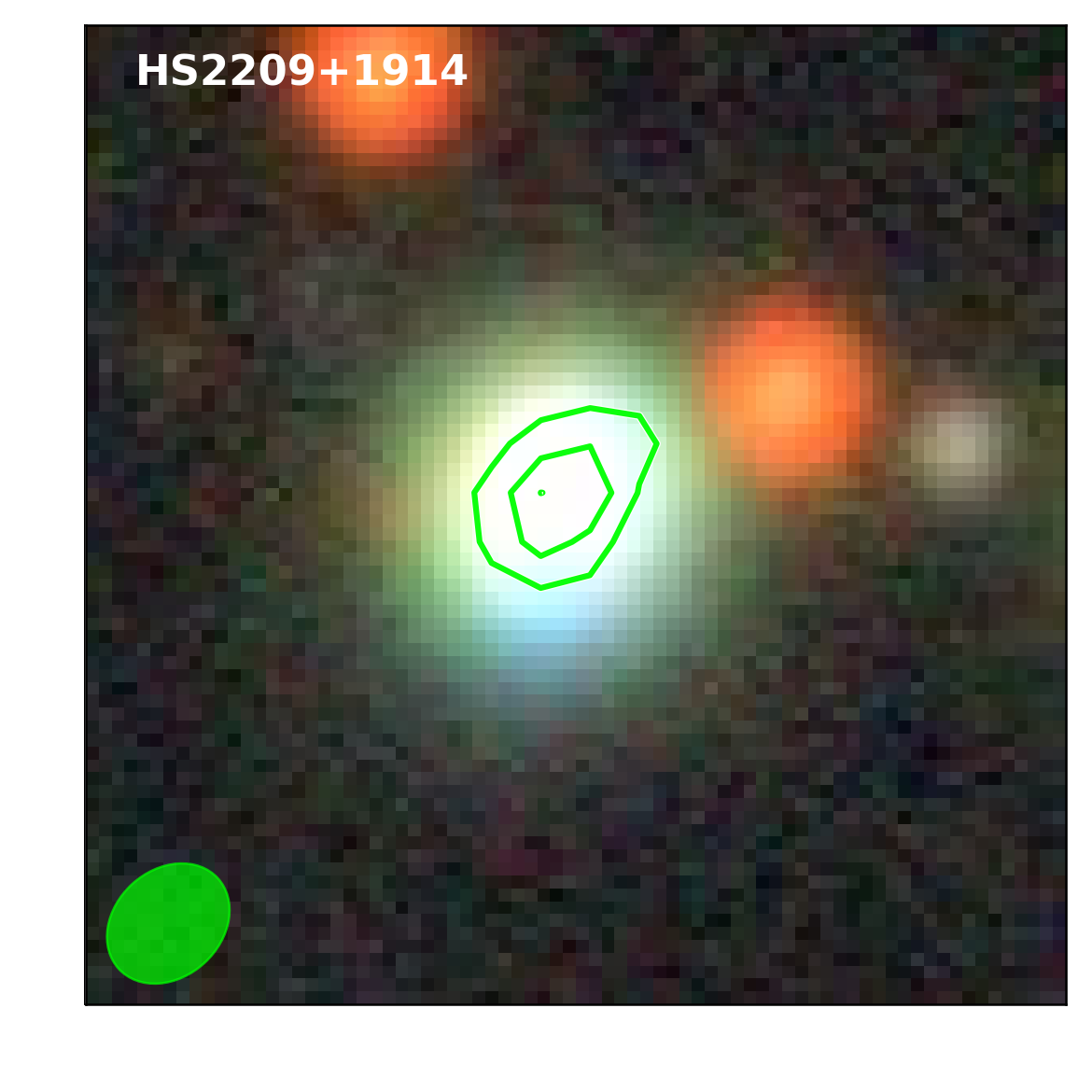}}
    \hspace{1mm}
    \subfigure{\includegraphics[trim={10mm 10mm 0 0}, clip, width=0.23\textwidth]{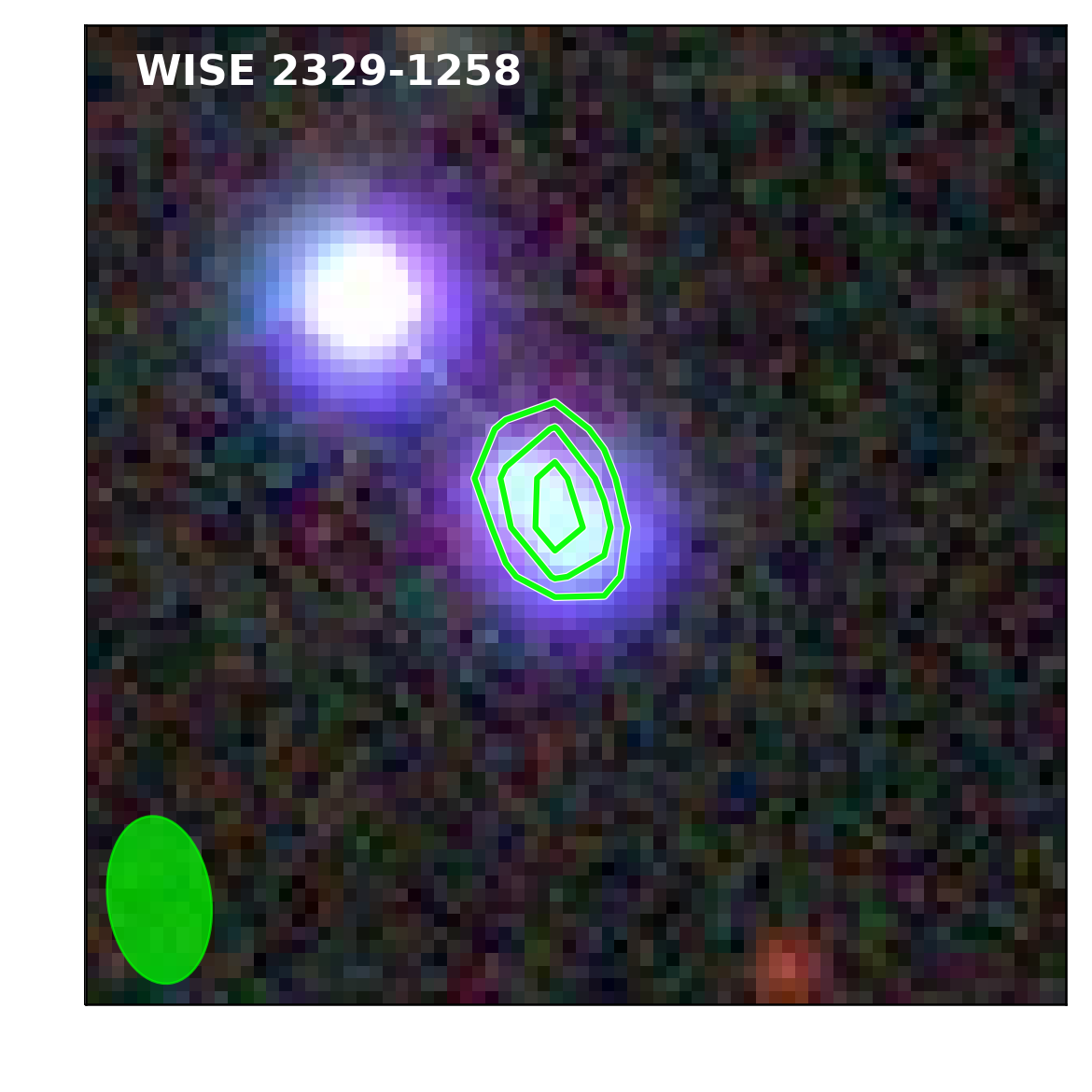}}
    \hspace{1mm}
    \subfigure{\includegraphics[trim={10mm 10mm 0 0}, clip, width=0.23\textwidth]{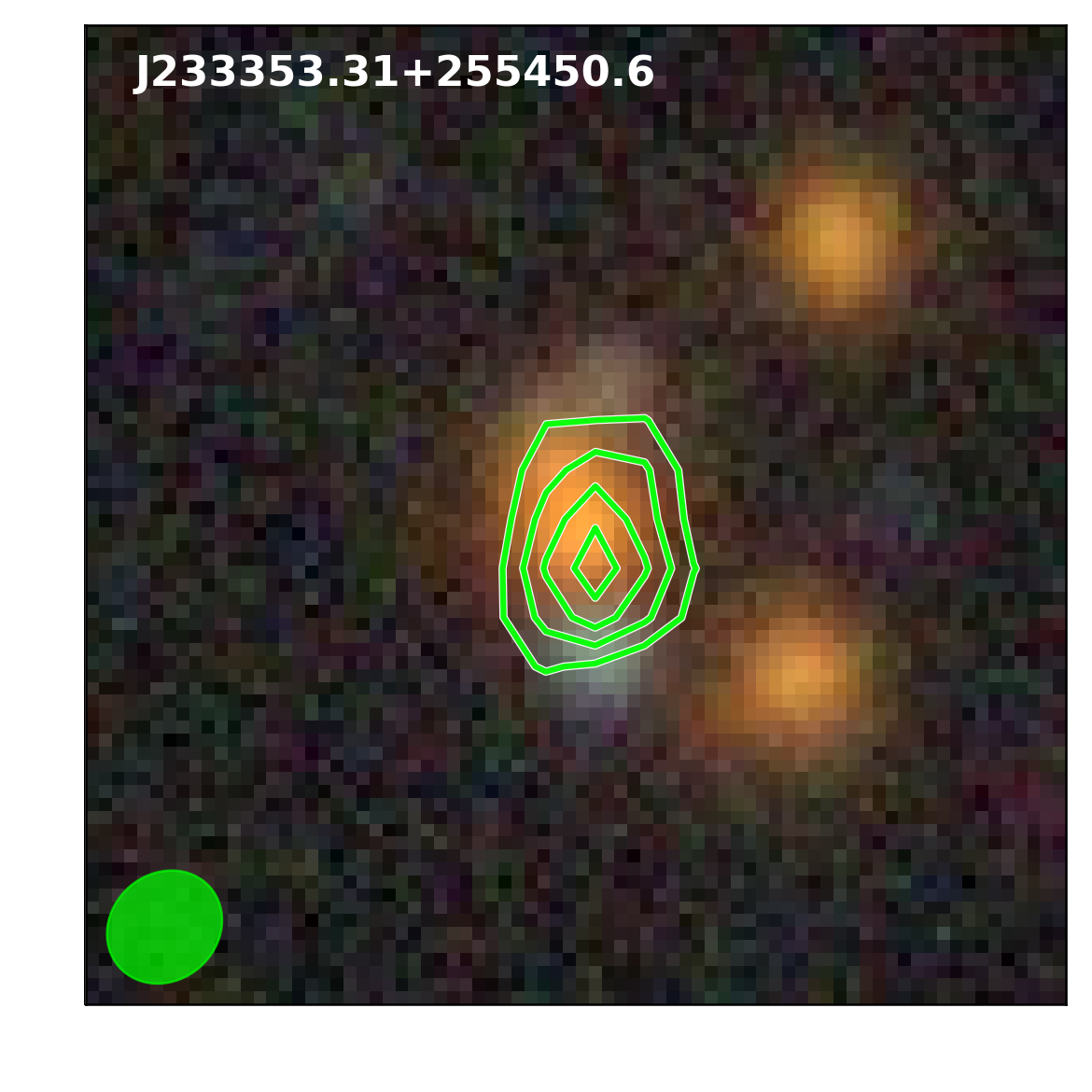}}
    
    \caption{Postage stamp optical cutouts ($20'' \times 20''$) of candidate lensed radio sources with VLASS contours overlaid (green).
    The optical images are three color (grz) images from LS-DR9 except for J0013$+$5119, J1817$+$2729 and J2145$+$6345 where PanSTARRS gri images are used instead.
    VLASS contour levels start at $0.5\,\text{mJy}\,\text{beam}^{-1}$ and increase in linear increments of $0.5\,\text{mJy}\,\text{beam}^{-1}$, except for the fainter radio sources DES J0412$-$2646, J2145$+$6345and WISE 2329$-$1258 where contours increase by $0.3\,\text{mJy}\,{beam}^{-1}$.
    The green ellipse in the lower left of each panel shows the VLASS beam.}
    \label{fig:eg_targets}
\end{figure*}

We find $\sim 1,700$ candidate radio lenses using this method.
We visually inspect these candidates to identify plausible targets, with $<1/60$ expected to yield a genuine lensed source \citep[][example rejects from this inspection are shown in Figure \ref{fig:eg_rejects}]{JB07}.
After visual inspection, we are left with only a handful of viable candidates.
Motivated by the desire to test the efficiency of our method while only using limited amounts of telescope time, we select the five most promising candidates for VLA follow-up.
%


\begin{deluxetable*}{lcccccr}
    \tabletypesize{\footnotesize}
    \tablecaption{Candidate lensed radio sources observed in this work, and a summary of VLA observations.
    \label{tab:observations}}
    \tablehead{\colhead{Name} & \colhead{$S_{\text{VLASS}}$} & \colhead{$\alpha_{1.4}^{3}$} & \colhead{Integration time} & \colhead{Image RMS} & \colhead{Calibrator} & \colhead{Identification method}\\
    \colhead{} & \colhead{[mJy]} & \colhead{} & \colhead{[s]} & \colhead{[$\mu\text{Jy}\,\text{beam}^{-1}$]} & \colhead{} & \colhead{}}
    \decimalcolnumbers
    \startdata
    J000835.17$-$073405.6 & 3.5 & +0.13 & 80 & 25 & J0006$-$0623 & JB07 method \\
    J0013+5119 & 2.4 & -0.53 & 130 & 21 & J2355+4950 & Gaia lensed QSO \\
    DES J0412$-$2646 & 2.8 & -0.30 & 90 & 30 & J0416$-$1851 & DES lensed galaxy \\
    J120157.21+421703.4 & 4.2 & +0.16 & 90 & 119 & J1146+3958 &  JB07 method \\
    J135413.22+325937.1 & 3.0 & -0.11 & 90 & 107 & J1416+3444 & JB07 method \\
    J171527.20+280452.4 & 2.6 & -0.31 & 100 & 107 & J1753+2848 & JB07 method \\
    J1817+2729 & 3.0 & +0.09 & 90 & 115 & J1753+2848 & Gaia lensed QSO \\
    J2145+6345 & 1.3 & $>$-0.60 \tablenotemark{a} & 299 & 15 & J2022+6136 & Gaia lensed QSO \\
    HS B2209+1914 & 2.3 & -0.88 & 219 & 17 & J2212+2355 & Gaia lensed QSO \\
    WISE J2329$-$1258 & 1.2 & -0.96 & 448 & 11 & J2331$-$1556 & Gaia lensed QSO \\
    J233353.31+255450.6 & 4.9 & -0.22 & 75 & \,\,\, 52 \tablenotemark{b} & J2340+2641 & JB07 method\\
    \enddata
    \tablecomments{This table lists (1) the name of the candidate lensed radio source; (2) the flux density in VLASS epoch 1; (3) the estimated spectral index between $1.4\,$GHz and $3\,$GHz based on measurements from VLASS and either FIRST or NVSS depending on sky location; (4) the time for which we observed the target; (5) the RMS noise of our cleaned image; and (6) the complex gain calibrator for that source.
    Column (7) notes whether the target was identified from known lensed quasars in Gaia \citep{Lemon19}, lensed galaxies in DES \citep{Jacobs2019} or by applying the method of \citet{JB07} to the VLASS and LS DR9 catalogs.}
    \tablenotetext{a}{J2145+6345 is not detected in NVSS and is outside the footprint of FIRST. As such we estimate a spectral index limit based on the $2\,$mJy detection limit of NVSS.}
    \tablenotetext{b}{The image RMS for J233353+255450 is given for the uv-tapered image, see Section \ref{2333}.}
\end{deluxetable*}

\section{VLA Observations} \label{sec:obs}

Using our two selection approaches and removing those previously known lensed radio sources and those targets for which there are high resolution observations in the VLA archive, we are left with 11 targets.
We show optical images overlaid with VLASS contours in Figure \ref{fig:eg_targets}.
We observe these targets with the Karl G. Jansky Very Large Array (VLA) in A-configuration (VLA Proposal: 23A-249).
Observations were conducted in the X-band using NRAO default correlator setup \texttt{X32f2A}, corresponding to 3-bit sampling and 2 second integration times, and basebands centered at 9 and 11 GHz and 2GHz bandwidth. 
The primary calibrator used was 3C 48 for all targets except J120157.21+421703.4, 135413.22+325937.1, 171527.20+280452.4, and J1817+2729, which used 3C 286.
Table \ref{tab:observations} shows our target list with the VLA integration times and complex calibrators used, alongside some of their selection criteria. 
The raw data was calibrated by the NRAO as part of the Science Ready Data Products (SRDP) initiative, which creates calibrated measurement sets optimized for continuum (Stokes I) imaging.

We imaged our visibilities using the \texttt{tclean} task in the NRAO's Common Astronomy Software Applications (CASA) suite of processing tools \citep{casa}.
As lensed radio quasar systems tend to be composed mainly of point sources, we used the \citet{hogbom} deconvolution method with single-term multi-frequency synthesis \citep{mfs}.
Given our targets are never more than a few arcseconds across, we did not use any wide-field imaging procedures.
Cleaning was done using an interactive mask, with a stopping threshold of 0.1mJy, which was usually between two and five times the noise floor.
After imaging, model visibilities were examined in order to assess the efficacy of increasing dynamic range via self-calibration \citep{selfcal}, but in each case our snapshot observation signal-to-noise was too low for a useful gain solution.
In a few cases this general imaging procedure was augmented with extra steps as required by the situation, these will be discussed individually in the following section.

\section{Results} \label{sec:results}

Table \ref{tab:photometry} summarizes the targeted VLA observation results, including the position and flux of each detected radio component.
These were calculated with the CASA task \texttt{imfit}, which fits elliptical gaussians to image-plane radio maps.
As expected for AGN cores, most observed components were fit as point sources with some exceptions noted in the table and discussed below.
Of the 11 targets observed, we found evidence of lensed radio emission in 4 previously known lensed quasar systems. The fifth previously known system studied was not detected but was found to be a lensed radio source by \citet{dobie23}. In 4 other sources, we found unlensed radio emission; we attribute the VLASS emission to either the putative lens galaxy or an unlensed quasar. Another source had no significant detection whatsoever, and the final source is an ambiguous case discussed further in Section \ref{0412}.

\begin{deluxetable*}{lcccccC}
\tabletypesize{\footnotesize}
\tablecaption{Position and flux measurements of detected radio components from targeted VLA observations. \label{tab:photometry}}
\tablehead{\colhead{Target} & \colhead{Component} & \colhead{RA} & \colhead{$\sigma$RA} & \colhead{Dec} & \colhead{$\sigma$Dec} & \colhead{Flux Density} \\
\colhead{} & \colhead{} & \colhead{[deg]} & \colhead{[mas]} & \colhead{[deg]} & \colhead{[mas]} & \colhead{[$\mu$Jy]}}
\startdata
J000835.17-073405.6 & Single Quasar & 2.1465188 & 2 & -7.5683175 & 2 & 2370 \pm 40 \\
J0013+5119 & A & 3.348415 & 14 & 51.318736 & 10 & 240 \pm 40 \\
 & B & 3.348112 & 5 & 51.3179497 & 3 & 250 \pm 30 \\
 & Lens Galaxy & 3.348073 & 4 & 51.3182923 & 3 & 490 \pm 30 \\
DES J0412-2646 & North & 63.179016 & 11 & -26.775585 & 23 & 160 \pm 40 \\
 & South \tablenotemark{a} & 63.179 & 26 & -26.77575 & 66 & 260 \pm 90 \\
J120157.21+421703.4 & Single Quasar & 180.4883875 & 1 & 42.2842668 & 12 & 7000 \pm 200 \\
J135413.22+325937.1 & Single Quasar & 208.5550522 & 2 & 32.993648 & 4 & 2800 \pm 200 \\
J171527.20+280452.4 & Not Detected &  &  &  &  &  \\
J1817+2729 & Not Detected &  &  &  &  &  \\
J2145+6345 & A & 326.2717218 & 7 & 63.7613599 & 2 & 430 \pm 30 \\
 & B & 326.27193 & 10 & 63.76152 & 4 & 250 \pm 30 \\
 & C & 326.270737 & 25 & 63.761261 & 7 & 130 \pm 30 \\
HS B2209+1914 & A & 332.876315 & 13 & 19.487111 & 7 & 290 \pm 30 \\
 & B & 332.876415 & 21 & 19.48684 & 12 & 270 \pm 40 \\
WISE J2329-1258 & A & 352.49105 & 47 & -12.98315 & 96 & 160 \pm 50 \\
 & Northeast \tablenotemark{a} & 352.491016 & 30 & -12.98286 & 63 & 80 \pm 20 \\
 & Southwest \tablenotemark{a} & 352.49132 & 99 & -12.98298 & 223 & 150 \pm 70 \\
J233353.31+255450.6 & Extended Emission \tablenotemark{a} & 353.47219 & 26 & 25.91395 & 146 & 1700 \pm 200 \\
\enddata
\tablenotetext{a}{This component was fit as an extended source by \texttt{imfit} rather than a point source. }
\tablecomments{Lensed quasar images are indicated by capital letters.}
\end{deluxetable*}

\subsection{Statistical Considerations} \label{statistics}
For our observations, especially those which we claim are indeed radio-loud lenses, we wish to reject the possibility that the radio emission is indeed from the quasar and lens separately, rather than a chance alignment of radio sources and optical ones.
We adopt a frequentist approach based on \citet{skacatalog} to give the probability each radio detection is associated with its corresponding optical detection.
The \textit{Gaia} survey has the sky coverage, sensitivity, and resolution necessary to detect in the optical all the radio quasar images we observed, and so was used as our optical survey for this analysis.
Let $\rho_0$ be the density of optically detected sources, which in the case of \textit{Gaia} DR3 \citep{Gaiadr3} is approximately $45,000\,\text{deg}^{-2}$.
Assuming no correlation between radio and optical, the number of expected optical detections within $r$ seconds of arc of a given radio detection is given by
$\int_0^r \rho_0(2\pi r') dr'$, or $\rho_0(\pi r^2)$. 
As \textit{Gaia}'s astrometric precision is typically less than one milliarcsecond, and our VLA precision (in A-config X-band) is on the order of $20\sim40$ milliarcseconds, a typical value of $r$ is expected to be tens of milliarcseconds for a real match, corresponding to an individual source random probability of between $10^{-5}$ and $10^{-7}$.
By contrast, two unrelated sources separated by $1''$ would give a random probability of closer to 1/100.
We expect for a real radio lens to observe emission from each quasar image, and thus will measure a random probability for each of them.
Multiplying these probabilities together gives an estimated total probability that the radio sources are chance alignments with the lensed optical images, and we will report this number for each claimed radio-loud gravitational lens in the next section.

\begin{figure*}
    \centering
    \subfigure{\includegraphics[trim={36mm, 12mm, 33mm, 12mm}, clip, width=0.35\textwidth]{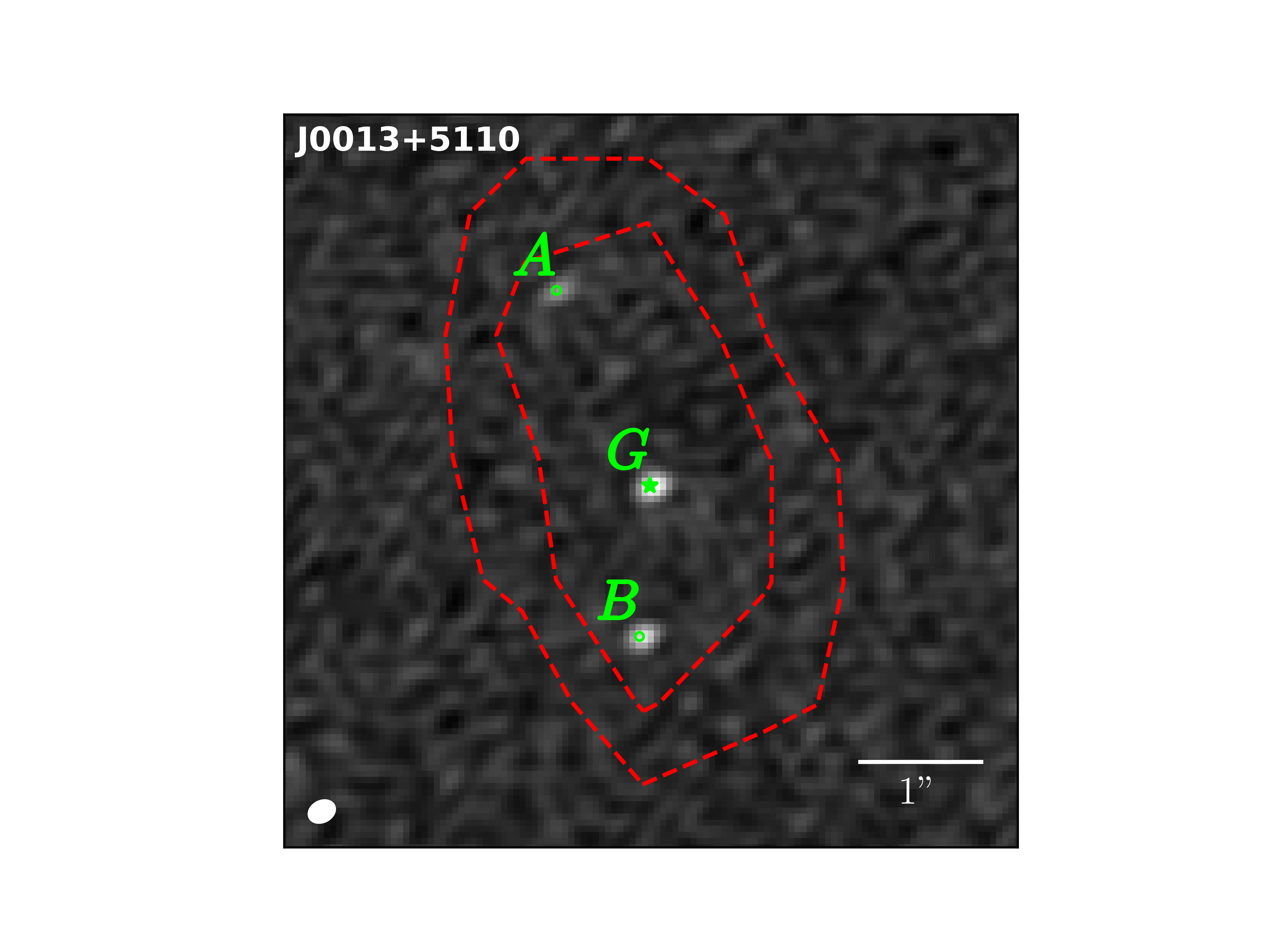}}
    \hspace{4mm}
    \subfigure{\includegraphics[trim={36mm, 12mm, 33mm, 12mm}, clip, width=0.35\textwidth]{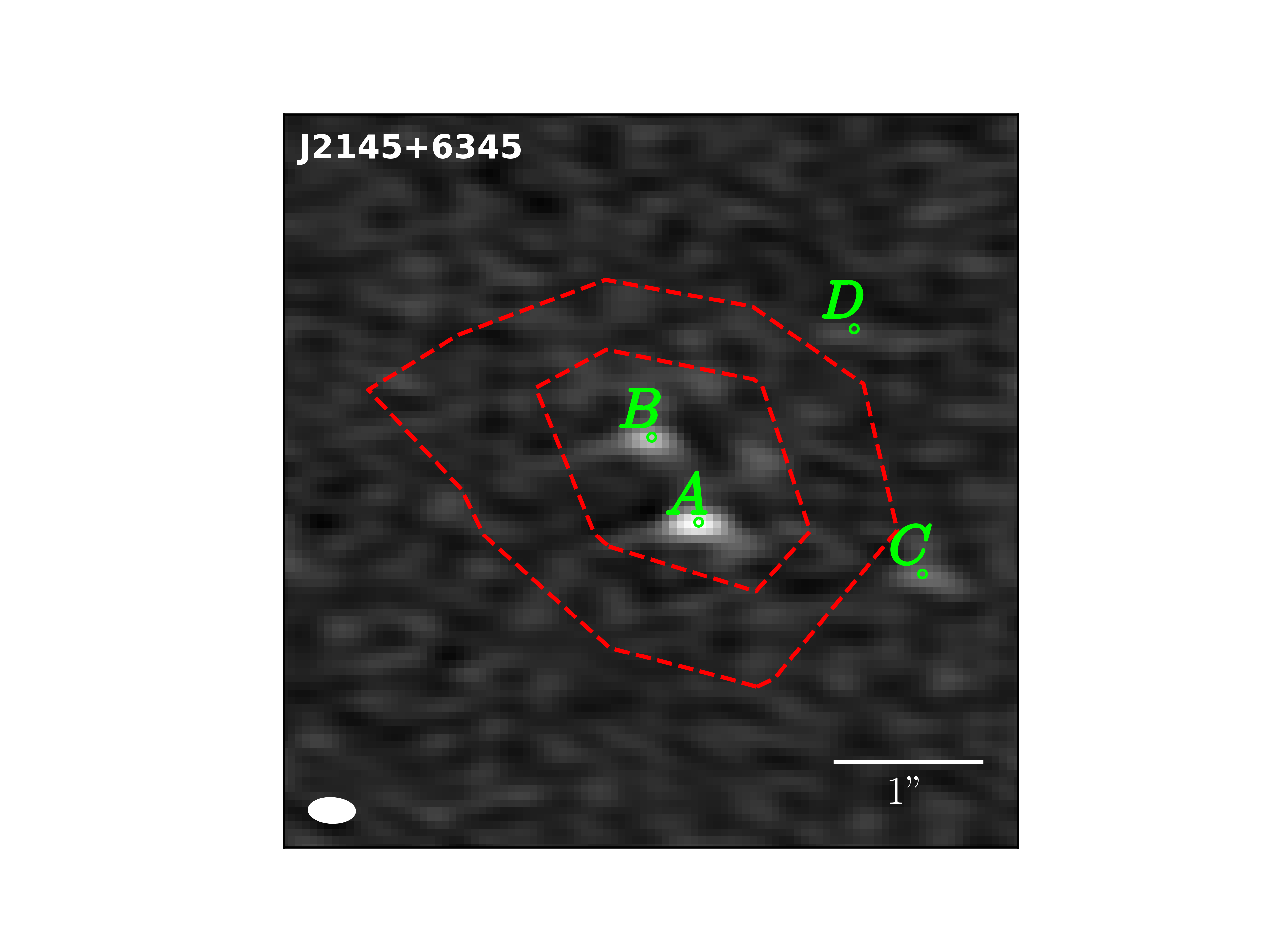}}
    \subfigure{\includegraphics[trim={36mm, 12mm, 33mm, 12mm}, clip, width=0.35\textwidth]{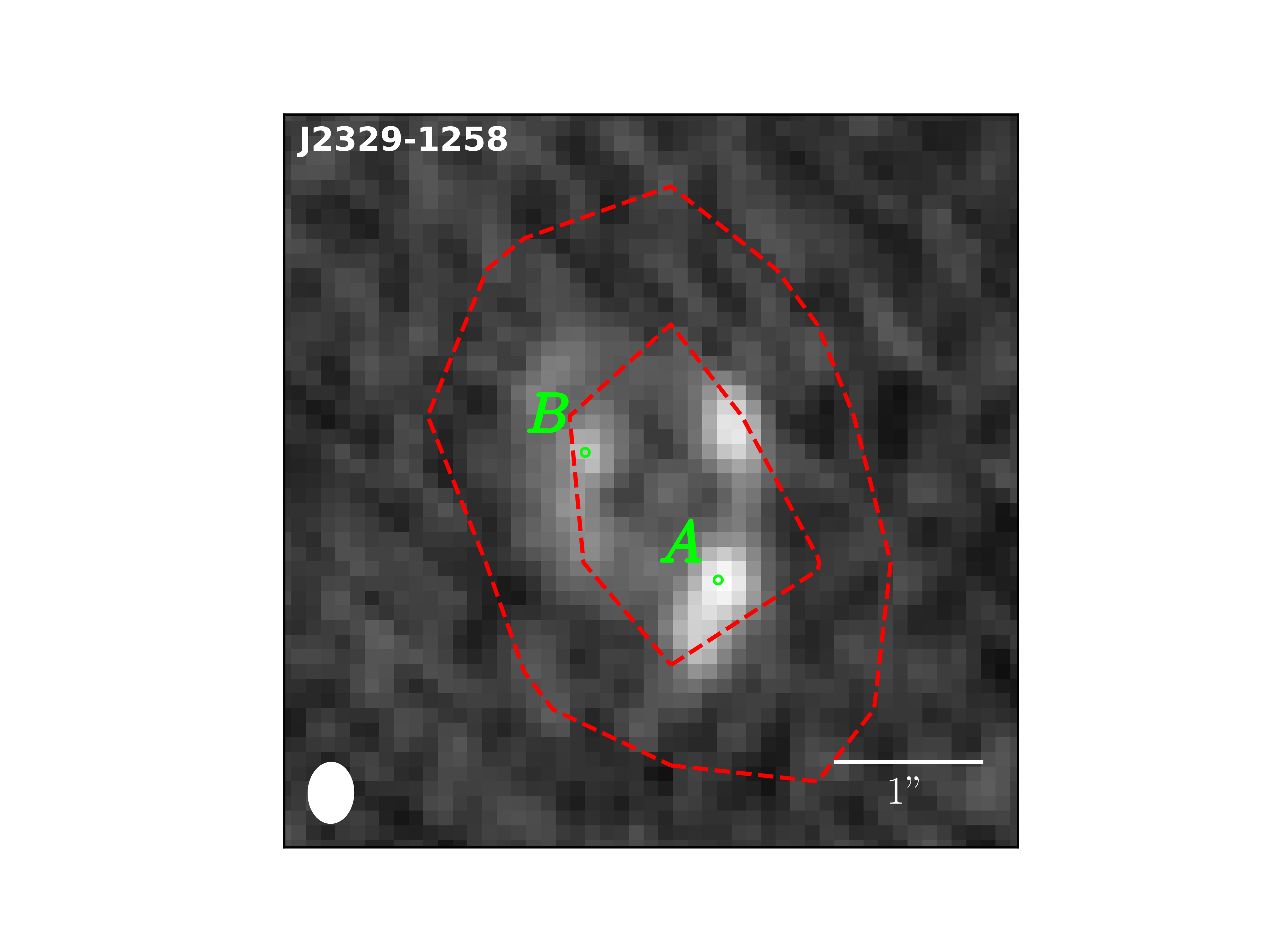}}
    \hspace{4mm}
    \subfigure{\includegraphics[trim={36mm, 12mm, 33mm, 12mm}, clip, width=0.35\textwidth]{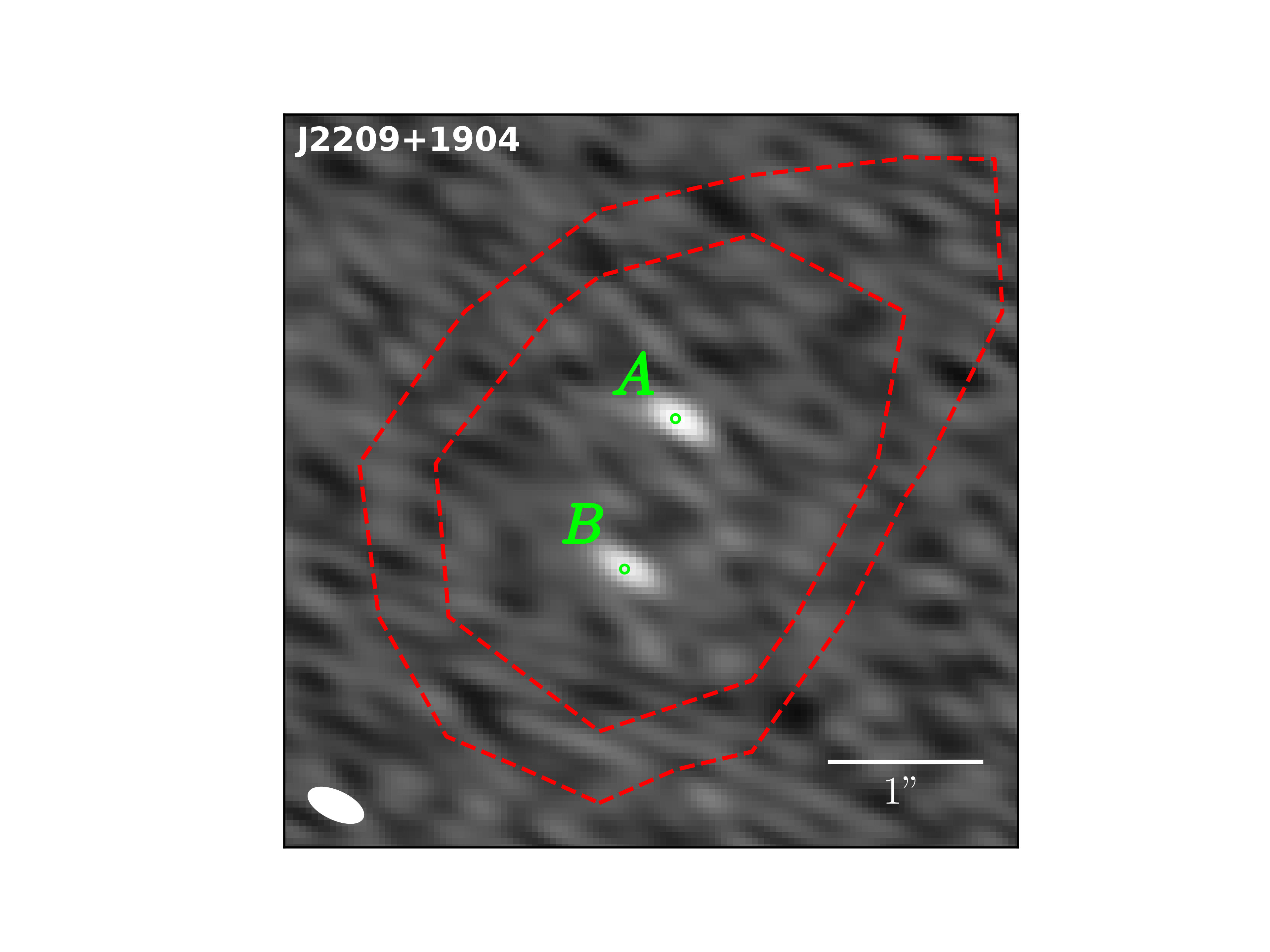}}
    \caption{Radio images of the five lenses discussed in this paper. Clockwise from top left: J0013+5119 (see Section \ref{0013}), J2145+6345 (Section \ref{2145}), HS B2209+1904 (Section \ref{2209}), and J2329$-$1258 (Section \ref{2329}). The final image is constructed using additional archival data as described in the text. These systems are not resolved in the VLASS epoch one quick-look images, as shown by the red contours (3- and 5-$\sigma$ levels shown). The targeted VLA observations show multiple source images coinciding with \textit{Gaia} positions, shown in lime. For the source J0013+5119, the PanSTARRS position of the lens galaxy is also shown.}
    \label{fig:lensfig}
\end{figure*}

\subsection{New Radio-Loud Lenses}

The radio loud lenses presented below are displayed in Figure \ref{fig:lensfig}.

\subsubsection{J0013+5119} \label{0013}
J0013+5119 was discovered as a doubly imaged quasar by \citet{Lemon19} using the Wide-field Infrared Survey Explorer (WISE) and \textit{Gaia} DR2 catalogues. The quasar source is at redshift $z=2.63$ and the two images are separated by $2''.92$.
Our VLA A-config observations revealed radio emission from both the lens and the quasar images, which are all fit to point sources in X-band using the CASA task \texttt{imfit} but appear as one source in VLASS.
The flux densities for the lensed images $A$ and $B$ taken from \texttt{imfit} give a flux ratio $ A/B = 1.0 \pm 0.2$, similar to the optical flux ratio of $1.2$ based on the \textit{Gaia} \text{$g$-band} measurements.
The lens galaxy was not detected in \textit{Gaia} and was too blended with quasar light to get a precise position measurement in any other available optical survey, so our probability consideration only includes the two measured quasar positions.
The system probability of random coincidence is then $3 \times 10^{-11}$.

As this is the only one of our new lenses to not have a published lens model, we made an effort to provide one in this paper. 
Using the Lenstronomy \citep{lenstronomy, lenstronomy2} software suite, we fit a simple Singular Isothermal Ellipsoid \citep{SIE} model with external shear to both VLA data and data from the PanSTARRS 1 survey  \citep[PS1,][]{Chambers2016}.
However, when testing our best-fit results from this method, we found the source plane positions of images A and B did not match, i.e. the model was not accurately reproducing observations.
We suspect this is due to the environment of the lens, and examining wider-field survey images of the J0013+5119 system show other galaxies of similar redshift in the vicinity of the lens, which could lead to a more complex lens model.
Modeling such a lens system would require deeper and sharper optical data and is beyond the scope of this paper.

\subsubsection{J2145+6345} \label{2145}
Quad lens J2145+6345 was also discovered by \citet{Lemon19} using the same method as J0013+5100, and was singled out by the authors as being ideal for time-delay studies given its reasonably large image separation (a max of $2".07$) and bright images. The quasar is located at $z=1.56$, and \citet{Lemon19} report no detection of a lens galaxy in the PanSTARRS survey.

We significantly detected the three brightest images of J2145+6345 in X-band as point sources, and also detected a noise bump coincident with the \textit{Gaia} position of the fourth quasar image. In VLASS, the system is blurred together into one component.
Excluding the faintest image, which was not significantly detected, we obtained a system chance of random of $3\times10^{-15}$.
We calculated the flux ratios between our significantly detected images as $A/B = 1.7 \pm 0.2$ and $A/C = 3.3 \pm 0.5$.
These radio flux ratios do not differ significantly from the \textit{Gaia} \text{$g$-band} flux ratios of $A/B = 1.4$ and $A/C = 3.9$.

\subsubsection{HS B2209+1904} \label{2209}

B2209+1904 (aka J2211+1929), a doubly-imaged quasar at $z=1.07$, was catalogued, along with its lens galaxy, in the Hamburg Quasar Survey \citep{hamburg}.
Our X-band observations detected both quasar images as point sources with a flux ratio of $A/B = 1.1 \pm 0.2$.
This is slightly, but not significantly, lower than the optical flux ratio observed by \textit{Gaia} in the \text{$g$-band} of $1.5$.
The chance of two random radio sources being in these positions is $7\times 10^{-12}$.

\subsubsection{J1817+2729} \label{1817}

J1817+2729, a quadruply imaged source at $z=3.07$ \citep{Lemon19}, was discovered by \citet{1817} using a blind catalog search in \textit{Gaia} DR2. 
Despite a strong detection in VLASS, our X-band observations report no significant emission at $10\,$GHz, and a manual re-reduction of the data showed the same.
Fortunately, the target was also observed by \citet{dobie23} in C band (6 GHz), and was confirmed as a lensed radio source therein.
J1817+2729 shows no variability between epochs 1 (May 2019) and 2 (Sept 2021) of VLASS, so we assume no significant variability for the source. From VLASS epoch 1 and the summed flux densities of all images in the C-band by \citet{dobie23}, we estimate a spectral index between 3 GHz and 6 GHz of $\alpha_{3\,\text{GHz}}^{6\,\text{GHz}} = -1.6\pm0.2$, substantially steeper than the relatively flat spectrum estimated from NVSS and VLASS.
This may be the result of genuine spectral curvature\textemdash for instance the spectral index between $1.4\,$GHz and $3\,$GHz might be capturing the spectral turnover of a peaked spectrum radio source \citep[e.g.,][]{ODea2021}. 
Extrapolating the C-band flux density to X-band using $\alpha_{3\,\text{GHz}}^{6\,\text{GHz}}$, we would expect the sum of the lensed images to have $S_{10\,\text{GHz}} \approx 440\,\mu$Jy. With the distribution of image brightness reported in \citet{dobie23}, we would expect the brightest lensed image to have a 10 GHz flux density of $\approx 230\,\mu$Jy, corresponding to a $<2\sigma$ detection in our image.
We conclude that our observations were simply not sensitive enough to detect the lensed images in X-band, a consequence of estimating the required integration time based on a lower-frequency spectral index and assuming no spectral curvature.

\begin{figure*}
    \centering
    \subfigure{\includegraphics[trim = {36mm, 12mm, 33mm, 12mm}, clip, width=0.30\textwidth]{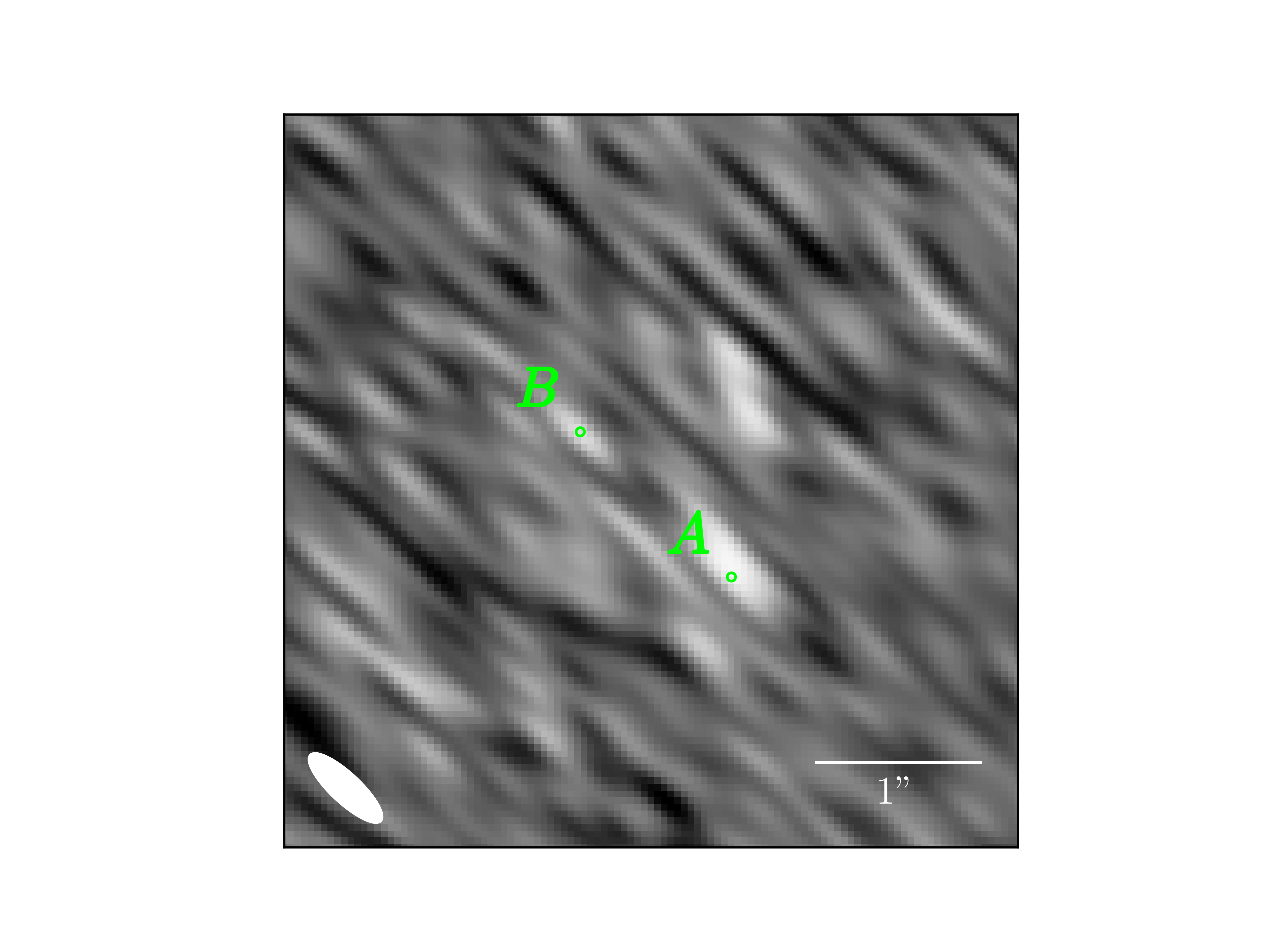}}
    \hspace{1mm}
    \subfigure{\includegraphics[trim = {36mm, 12mm, 33mm, 12mm}, clip, width=0.30\textwidth]{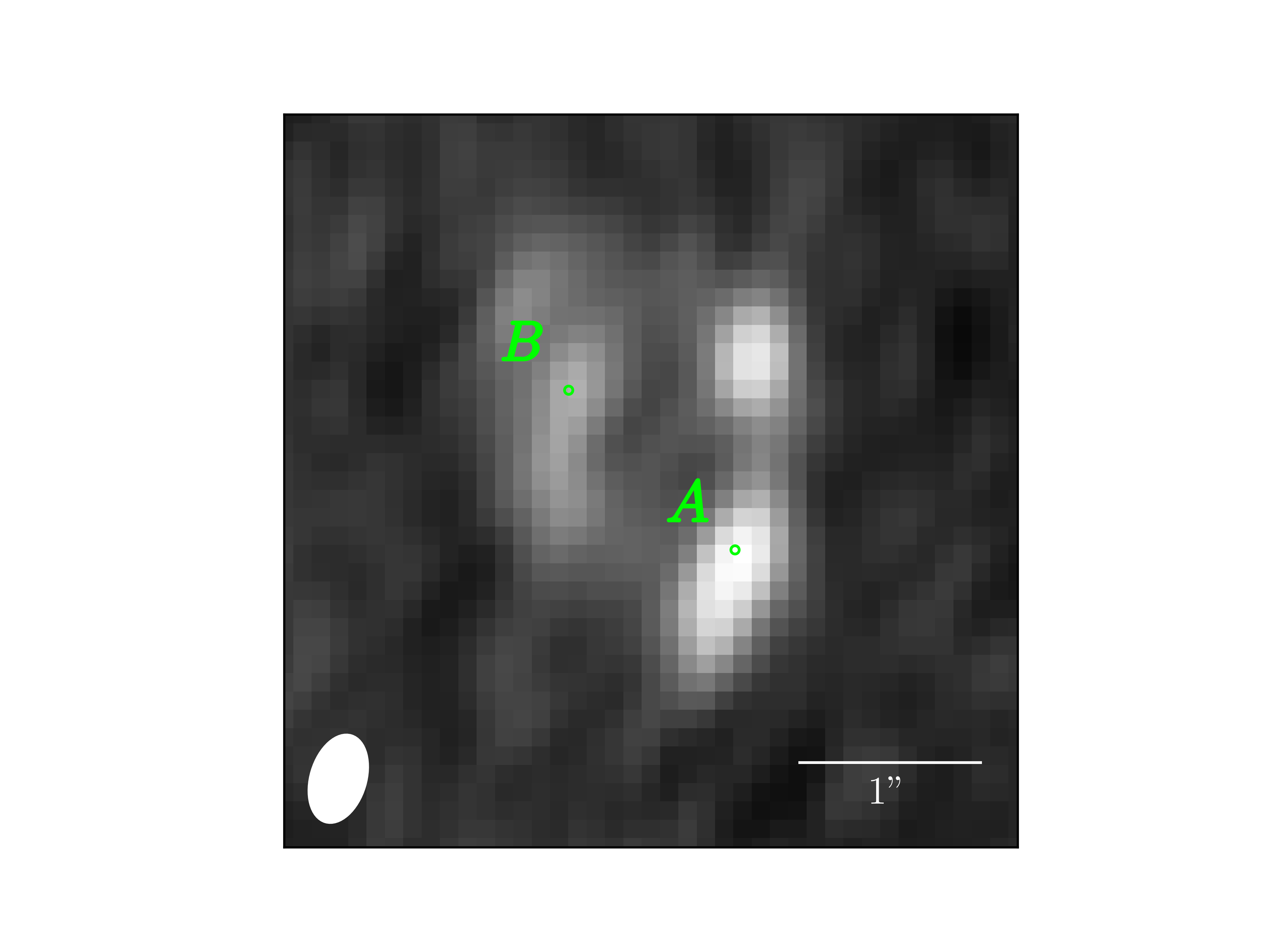}}
    \hspace{1mm}    
    \subfigure{\includegraphics[trim = {28mm, 12.5mm, 27mm, 13mm}, clip, width=0.30\textwidth]{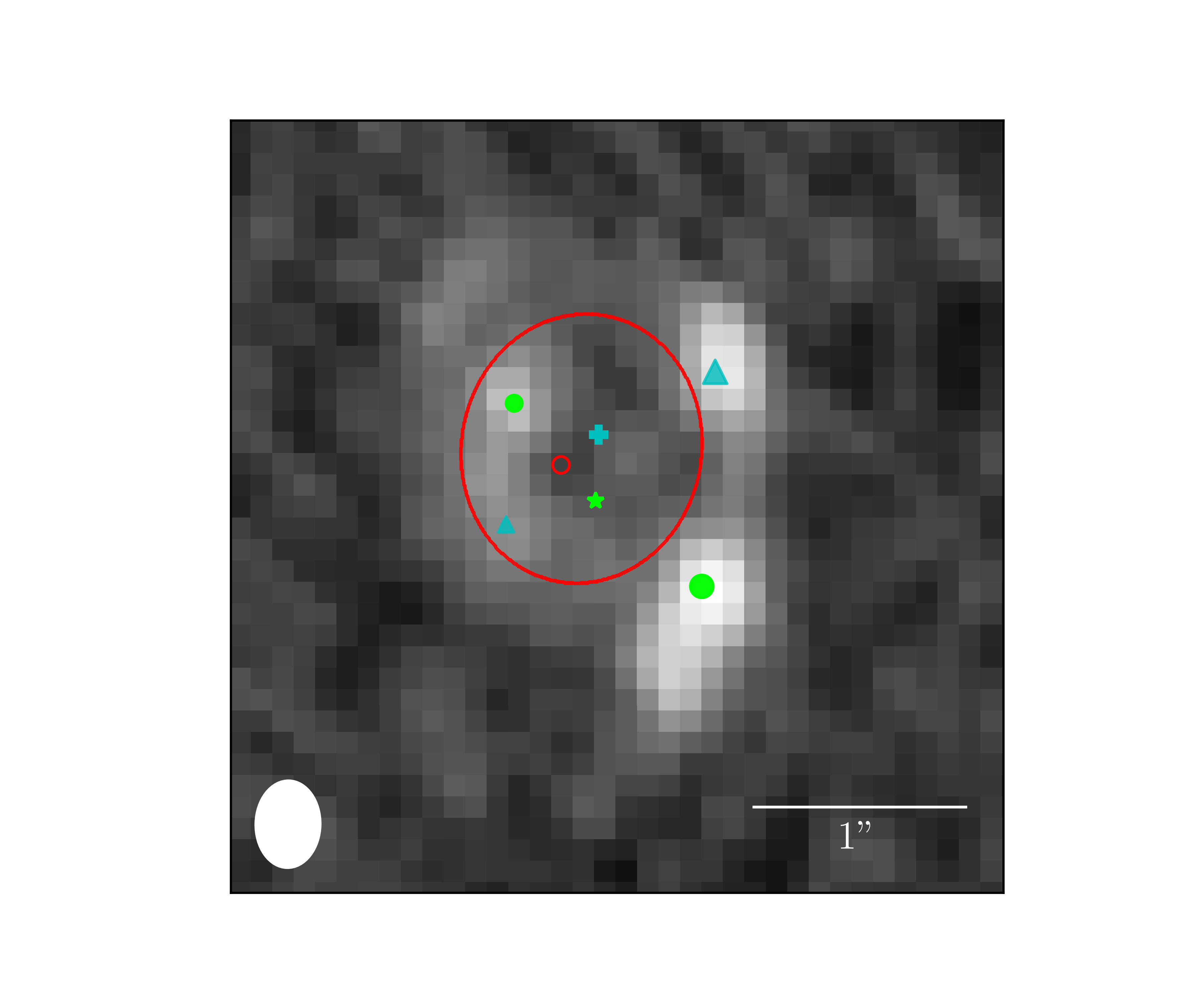}}
    \caption{Left: VLA X-band observation of J2329$-$1258, with \textit{Gaia} positions of quasar images A and B overlaid. 
    Center: The same system in combined S-band and C-band, from VLA project 19A-176.
    Right: Stacked $2-12\,GHz$ image of J2329$-$1258 with the optical-based lens model of \citet{shajib}. The model's critical curve is shown in red, and pairs of features which correspond according to the lens model are shown, with the quasar images still in lime circles and the extended emission in cyan triangles. Source positions for the quasar and extended emission are shown as a star and plus, respectively.}
    \label{fig:2329lensmodel}
\end{figure*}

\subsubsection{J2329$-$1258} \label{2329}

J2329-0734 was discovered by \citet{2329disc}, who used a WISE $W1 - W2$ color cut to select potential blended quasar pairs and crossmatched with the ATLAS survey.
Candidates were checked for consistency in putative image colors and visually inspected before spectroscopic follow-up, which confirmed this object as a lensed quasar at $z=1.31$.
Our X-band observations detected the brighter image, as well as extended emission coming from just above that image.
We detected a noise bump at the position of the other quasar image, but our image fitting procedure favored extended rather than point-source emission at this location.
We do not attempt to set a limit on the flux ratio in this system due to this extended emission.
To further investigate the nature of this source, we turned to archival data from VLA project 19A-176, who obtained A-config observations of the object in the S and C bands.
This data also shows pointlike features at the quasar image location and even more diffuse emission than the X-band data.
A 2-term Multi-Frequency Synthesis \citep{mfs} image created from visibility-space stacking both our data and the archival data is shown in the bottom left panel of Figure \ref{fig:lensfig}.
Due to our only matching one quasar image, our statistical chance of random coincidence from our observations is much lower, at only $8\times 10^{-5}$.

To further investigate the nature of the extended emission present in this lens system, we utilized a lens model created by \citet{shajib}. This model, constructed using $K$-band ($2.2\,\mu m$) Adaptive Optics observations on the Keck Telescope's NIRC2 instrument, only incorporates near-infrared data and thus is an independent test for our radio observations. 
Figure \ref{fig:2329lensmodel} shows our X-band data, the archival S and C band data, and their combination, as well as the critical curve of \citet{shajib}'s lens model.
We propagate the locations of image A and the northeast extended component through the lens model and plot their predicted positions.
Image A's counterpart is located at image B, as expected, and the northeast component's counterimage is predicted to appear at the location of the southwest component.  
It is therefore likely that at least one component in radio map besides the AGN core is strongly lensed.
This extended lensed emission may be useful to for a gravitational imaging analysis similar to that of \citet{2019MNRAS.483.2125S} with VLBI follow-up. 
However, given the faintness of this source, such an analysis may not be possible without the enhanced sensitivity of the next generation of radio telescopes \citep[priv. comm.]{mckeancomm}.

\subsection{Non-Lensing Results}


\begin{figure*}
    \centering
    \subfigure{\includegraphics[trim = {36mm, 12mm, 33mm, 12mm}, clip, width=0.35\textwidth]{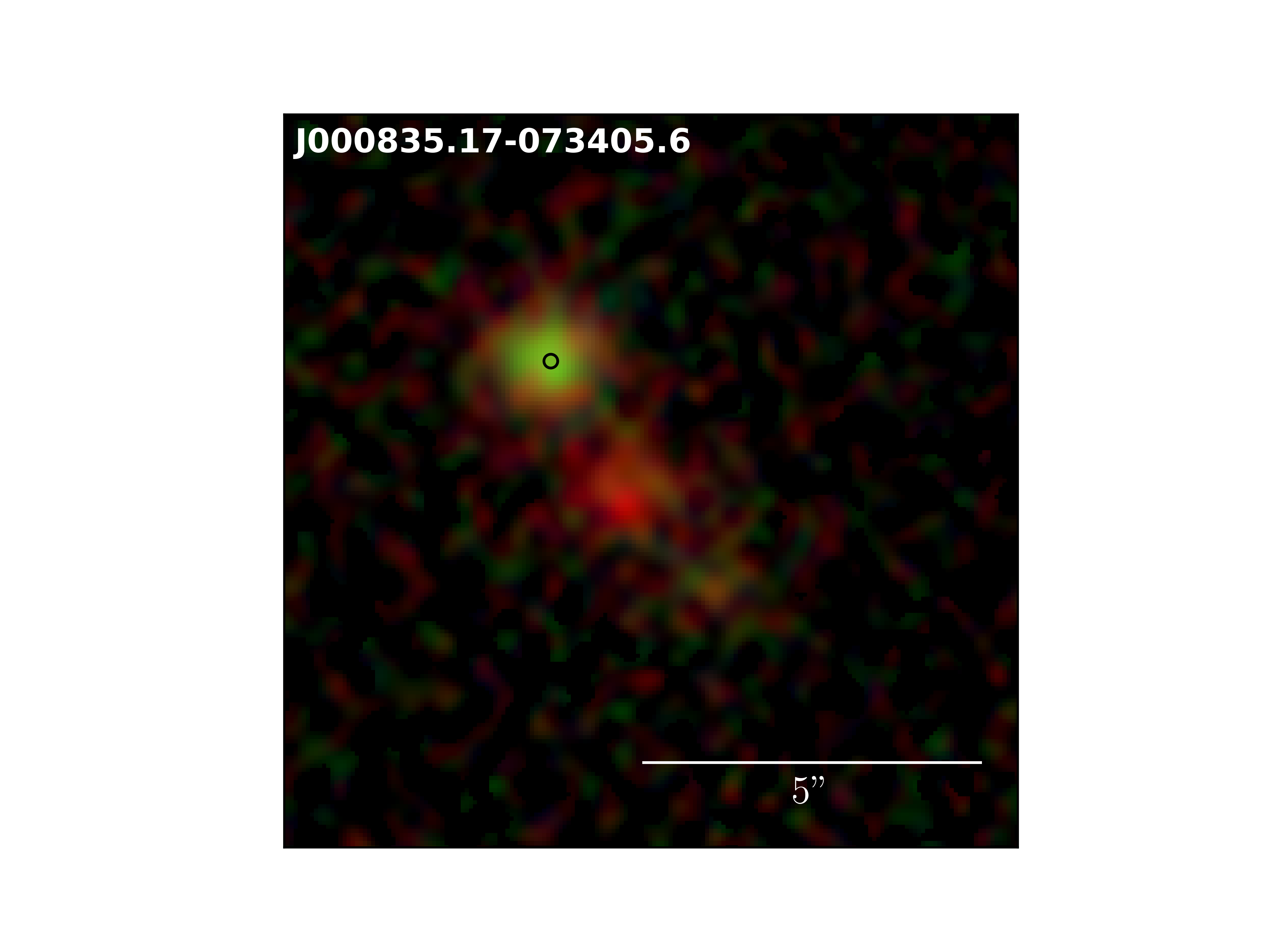}}
    \hspace{4mm}
    \subfigure{\includegraphics[trim = {36mm, 12mm, 33mm, 12mm}, clip, width=0.35\textwidth]{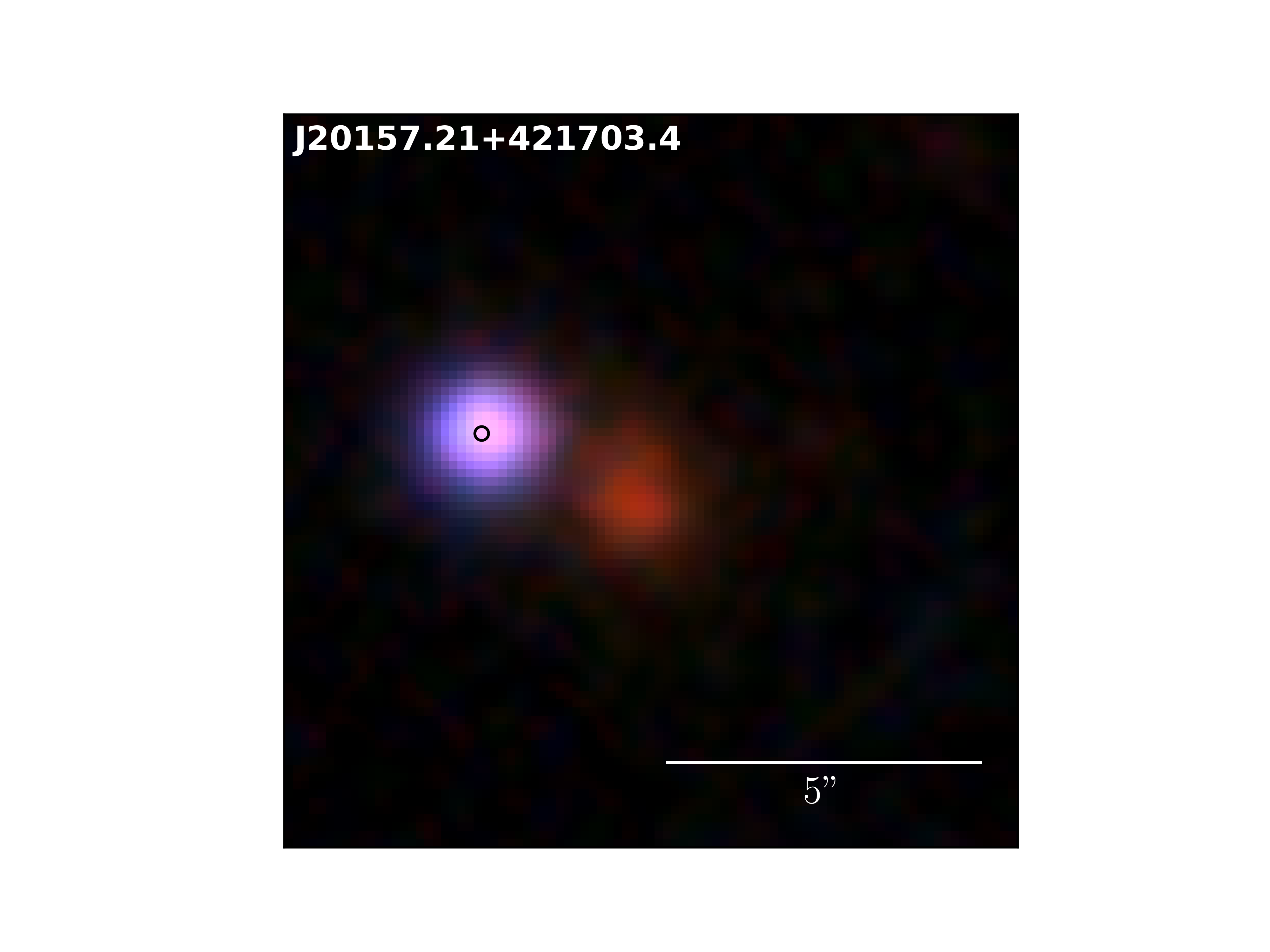}}
    \subfigure{\includegraphics[trim = {36mm, 12mm, 33mm, 12mm}, clip, width=0.35\textwidth]{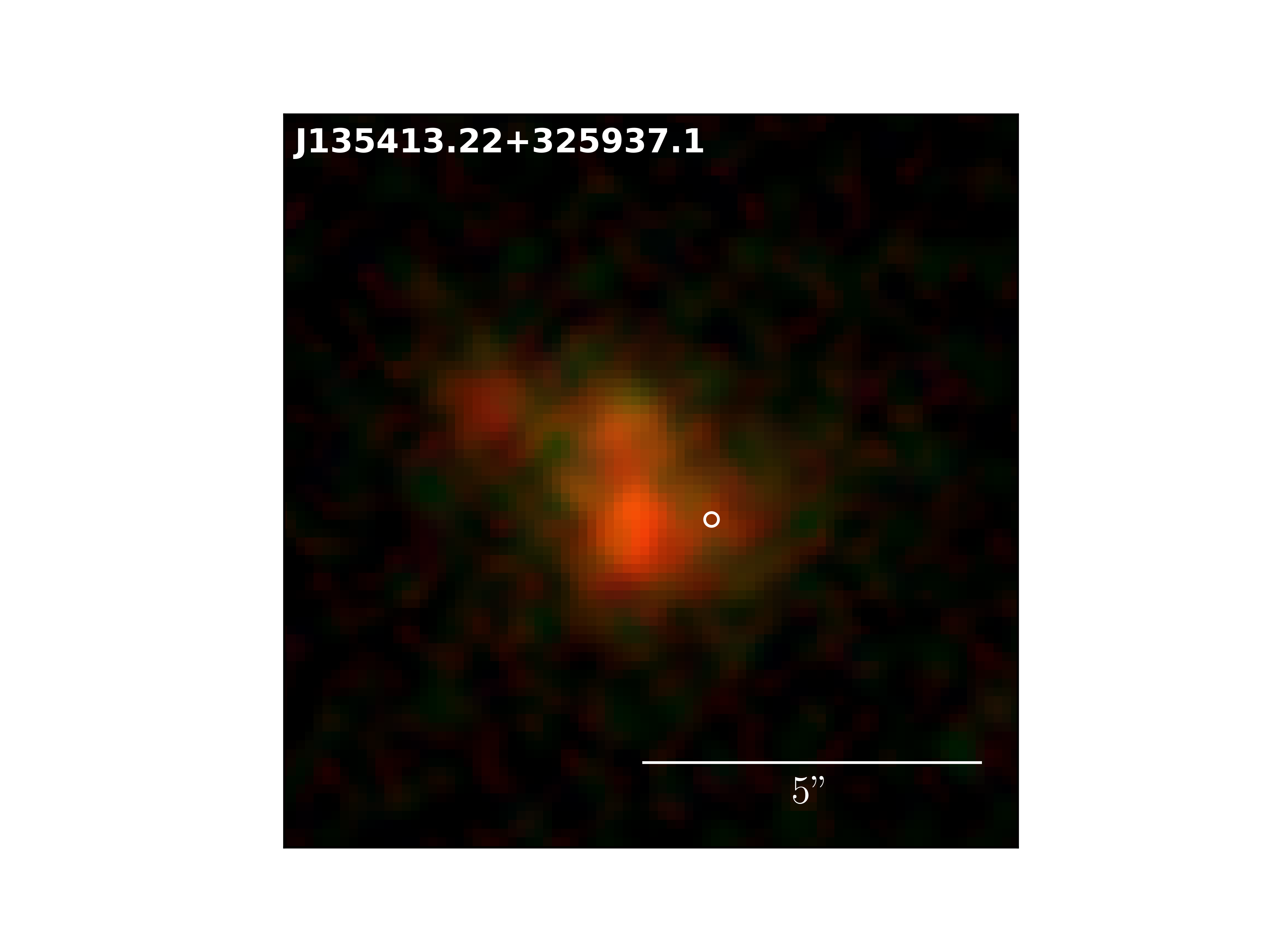}}
    \hspace{4mm}
    \subfigure{\includegraphics[trim = {36mm, 12mm, 33mm, 12mm}, clip, width=0.35\textwidth]{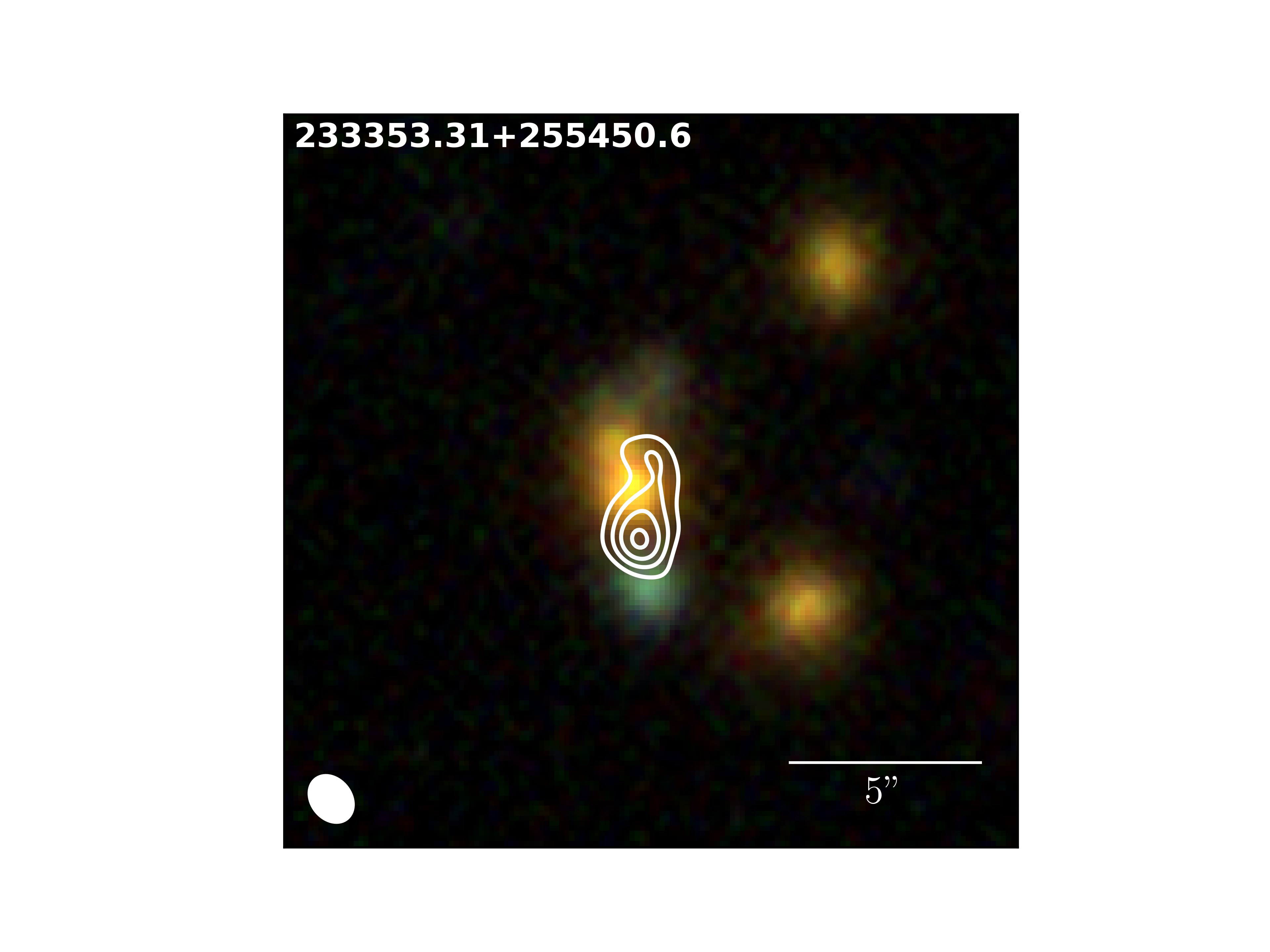}}
    \caption{Non-lensing targets. Top row, left to right: J0008$-$0734 (Section \ref{0008}),  J120157+421703 (Section \ref{120157}).
    Bottom row, left to right: J135413+325937 (Section \ref{135413}), J233353+255450 (Section \ref{2333}). 
    The location of the point-source radio detection is shown by a circle for all panels except for J233353+255450 where we detected extended emission. For J2333353+255450 we show radio contours correspond to 3-,5-,7-, and 9-$\sigma$ flux densities in the uv-tapered radio image.}
    \label{fig:nonlens}
\end{figure*}

\subsubsection{J0008$-$0734} \label{0008}

This source was identified as a potential radio lens using the JB07 method, and was singled out for observation due to the relatively bright VLASS detection, green color of the potential source, and possible counterimage in DECaLS. 
However, our X-band follow-up revealed only a $2.37\,$mJy point source coincident with the optical quasar and no counterimage.
Our VLA observations of this target have an rms noise of $25\,\mu\text{Jy}\,\text{beam}^{-1}$, and at the $5\sigma$ level we should be sensitive to point sources brighter than $125\,\mu$Jy.
That we detect no radio counterimage suggest that if there were such a counterimage, the flux ratio of the lensed radio source would be a seemingly unrealistic $>20$.
Moreover, the optical flux ratio of the sources immediately north-east and south-west of the the LRG is $\approx8$, so should this be a lensed source then there would be a substantial discrepancy between the optical and radio flux ratios.
While it is not impossible that this source is a lensed quasar, our observations don't support such a conclusion, and we posit that these are likely two unrelated sources.


\subsubsection{J120157+421703} \label{120157}

This source was identified as a possible radio lens using the JB07 method.
The DECaLS image of this source shows a possible very faint arc to the lower right of the LRG. 
The VLA X-band data showed a $7.0$ mJy point source coincident with the optical point source from DECaLS, but no counterimage.
This presents two possibilities when taking the possible arc into account: either the arc is simply an image artifact or other phenomenon and there is no lensing present at all, or the quasar is at or near the lens redshift and is therefore not strongly lensed.

\subsubsection{J135413+326937} \label{135413}

This source was identified as a potential radio lens using the JB07 method.
The X-band observations show a $2.8$ mJy point source offset from the LRG and coincident with the VLASS detection, but no counterimage. Therefore we conclude the quasar is not multiply imaged.

\subsubsection{J233353+255450} \label{2333}

This source was identified as a possible radio lens using the JB07 method.
Our initial X-band data reduction showed hints of extended emission near the lens location, and so we re-imaged the data with a $1''$ uv-plane taper to increase sensitivity at the cost of resolution.
We found an extended $1.7\,$mJy source located between the supposed lens and source, which we interpret as a radio lobe from the LRG rather than a lensed radio source, a hypothesis that is consistent with the steep spectrum ($\alpha=-0.8$) we measure from the VLASS and X-band flux densities.

\subsection{Other Results}


\begin{figure}
    \centering
    \subfigure{\includegraphics[trim = {36mm, 12mm, 32mm, 12mm}, clip, width=0.3\textwidth]{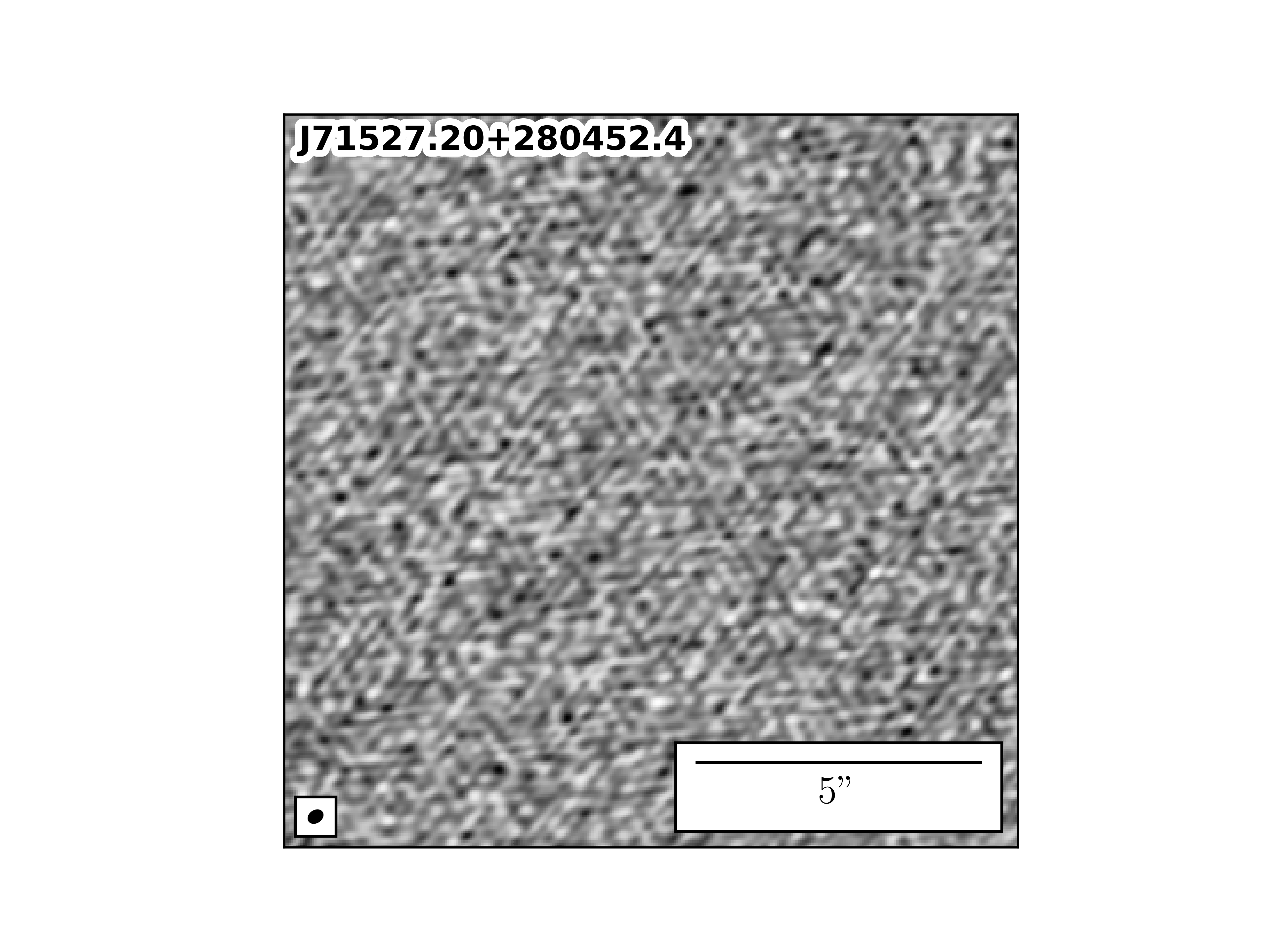}}
    \subfigure{\includegraphics[trim = {36mm, 12mm, 32mm, 12mm}, clip, width=0.3\textwidth]{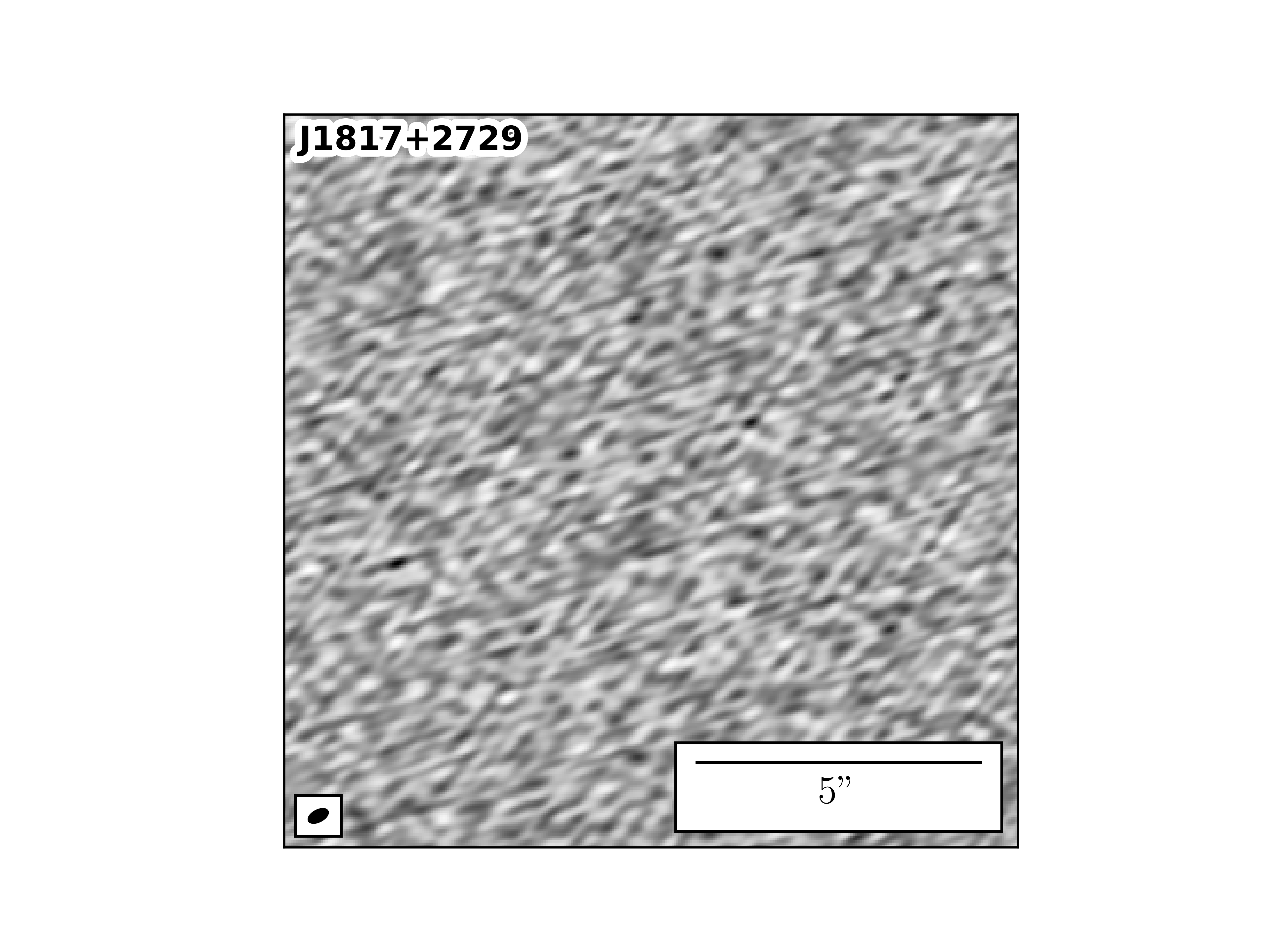}}
    \caption{The two non-detections from our observations. Top: J171527+280452 (Section \ref{1715}), and Bottom: J1817+2729 (Section \ref{1817}).
    }
    \label{fig:nondetection}
\end{figure}

\subsubsection{J171527+280452} \label{1715}

This source was identified as a possible radio lens using the JB07 method. 
However, we detected no significant emission in our X-band observations.
Based on the $2.6$mJy VLASS epoch 1 flux of the source and a $3\sigma$ nondetection threshold, we estimate a spectral index between 3 and 10 GHz for this source of -2.2, much higher than its VLASS-NVSS spectral index of -0.31.
The source shows no significant variability between VLASS epochs 1 and 2, leading us to suspect the target is either a peaked-spectrum compact source which is undetected at 10GHz, or an extended source which we do not detect due to resolution or sensitivity.
In either case, we cannot rule out the possibility of lensing.

\subsubsection{DES J0412$-$2646} \label{0412}

\begin{figure}
    \centering
    \includegraphics[trim = {36mm, 12mm, 33mm, 12mm}, clip, width=0.7\columnwidth]{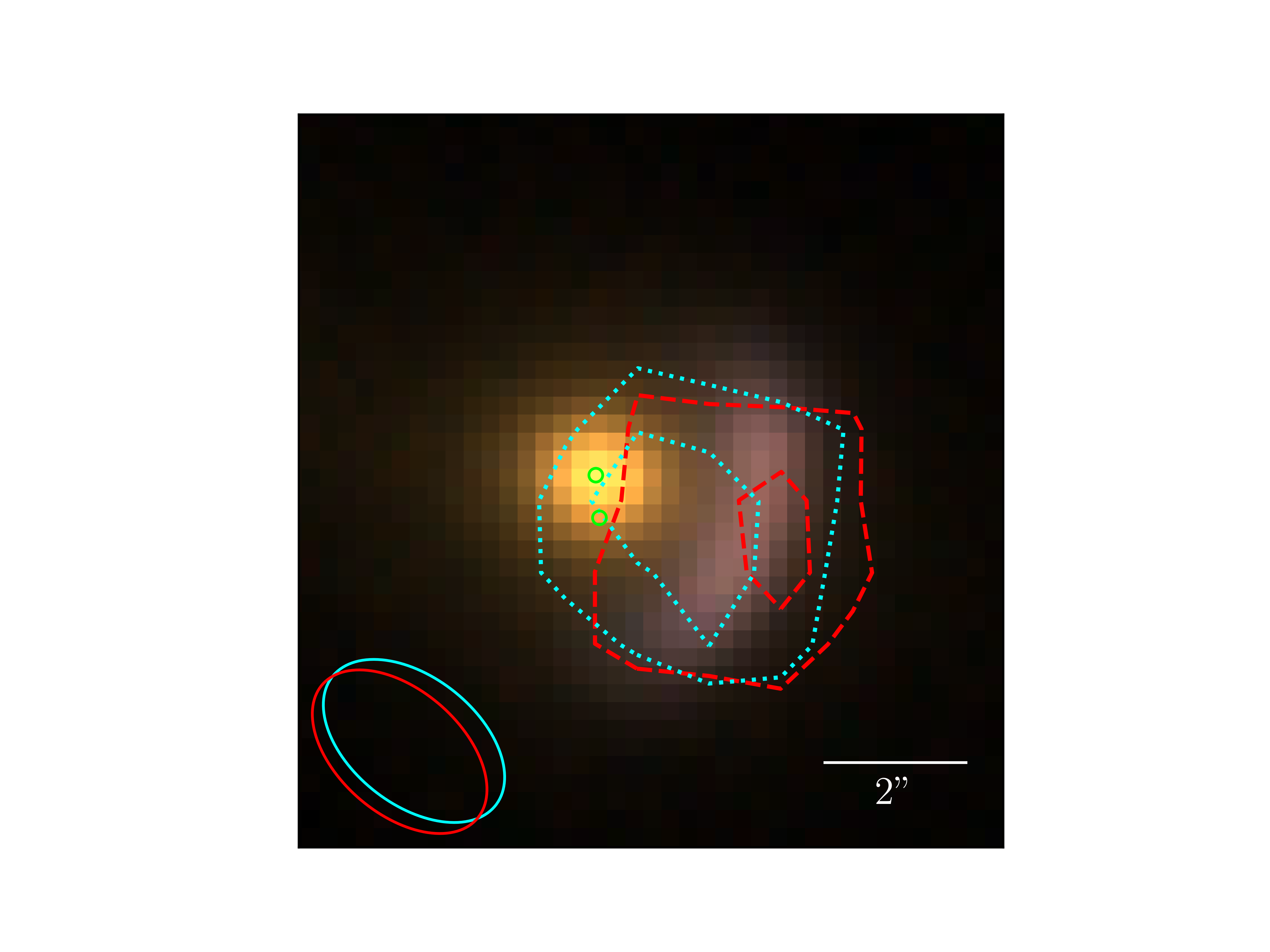}
    \caption{DECaLS $grz$ image of target J0412-2646. 4- and $6-\sigma$ contours are overlaid for VLASS epochs 1 (dashed red lines) and 2 (dotted cyan). The locations of the two radio components from by our observations (see Table \ref{tab:photometry}) are shown in green.}
    \label{fig:0412}
\end{figure}

This source was identified as a lensed galaxy by \citet{Jacobs2019} using a Neural Network-based search of DES. While VLASS images from both epochs seem to be centered away from the lens, our follow-up data shows a $160\,\mu$Jy point source at the location of the lens galaxy and a $260\,\mu$Jy diffuse component to the south of that, possibly indicative of a core+jet or core+lobe morphology.
Figure \ref{fig:0412} shows a DECaLS image of this source with the locations of our VLA detections and contours of two VLASS epochs.
While one epoch has the peak of emission located on top of the arc, the other places it between the arc and the lens galaxy.
It is possible diffuse emission from the source galaxy is responsible for shifting the VLASS detection over, and that this emission is too low surface brightness for or resolved out of our observations at 10 GHz.
However, further observations would be needed to address this hypothesis.

\section{The Known Population of Lensed Radio Sources} \label{sec:discussion}


\subsection{Variability and Spectral Indices of Lensed Radio Sources}


Until recently, only a handful of lensed radio sources were known, with most of these being identified through dedicated searches such as CLASS and JVAS.
The advent of deep and high resolution wide-area sky surveys such as VLASS is now resulting in more detections and correct associations of radio emission from lensed systems, especially lensed quasars.
Additionally, the latest generation of optical surveys with high astrometric precision, such as Gaia, are allowing for the identification of hundreds of new lensed quasars \citep[e.g.,][]{Jacobs2019}.
The result is such that there are now $\approx 80$ lensed radio sources known, more than double the number known less than a decade ago \citep{McKean2015}.
We list all the published gravitational lenses with emission detected at frequencies lower than 100 GHz ($\lambda > 3\,$mm) in Table \ref{tab:allknown}.
This cutoff was chosen to correspond roughly with both the point where dust begins to dominate the SED of a normal galaxy rather than synchrotron emission \citep{condon92} and the highest observable frequencies of the ngVLA \citep{Carilli2015}.
In this Section of the paper we use these $80$ objects to broadly characterise the observational properties of the lensed radio source population.

Some previous dedicated searches for lensed radio sources have specifically looked for flat-spectrum radio sources \citep[e.g.,][]{JB07, Myers2003}.
In principle such a strategy should reduce contamination from the lobes of radio galaxies that can appear offset from their host galaxies, often LRGs, and thus potentially mimic a lensed object in catalog space.
With a reasonably large sample of lensed radio sources now in hand we can potentially explore the spectral index distribution of the population.
Doing so has several benefits, the spectral index can i) provide insights into the type of source being lensed (e.g., quasar, lobe-dominated radio galaxy etc.); ii) potentially guide future search strategies for lensed radio sources; and iii) be used to show the flux distribution of lensed radio sources at a single observer-frame frequency, as opposed to comparing flux densities from different observations at e.g., $1.4\,$GHz and $10\,$GHz.

\begin{figure}
    \centering
    \includegraphics[width=\columnwidth]{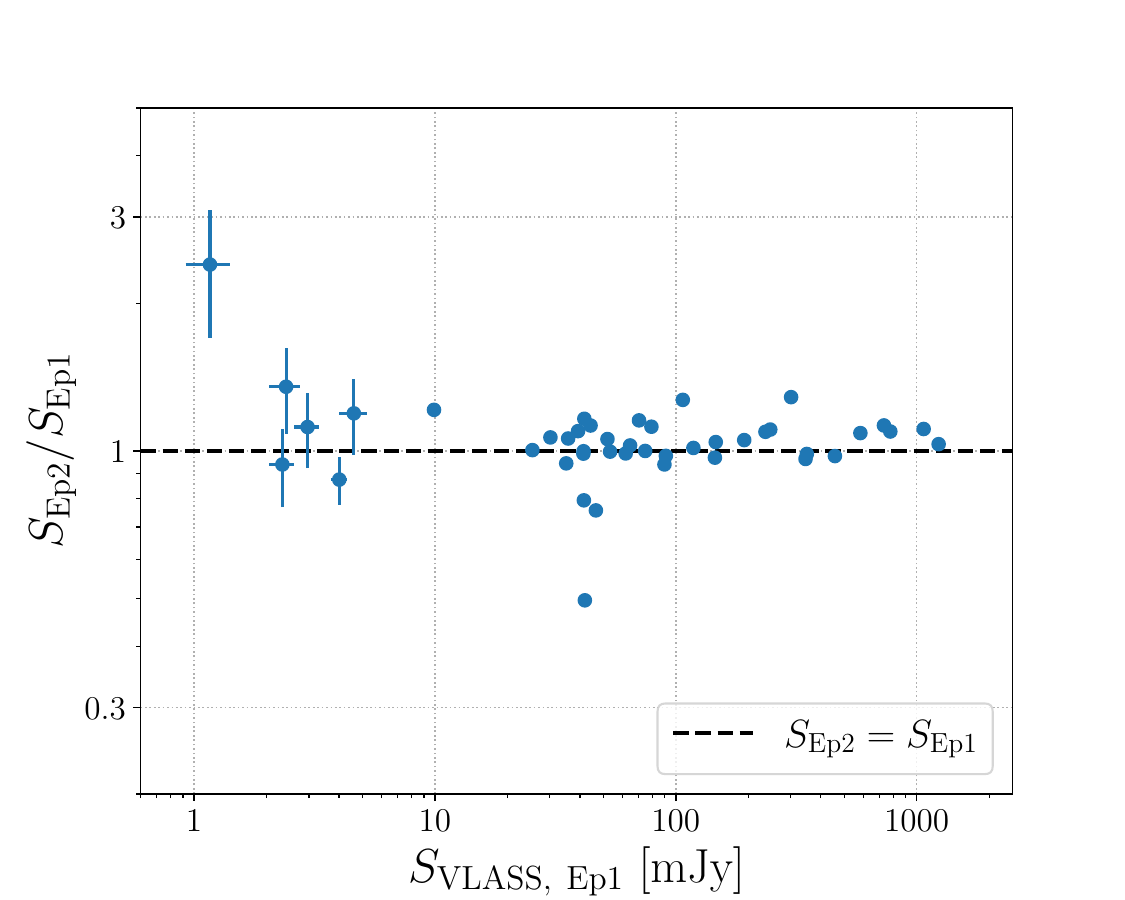}
    \caption{The variability of lensed radio sources in VLASS ($S_{\text{Epoch 1}}/S_{\text{Epoch 2}}$) as a function of brightness in Epoch 1, with the black dashed line denoting zero variability between the two epochs.
    The slight but systematic trend for brighter fluxes in epoch 2 is likely driven by the limited quality of the VLASS `Quick Look' images.
    }
    \label{fig:vlass-variability}
\end{figure}

Ideally the spectral index for radio sources should be calculated using flux measurements from different frequencies obtained at the same time to avoid the potential for source variability biasing the measurement. 
For their $8$ lensed quasars observed with the Australia Telescope Compact Array, \citet{dobie23} provide contemporaneous measurements at $5.5\,$GHz and $9\,$GHz which we use to calculate the spectral index for these sources.
For the remaining sources we do not have contemporaneous multi-band flux density measurements, and are thus dependent on measurements that might be subject to variability.
Using the catalogs from the first two epochs of VLASS \citep[][B. Sebastian et al. in prep.]{Gordon2021} we characterise the variability of the $44$ lensed sources detected in the first epoch of VLASS over timescales of $\sim2$ years in Figure \ref{fig:vlass-variability}.
With only a few exceptions, most lensed radio show little variability between Epochs 1 and 2 of VLASS, with a median and standard deviation for $S_{\text{Ep 2}}/S_{\text{Ep 1}}$ of $1.0$ and $0.2$ respectively.
Knowing that most lensed radio sources aren't strongly variable strengthens the argument for using flux density measurements taken at different times to estimate the spectral index of these sources.
For $24$ lensed radio sources we have flux density measurements from both VLASS ($3\,$GHz) and FIRST ($1.4\,$GHz).
For the $32$ lensed sources for which we have spectral information, we find the median spectral index to be $\alpha_{\text{median}} = -0.7$, similar to the typical spectral index for the general radio source population \citep[e.g.,][]{Condon1998, Gordon2021}.

\subsection{Future Searches}



Notably, $100\,\%$ of our VLASS detected targets that are lensed optical quasars have radio emission from the lensed source, suggesting that lensed quasars conincident with legacy detections in radio surveys present an efficient approach to identifying candidate lensed radio sources.
Moreover, those lensed radio sources detected in flux-limited surveys are likely the most scientifically useful targets due to their typically higher brightness than many sources detected through blind, deep radio observations of lensed optical quasars.
With a suite of deep wide-area optical and near-IR imaging surveys from ground and space, such as the Vera C. Rubin Observatory \citep{Ivezic2019}, the Nancy Grace Roman Space Telescope \citep{Spergel2015}, and the Euclid telescope \citep{Laureijs2011}, coming online over the next few years, thousands of lensed quasars will be discovered \citep[e.g.][]{Yue2022}.
It is interesting to consider how many of these sources will have complementary radio observations.
Some of the new lensed quasars may have already been detected at radio wavelengths, but due to the multi-arcsecond PSF of current wide-area radio surveys their status as lensed radio sources remains unknown.
Moreover, for a multi-epoch survey such as VLASS, the ability to combine the individual observations from each epoch enables deeper imaging than one epoch of observations alone, increasing their power as a legacy reference catalog to identify radio emission from newly discovered lensed quasars. 
After the end of the planned survey, combined three-epoch VLASS images are expected to have a point source depth of $S_{3\,\text{GHz}}\approx 350\,\mu$Jy \citep{VLASS}, substantially deeper than the $\sim 1\,$mJy depth of the Quick Look images from a single epoch currently available\footnote{A recently proposed fourth VLASS epoch would push the point source sensitivity of combined images down to $300\,\mu$Jy \citep{Nyland2023}.}.

To make predictions for the number of lensed radio sources that might be detected in VLASS we first determine the $3\,$GHz flux density distribution of the known lensed radio sources.
For bright lensed sources within the VLASS footprint we take the $3\,$GHz flux density measurement from the VLASS Epoch 1 Quick Look catalog \citep{Gordon2021}.
For those sources too faint to be detected by VLASS or lying outside the survey footprint, we estimate their $3\,$GHz flux density by extrapolating from available measurements at other frequencies using their measured spectral index where available.
For those sources for without a spectral index we assume $\alpha=-0.7$ in line with the typical spectral index for the lensed radio source population.
Where published flux densities for individual lensed images are used, these are summed to provide a total flux for the lensed system, a better approximation of what will observed by a single VLASS beam.
We note here that we do not estimate the $3\,$GHz flux density for PSS 2322$+$1944, as the observed $45\,$GHz emission is attributed to CO($J=2\rightarrow1$) line emission rather than being continuum emission \citep{2008ApJ...686..851R}, and thus extrapolating to $3\,$GHz based on an assumed spectral index is inappropriate in this instance.

\begin{figure}
    \centering
    \includegraphics[trim=1mm 1mm 0 0 clip, width=\columnwidth]{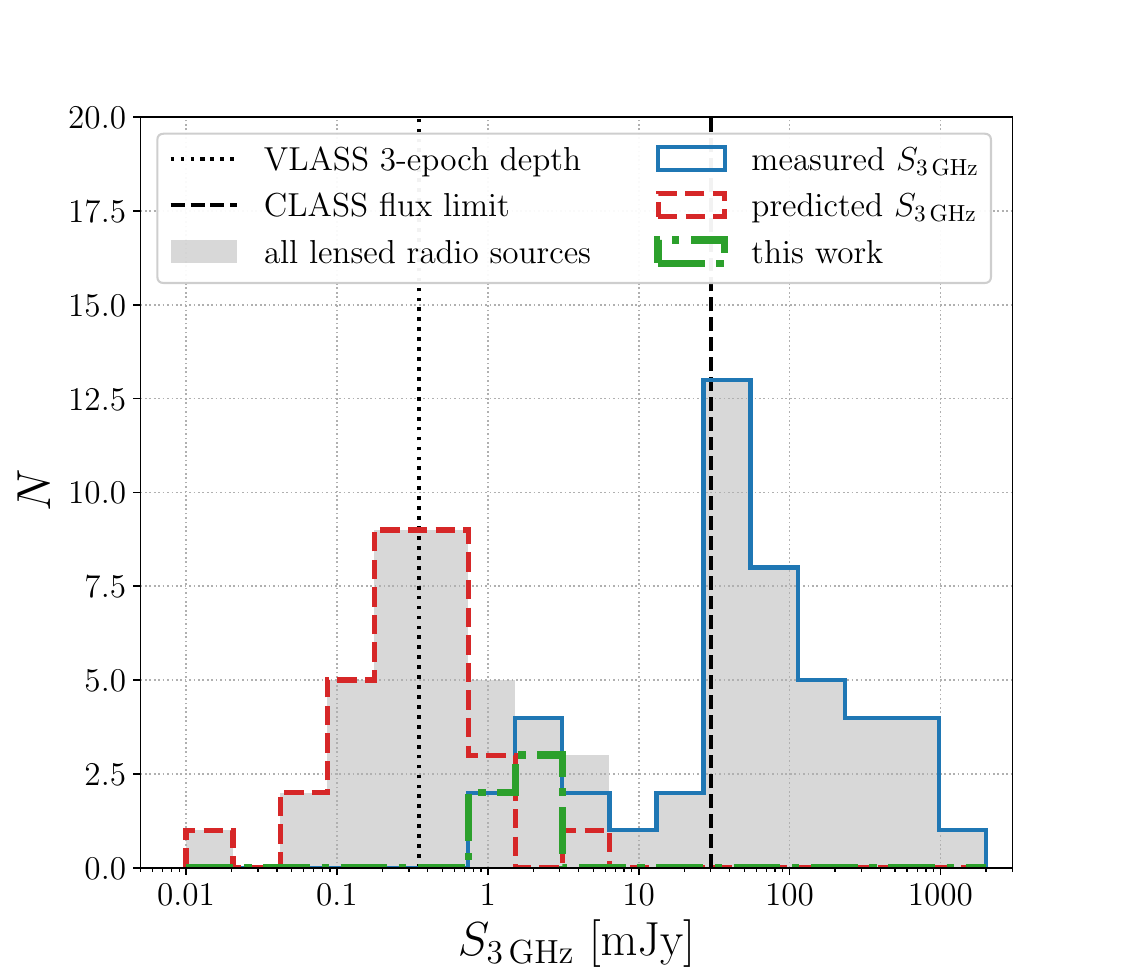}
    \caption{Distributions of integrated $3\,$GHz flux densities for lensed radio sources (grey solid histogram).
    Flux measurements are taken from VLASS where possible (solid blue line) and estimated using spectral index information otherwise (red dashed line).
    The green dot-dashed line shows the five previously unreported lenses we identified with VLASS in this work.
    The dashed black vertical line shows the flux limit of the CLASS survey, highlighting the additional sources that can be identified by combining optical and radio information rather than just relying on a dedicated flux-limited radio search.
    The black dotted line shows the $350\,\mu$Jy point source depth that VLASS will reach after three epochs.
    }
    \label{fig:fluxdist}
\end{figure}

The $S_{3\,\text{GHz}}$ distribution for lensed radio sources is shown in Figure \ref{fig:fluxdist}, with predicted and measured flux densities shown by the red dashed and blue sold lines respectively.
An important feature of Figure \ref{fig:fluxdist} is the apparent bimodality of the radio flux distribution of lensed sources.
This can be explained by the two broad selection approaches used over the years.
The brighter peak (centered around $100\,$mJy) is mostly the result of the targeted searches for lensed radio sources conducted by CLASS, JVAS, and MG-VLA \citep{MGVLA}.
Indeed, the flux limited nature of these searches is evident in Figure \ref{fig:fluxdist} as the sudden drop in sources below \text{$S_{3\,\text{GHz}}\approx30\,$mJy}.
The fainter peak (centered around $300\,\mu$Jy) is the result of radio observations of newly identified lensed quasars in optical imaging. 
Notably, about half of these objects should be detectable in future multi-epoch combined VLASS images, providing a potential pathway to more efficient target selection for future in depth radio observations.
Large numbers of lensed radio sources will be detected in forthcoming optical surveys.
For instance \citet{Yue2022} predict $2,400$ lensed quasars will be identified in the Legacy Survey of Space and Time \citep[LSST,][]{Ivezic2019}, $\sim1,000$ of which will be at depths detectable by current optical surveys.
Approximately $14,000\,\text{deg}^{2}$ ($70\,$\%) of the $20,000\,\text{deg}^{2}$ footprint of LSST will be covered by VLASS, it follows that hundreds of the lensed sources may be detectable in the final-depth VLASS images.

In this work we have focused on using VLASS to identify lensed radio sources, and indeed the high resolution and time domain aspects of the survey provide unique advantages over previous radio surveys for this challenge.
The next generation of radio telescopes however will be even more well suited for identifying lensed radio sources.
The high angular resolution and survey speeds of the Square Kilometer Array (SKA) and the ngVLA will enable the ready identification of the multiple images of radio sources separated on sub arcsecond scales. 
This can provide two key advantages over current approaches.
First, not being dependent on optical observations to identify the lensing configuration has the potential to identify systems where the lensed galaxy is has an intrinsically high radio-to-optical luminosity such that it is only detected in radio.
Second, in systems where only the lensed background object is radio loud, the low level of radio contamination may allow for more tightly constrained lens models than would be possible from optical observations where light from the foreground lens galaxy may become problematic.


\section{Conclusions} \label{sec:summary}

We report first results from a pilot study seeking to efficiently identify strongly lensed radio sources by combining wide-area optical and radio survey data.
We find that a high fraction of optically selected lensed quasars with radio counterparts in VLASS at mJy-level flux densities are in fact high-confidence lensed radio sources.
The results here suggest that large samples of radio strong lenses could be efficiently identified via targeted follow-up of radio counterparts to lenses found in near-future optical and NIR imaging surveys with the Vera C. Rubin Observatory, Euclid, and Nancy Grace Roman Space Telescope.
Importantly, the radio lens systems from VLASS are bright enough to allow detailed characterization.
Our findings are reinforced by complementary recent results from \citet{dobie23} and \citet{Jackson2024}.

We observed 11 radio lens candidates based on two selection methods.
The method based on \citet{JB07} aiming to discover entirely new lens systems yielded no new radio lenses.
However, given the rarity of gravitational lensing in general, this result was not unexpected, and we note that JB07 themselves found no candidates among a larger follow-up sample.
A successful catalog-based method would require a more sophisticated approach than the one we utilized, and such an approach become much more necessary in the future thanks to upcoming large and deep surveys in both the radio and optical.

The second method, which utilized existing catalogs of lensed quasars and galaxy-scale arcs, was much more successful.
Five out of the five existing lensed quasars we observed had radio emission from the quasars, rather than the lens, and in only one case did the lens galaxy also emit in the radio.
Furthermore, our single lensed galaxy target is still a possible radio lens given the mismatch between VLASS and VLA positions, although its emission seems to be much fainter than suggested by the VLASS epoch 1 data.
These results suggest that survey-resolution radio emission from lensed quasar systems is more likely to come from the quasar rather than the lens, and presents a possible method to identify more lensed radio sources in the future.

Our candidate selection for this method utilized a list of lensed quasars published in 2019, containing 220 systems. Since then, the publication of hundreds of lensed quasar \citep[e.g.][]{lemon2023, he23} and galaxy-galaxy lens \citep[e.g.][]{dawes23, zaborowski23} candidates has greatly expanded the number of possible targets, suggesting that a new search incorporating the same methodology is likely to discover many more systems.

During the final preparations of this manuscript, \citet{Jackson2024} reported independent observations for a sample of radio lens candidates, including 3 of the 4 previously unreported radio lens systems presented in Section \ref{sec:results}, as well as an additional 30 sources not considered here.
These two works underscore the opportunities for expanding the catalog of known lensed radio sources through target selection based on lenses identified at other wavelengths.
Similar to \citet{dobie23}, we both find that radio emission from systems involving optically-selected lensed quasars is typically dominated by emission from the lensed quasar rather than the main deflector galaxy.
Our target selection differs from \citet{dobie23} and \citet{Jackson2024} in that we required a spatially coincident VLASS source, and thus all of the new radio lenses discussed here have integrated flux density brighter than $\sim1$ mJy at 3 GHz (Figure~\ref{fig:fluxdist}).

\section{Acknowledgments}.
M.N.M., Y.A.G, K.B., and P.S.F. are supported by U.S. National Science Foundation grants AST 20-09441 and AST 22-06053.

The National Radio Astronomy Observatory is a facility of the National Science Foundation operated under cooperative agreement by Associated Universities, Inc.

The Legacy Surveys consist of three individual and complementary projects: the Dark Energy Camera Legacy Survey (DECaLS; Proposal ID \#2014B-0404; PIs: David Schlegel and Arjun Dey), the Beijing-Arizona Sky Survey (BASS; NOAO Prop. ID \#2015A-0801; PIs: Zhou Xu and Xiaohui Fan), and the Mayall z-band Legacy Survey (MzLS; Prop. ID \#2016A-0453; PI: Arjun Dey). DECaLS, BASS and MzLS together include data obtained, respectively, at the Blanco telescope, Cerro Tololo Inter-American Observatory, NSF’s NOIRLab; the Bok telescope, Steward Observatory, University of Arizona; and the Mayall telescope, Kitt Peak National Observatory, NOIRLab. Pipeline processing and analyses of the data were supported by NOIRLab and the Lawrence Berkeley National Laboratory (LBNL). The Legacy Surveys project is honored to be permitted to conduct astronomical research on Iolkam Du’ag (Kitt Peak), a mountain with particular significance to the Tohono O’odham Nation.

NOIRLab is operated by the Association of Universities for Research in Astronomy (AURA) under a cooperative agreement with the National Science Foundation. LBNL is managed by the Regents of the University of California under contract to the U.S. Department of Energy.

This project used data obtained with the Dark Energy Camera (DECam), which was constructed by the Dark Energy Survey (DES) collaboration. Funding for the DES Projects has been provided by the U.S. Department of Energy, the U.S. National Science Foundation, the Ministry of Science and Education of Spain, the Science and Technology Facilities Council of the United Kingdom, the Higher Education Funding Council for England, the National Center for Supercomputing Applications at the University of Illinois at Urbana-Champaign, the Kavli Institute of Cosmological Physics at the University of Chicago, Center for Cosmology and Astro-Particle Physics at the Ohio State University, the Mitchell Institute for Fundamental Physics and Astronomy at Texas A\&M University, Financiadora de Estudos e Projetos, Fundacao Carlos Chagas Filho de Amparo, Financiadora de Estudos e Projetos, Fundacao Carlos Chagas Filho de Amparo a Pesquisa do Estado do Rio de Janeiro, Conselho Nacional de Desenvolvimento Cientifico e Tecnologico and the Ministerio da Ciencia, Tecnologia e Inovacao, the Deutsche Forschungsgemeinschaft and the Collaborating Institutions in the Dark Energy Survey. The Collaborating Institutions are Argonne National Laboratory, the University of California at Santa Cruz, the University of Cambridge, Centro de Investigaciones Energeticas, Medioambientales y Tecnologicas-Madrid, the University of Chicago, University College London, the DES-Brazil Consortium, the University of Edinburgh, the Eidgenossische Technische Hochschule (ETH) Zurich, Fermi National Accelerator Laboratory, the University of Illinois at Urbana-Champaign, the Institut de Ciencies de l’Espai (IEEC/CSIC), the Institut de Fisica d’Altes Energies, Lawrence Berkeley National Laboratory, the Ludwig Maximilians Universitat Munchen and the associated Excellence Cluster Universe, the University of Michigan, NSF’s NOIRLab, the University of Nottingham, the Ohio State University, the University of Pennsylvania, the University of Portsmouth, SLAC National Accelerator Laboratory, Stanford University, the University of Sussex, and Texas A\&M University.

BASS is a key project of the Telescope Access Program (TAP), which has been funded by the National Astronomical Observatories of China, the Chinese Academy of Sciences (the Strategic Priority Research Program “The Emergence of Cosmological Structures” Grant \# XDB09000000), and the Special Fund for Astronomy from the Ministry of Finance. The BASS is also supported by the External Cooperation Program of Chinese Academy of Sciences (Grant \# 114A11KYSB20160057), and Chinese National Natural Science Foundation (Grant \# 12120101003, \# 11433005).

The Legacy Survey team makes use of data products from the Near-Earth Object Wide-field Infrared Survey Explorer (NEOWISE), which is a project of the Jet Propulsion Laboratory/California Institute of Technology. NEOWISE is funded by the National Aeronautics and Space Administration.

The Legacy Surveys imaging of the DESI footprint is supported by the Director, Office of Science, Office of High Energy Physics of the U.S. Department of Energy under Contract No. DE-AC02-05CH1123, by the National Energy Research Scientific Computing Center, a DOE Office of Science User Facility under the same contract; and by the U.S. National Science Foundation, Division of Astronomical Sciences under Contract No. AST-0950945 to NOAO.

The PanSTARRS 1 Surveys (PS1) and the PS1 public science archive have been made possible through contributions by the Institute for Astronomy, the University of Hawaii, the PanSTARRS Project Office, the Max-Planck Society and its participating institutes, the Max Planck Institute for Astronomy, Heidelberg and the Max Planck Institute for Extraterrestrial Physics, Garching, The Johns Hopkins University, Durham University, the University of Edinburgh, the Queen's University Belfast, the Harvard-Smithsonian Center for Astrophysics, the Las Cumbres Observatory Global Telescope Network Incorporated, the National Central University of Taiwan, the Space Telescope Science Institute, the National Aeronautics and Space Administration under Grant No. NNX08AR22G issued through the Planetary Science Division of the NASA Science Mission Directorate, the National Science Foundation Grant No. AST-1238877, the University of Maryland, Eotvos Lorand University (ELTE), the Los Alamos National Laboratory, and the Gordon and Betty Moore Foundation.

%

\vspace{5mm}
\facilities{Karl G. Jansky Very Large Array}


\software{astropy \citep{2013A&A...558A..33A,2018AJ....156..123A}
          CASA \citep{casa}
          lenstronomy \citep{lenstronomy, lenstronomy2}
          }

\begin{longrotatetable}
\tabletypesize{\footnotesize}
\begin{deluxetable*}{lccccccp{118pt}c}
\tablecaption{List of published radio gravitational lenses.}
\label{tab:allknown}
\movetabledown=15mm
\tablehead{\colhead{Name} & \colhead{Method} & \colhead{RA} & \colhead{Dec} & \colhead{VLASS Flux} & \colhead{Images} & \colhead{Sep.} & \colhead{References} & \colhead{VLASS Component} \\ 
\colhead{} & \colhead{} & \colhead{[deg]} & \colhead{[deg]} & \colhead{[mJy]} & \colhead{} & \colhead{['']}& \colhead{} & \colhead{} }
\decimalcolnumbers
\startdata
QSO B097+5608 & RADIO & 150.3369 & 55.8974 & 317.483& 2 & 6.17 & \citet{walsh79} & J100120.93+555355.8 \\
PG B1115+080 & OPTICAL & 169.57062 & 7.7663 & $<1.0$ & 4 & 2.43 & \citet{2021MNRAS.508.4625H} \newline \citet{1980Natur.285..641W} &  \\
MG B2016+112 & RADIO & 304.8253 & 11.4537 & 93.026 & 3 & 2.56 & \citet{1984Sci...223...46L} & J201918.00+112712.2 \\
B2237+0305 & OPTICAL & 340.125975 & 3.358508 & $<1.0$ & 4 & 1.78 & \citet{1996AJ....112..897F} \newline \citet{1985AJ.....90..691H} &  \\
MG B1131+0456 & RADIO & 172.9854 & 4.9302 & 261.992 & 2 & 2.2 & \citet{1988Natur.333..537H} & J113156.44+045549.5 \\
PKS B1830$-$211 & RADIO & 278.4164 & $-$21.0609 & \tablenotemark{a} & 3 & 0.99 & \citet{1988MNRAS.231..229P} &  \\
B1413 + 117 & OPTICAL & 213.9426 & 11.4953 & 3.487 & 4 & 1.35 & \citet{2023MNRAS.524.3671Z} \newline \citet{1988Natur.334..325M} & J141546.22+112943.7 \\
MG B1654+1346 & RADIO & 253.6741 & 13.7726 & 205.031 & Lobe & 2.0 & \citet{1989AJ.....97.1283L} & J165441.79+134621.4 \\
MG B0414+0534 & RADIO & 63.6571 & 5.5786 & 930.234 & 4 & 2.4 & \citet{1992AJ....104..968H} & J041437.74+053443.0 \\
JVAS B1422+231 & JCP & 216.1587 & 22.9335 & 676.326 & 4 & 1.3 & \citet{1992MNRAS.259P...1P} & J142438.11+225600.7 \\
JVAS B0218+35.7 & JCP & 35.2729 \tablenotemark{d} & 35.9372 & 1073.526 & 2 & 0.335 & \citet{1993MNRAS.261..435P} & J022105.46+355613.8 \\
MG B1549+3047 & RADIO & 237.3014 & 30.7879 & 508.064 & Lobe & 2.0 & \citet{1993AJ....105..847L} & J154912.55+304714.9 \\
CLASS B1600+434 & JCP & 240.4187 \tablenotemark{d} & 43.2798 & 40.532 & 2 & 1.4 & \citet{1995MNRAS.274L..25J} & J160140.50+431647.2 \\
CLASS B1608+656 & JCP & 242.3082 & 65.5413 & 35.675 & 4 & 2.27 & \citet{1995ApJ...447L...5M} & J160914.03+653228.1 \\
FSC 10214+4724 & OPTICAL & 156.1437 & 47.1531 & $<1.0$ & 4/Lobe/SFG & 1.0 & \citet{2013MNRAS.434.3322D} \newline \citet{1995ApJ...449L..29G} &  \\
MG B0751+2716 & RADIO & 117.923 & 27.2755 & 304.438 & 4 & 0.8 & \citet{1997AJ....114...48L} & J075141.53+271631.8 \\
JVAS B1938+666 & JCP & 294.6055 \tablenotemark{d} & 66.8148 & 398.267 & 4 & 1.02 & \citet{1997MNRAS.289..450K} & J193825.26+664852.8 \\
RX J0911+0551 & OPTICAL & 137.86479 \tablenotemark{a} & 5.848 & $<1.0$ & 4 & 3.25 & \citet{2015MNRAS.454..287J} \newline \citet{1997AaA...317L..13B} &  \\
CLASS B0712+472 & JCP & 109.0152 & 47.1474 & 26.245 & 4 & 1.46 & \citet{1998MNRAS.296..483J} & J071603.59+470850.1 \\
FBQ B0951+2635 & XMATCH & 147.84412 & 26.58725 & $<1.0$ & 2 & 1.1 & \citet{1998AJ....115.1371S} &  \\
CLASS B1933+503 & JCP & 293.6293 & 50.4232 & 79.109 & 4 & 1.52 & \citet{1998MNRAS.301..310S} & J193430.92+502523.3 \\
APM B08279+5255 & OPTICAL & 127.9235 \tablenotemark{d} & 52.75486 & $<1.0$ & 3 & 0.38 & \citet{1999AJ....118.1922I} \newline \citet{1998ApJ...505..529I} &  \\
JVAS B1030+074 & JCP & 158.3918 \tablenotemark{d} & 7.1906 & 300.708 & 2 & 1.65 & \citet{1998MNRAS.300..649X} & J103334.02+071126.3 \\
CLASS B1127+385 & JCP & 172.5007 & 38.2005 & 35.965 & 2 & 0.7 & \citet{1999MNRAS.303..727K} & J113000.14+381203.1 \\
CLASS B1152+199 & JCP & 178.8264 & 19.6615 & 53.896 & 2 & 1.56 & \citet{1999AJ....117.2565M} & J115518.32+193942.0 \\
CLASS B1359+154 & JCP & 210.3981 \tablenotemark{d} & 15.2237 & 45.277 & 6 & 1.71 & \citet{1999AJ....117.2565M} & J140135.54+151324.8 \\
CLASS B1555+375 & JCP & 239.2998 \tablenotemark{d} & 37.36 & 34.178 & 4 & 0.42 & \citet{1999AJ....118..654M} & J155711.95+372136.0 \\
CLASS B2045+265 & JCP & 311.8349 & 26.7339 & 36.293 & 4 & 1.9 & \citet{1999AJ....117..658F} & J204720.27+264402.4 \\
JVAS B2114+022 & JCP & 319.2116 & 2.4297 & 127.486 & 2 & 2.56 & \citet{1999MNRAS.307..225K} & J211650.76+022546.7 \\
HS B2209+1914 & OPTICAL & 332.87625 \tablenotemark{d} & 19.4869 & 2.024 & 2 & 1.04 & This work \newline \citet{hamburg} & J221130.31+192913.3 \\
CLASS B0128+437 & JCP & 22.8059 \tablenotemark{d} & 43.9703 & 61.221 & 4 & 0.55 & \citet{2000MNRAS.319L...7P} & J013113.45$+$435812.9 \\
PMN J1838$-$3427 & JCP & 279.6187 & $-34.4618$ & 214.667 & 2 & 0.99 & \citet{2000AJ....120.2868W} & J183828.50$-$342741.2 \\
CLASS B0739+366 & JCP & 115.7132 \tablenotemark{d} & 36.5788 & 30.526 & 2 & 0.53 & \citet{2001AJ....121..619M} & J074251.20+363443.6 \\
FIRST J0816+5003 & XMATCH & 124.1618 & 50.0688 & 64.939 & Lobe & 2.0 & \citet{2001ApJ...547...60L} & J081638.73+500407.2 \\
FIRST J0823+3906 \tablenotemark{b} & XMATCH & 125.8496 & 39.11 & 56.205 & Lobe & 5.0 & \citet{2001ApJ...547...60L} & J082323.65+390638.4 \\
FIRST J1622+3531 \tablenotemark{b} & XMATCH & 245.6239 & 35.5257 & 102.976 & Lobe & 3.0 & \citet{2001ApJ...547...60L} & J162229.77+353134.3 \\
PMN J2004$-$1349 & JCP & 301.0294 & $-$13.8252 & 22.097 & 2 & 1.13 & \citet{2001AJ....121.1223W} & J200407.05$-$134931.0 \\
CLASS B2319+051 & JCP & 350.4201 & 5.4602 & 68.925 & 2 & 1.36 & \citet{2001AJ....122..591R} & J232140.81+052737.3 \\
PMN J0134$-$0931 & JCP & 23.6486 \tablenotemark{d} & $-$9.5175 & 636.323 & 5 & 0.68 & \citet{2002ApJ...564..143W} & J013435.67$-$093102.7 \\
CLASS B0445+123 & JCP & 72.0916 \tablenotemark{d} & 12.4654 & 31.086 & 2 & 1.35 & \citet{2003MNRAS.338..957A} & J044822.00+122755.5 \\
FIRST J1004+1229 & XMATCH & 151.1037 & 12.4894 & 8.624 & 2 & 1.54 & \citet{2002AJ....123.2925L} & J100424.87+122922.5 \\
PMN J1632$-$0033 & JCP & 248.2403 & $-$0.5559 & 167.349 & 3 & 1.47 & \citet{2002AJ....123...10W} & J163257.68$-$003320.9 \\
HE B0435$-$1223 & OPTICAL & 69.56198 & $-$12.28739 & $<1.0$ & 4 & 2.54 & \citet{2015MNRAS.454..287J} \newline \citet{2002AaA...395...17W} &  \\
HS B0810+2554 & OPTICAL & 123.38054 & 25.75092 & $<1.0$ & 4 & 0.91 & \citet{2015MNRAS.454..287J} \newline \citet{2002AaA...382L..26R} &  \\
CLASS B0631+519 & JCP & 98.8013 \tablenotemark{d} & 51.9505 & 46.425 & 2 & 1.16 & \citet{2003MNRAS.341...13B} & J063512.35+515701.2 \\
CLASS B0850+054 & JCP & 133.2232 & 5.2543 & 78.061 & 2 & 0.68 & \citet{2003MNRAS.338.1084B} & J085253.57+051515.8 \\
CLASS B2108+213 & JCP & 317.7256 & 21.5162 & 36.473 & 2 & 4.57 & \citet{2003MNRAS.341...13B} & J211054.07+213058.8 \\
RXS J1131$-$1231 & OPTICAL & 172.96461 & $-$12.53289 & 4.008 & 4/SFG \tablenotemark{e} & 3.23 & \citet{Wucknitz:2009xu} \newline \citet{2003AaA...406L..43S} & J113151.53$-$123158.0 \\
SDSS J1004+4112 & OPTICAL & 151.14546 & 41.21189 & $<1.0$ & 4 & 14.62 & \citet{2011ApJ...739L..28J} \newline \citet{2003Natur.426..810I} &  \\
SDSS J0924+0219 & OPTICAL & 141.2325771 & 2.3234747 & $<1.0$ & 4 & 1.81 & \citet{2015MNRAS.454..287J} \newline \citet{2003AJ....126..666I} &  \\
FOV J0743+1553 \tablenotemark{b} & XMATCH & 115.9744 & 15.8903 & 47.366 & Lobe & 1.8 & \citet{2005AJ....130.1977H} & J074353.85+155324.8 \\
SDSS J1259+1241 \tablenotemark{cd} & OPTICAL & 194.9811138 \tablenotemark{d} & 12.69751076 & $<1.0$ & 2 & 3.5 & \citet{dobie23} \newline \citet{2006AJ....131....1H} &  \\
CLASS J0316+4328 & JCP & 49.2122 \tablenotemark{d} & 43.472 & 126.464 & 2 & 0.5 & \citet{2007MNRAS.381L..55B} & J031650.88+432819.2 \\
PSS J2322+1944 & OPTICAL & 350.5298 & 19.7397 & $<1.0$ & SFG & 1.5 & \citet{2008ApJ...686..851R} &  \\
WISE J2329$-$1258 & OPTICAL & 352.491 & $-$12.98306 & 1.013 & 2 & 1.26 & This work \newline \citet{2329disc} & J232957.86$-$125859.1 \\
PS J1721+8842 & GAIA & 260.43437 & 88.70599 & 1.848 & 4/2 \tablenotemark{f} & 4.03 & \citet{2021MNRAS.508L..64M} \newline \citet{Lemon2018} & J172146.08+884221.9 \\
GRAL J1131$-$4419 \tablenotemark{c} & GAIA & 172.750041 \tablenotemark{d} & $-$44.3330556 & \tablenotemark{a} & 4 & 1.7 & \citet{dobie23} \newline \citet{2018AaA...616L..11K} &  \\
WGD J2038$-$4008 \tablenotemark{c} & GAIA & 309.511278 \tablenotemark{d} & $-$40.137107 & \tablenotemark{a} & 4 & 2.87 & \citet{dobie23} \newline \citet{2018MNRAS.479.4345A} &  \\
MJV 1255+1158 \tablenotemark{b} & RADIO & 193.874 \tablenotemark{d} & 11.9816 & 36.024 & 2 & 0.46 & \citet{2019MNRAS.483.2125S} & J125529.76+115854.2 \\
MJV J1330+3141 \tablenotemark{b} & RADIO & 202.5398 \tablenotemark{d} & 31.6846 & 38.521 & 2 & 0.54 & \citet{2019MNRAS.483.2125S} & J133009.54+314104.5 \\
J0013+5119 & GAIA & 3.348077 \tablenotemark{d} & 51.3183 & 2.097 & 2 & 1.89 & This work \newline \citet{Lemon19} & J001323.53+511905.9 \\
DES J0229+0320 \tablenotemark{c} & GAIA & 37.49255525 \tablenotemark{d} & 37.49255525 & $<1.0$ & 2 & 2.14 & \citet{dobie23} \newline \citet{2020MNRAS.494.3491L} &  \\
GRAL J0246$-$1845 \tablenotemark{c} & GAIA & 41.5508333 \tablenotemark{d} & $-$18.7514167 & $<1.0$ & 2 & 1.0 & \citet{dobie23} \newline \citet{2019arXiv191208977K} &  \\
GRAL J0248+1913 & GAIA & 42.2031 \tablenotemark{d} & 19.22528 & $<1.0$ & 4 & 1.76 & \citet{dobie23} \newline \citet{2019AaA...622A.165D} &  \\
GRAL J0346+2154 & GAIA & 56.5458 & 21.9124 & $<1.0$ & 2 & 0.99 & \citet{dobie23} \newline \citet{2019arXiv191208977K} &  \\
GRAL J0530$-$3730 \tablenotemark{c} & GAIA & 82.6541 \tablenotemark{d} & 82.6541 & $<1.0$ & 3 & 1.04 & \citet{dobie23} \newline \citet{2019AaA...622A.165D} &  \\
GRAL J0659+1629 & GAIA & 104.766823 \tablenotemark{d} & 16.485772 & $<1.0$ & 4 & 5.2 & \citet{dobie23} \newline \citet{2019AaA...622A.165D} &  \\
GRAL J0818+0601 & GAIA & 124.6269582 \tablenotemark{d} & 6.027244393 & $<1.0$ & 2 & 1.15 & \citet{dobie23} \newline \citet{2019arXiv191208977K} &  \\
GRAL J1556$-$1352 \tablenotemark{c} & GAIA & 239.23375 \tablenotemark{d} & $-$13.8694722 & $<1.0$ & 2 & 0.96 & \citet{dobie23} \newline \citet{2019arXiv191208977K} &  \\
J2145+6345 & GAIA & 326.2713 & 63.7614461 & 1.182 & 4 & 2.07 & This work \newline \citet{Lemon19} & J214505.20+634541.1 \\
GRAL 2343+0435 & GAIA & 355.8775 \tablenotemark{d} & 4.5994444 & $<1.0$ & 2 & 1.23 & \citet{dobie23} \newline \citet{2019arXiv191208977K} &  \\
GRAL J0607$-$2152 \tablenotemark{c} & GAIA & 91.795 \tablenotemark{d} & $-$21.8713889 & $<1.0$ & 4 & 1.7 & \citet{dobie23} \newline \citet{2021ApJ...921...42S} &  \\
GRAL J0608+4229 & GAIA & 92.1725 \tablenotemark{d} & 42.4936111 & $<1.0$ & 4 & 1.3 & \citet{dobie23} \newline \citet{2021ApJ...921...42S} &  \\
GRAL J0818$-$2613 \tablenotemark{c} & GAIA & 124.6179167 \tablenotemark{d} & $-$26.2236111 & $<1.0$ & 4 & 6.2 & \citet{dobie23} \newline \citet{2021ApJ...921...42S} &  \\
J1537$-$3010 \tablenotemark{c} & GAIA & 234.355598 & $-$30.171335 & $<1.0$ & 4 & 3.3 & \citet{dobie23} \newline \citet{Lemon19} &  \\
GRAL J1651$-$0417 & GAIA & 252.7720833 \tablenotemark{d} & $-$4.2902778 & $<1.0$ & 4 & 10.1 & \citet{dobie23} \newline \citet{2021ApJ...921...42S} &  \\
GRAL J1817+2729 & GAIA & 274.378545 \tablenotemark{d} & 27.494468 & 2.575 & 4 & 1.8 & This work \newline \citet{dobie23} \newline \citet{Lemon19} & J181730.82+272940.2 \\
GRAL J2014$-$3024 & GAIA & 303.7258333 \tablenotemark{d} & $-$30.4144444 & $<1.0$ & 4 & 2.5 & \citet{dobie23} \newline \citet{2019AaA...622A.165D} &  \\
GRAL J2103$-$0850 & GAIA & 315.8708333 \tablenotemark{d} & $-$8.8469444 & $<1.0$ & 4 & 1.0 & \citet{dobie23} \newline \citet{2021ApJ...921...42S} &  \\
SDSS J0823+2418 & GAIA & 125.9211496 & 24.3015122 & $<1.0$ & 2 & 0.64 & \citet{2023ApJ...956..117G} \newline \citet{2021ApJ...921...42S} &  \\
\enddata
\tablecomments{Objects are ordered by lens discovery year. (1): The name given to object in its discovery paper.
(2): Discovery method for the lens system, using the following key: JCP - Bright, flat-spectrum source search, as seen in the JVAS \citep{1999MNRAS.307..225K}, CLASS \citep{2003MNRAS.341...13B}, and PMN \citep{2000AJ....120.2868W} lens surveys; RADIO - Other radio-based lens search or serendipitous radio discovery; XMATCH - Joint optical+radio search; OPTICAL - Lens system discovered by an optical search and confirmed as a radio source later; GAIA - Lens discovered specifically utilizing \textit{Gaia} data and confirmed as a radio source later. RA, Dec: Coordinates are J2000 and correspond to the lens deflector in each system, unless otherwise noted. Many close quasar lenses, such as those in \citet{dobie23}, have faint or blended lenses with poor astrometry, and in these cases the coordinates of the brightest image have been given instead. 
(3): Total flux from the nearest component to the lens coordinates within $5''$, using the \citet{Gordon2021} VLASS quick-look catalog. Non-detections are marked $<1.0mJy$ corresponding to that catalog's limiting flux.
(4): Number of images of the radio AGN visible in the system. Sources where the radio emission is from a lensed radio lobe rather than an AGN core are marked ``Lobe", and those where the emission is from a lensed high redshift, ultra-luminous star-forming galaxy are marked ``SFG". 
(5): Maximum image separation for lensed AGN cores. For SFG and Lobe sources the Einstein radius is given. 
(6): When multiple references are given, the first corresponds to the discovery of radio emission and the others to the (original) discovery of lensing at another wavelength.
(7): VLASS source associated with this lens system, left blank if no systems were matched within $5''$.}
\tablenotetext{a}{Source is outside of VLASS footprint ($\delta < 40^{\circ}$) or otherwise masked}
\tablenotetext{b}{Listed as a strong candidate for lensing but not spectroscopically confirmed}
\tablenotetext{c}{Lens system bright in radio but at too low resolution to confirm emission from source}
\tablenotetext{d}{Position of lens unreliable/unknown, position of brightest source image given instead}
\tablenotetext{e}{RXS J1131$-$1231 emits from both star-forming regions and the quadruply lensed AGN core.}
\tablenotetext{f}{PS J1721+8842 is a lensed dual AGN system, with one core quadruply lensed and the other doubly lensed.}
\end{deluxetable*}
\end{longrotatetable}

\bibliography{main}{}

\begin{thebibliography}{}
\expandafter\ifx\csname natexlab\endcsname\relax\def\natexlab#1{#1}\fi
\providecommand{\url}[1]{\href{#1}{#1}}
\providecommand{\dodoi}[1]{doi:~\href{http://doi.org/#1}{\nolinkurl{#1}}}
\providecommand{\doeprint}[1]{\href{http://ascl.net/#1}{\nolinkurl{http://ascl.net/#1}}}
\providecommand{\doarXiv}[1]{\href{https://arxiv.org/abs/#1}{\nolinkurl{https://arxiv.org/abs/#1}}}

\bibitem[{{Abazajian} {et~al.}(2009){Abazajian}, {Adelman-McCarthy}, {Ag{\"u}eros}, {Allam}, {Allende Prieto}, {An}, {Anderson}, {Anderson}, {Annis}, {Bahcall}, {Bailer-Jones}, {Barentine}, {Bassett}, {Becker}, {Beers}, {Bell}, {Belokurov}, {Berlind}, {Berman}, {Bernardi}, {Bickerton}, {Bizyaev}, {Blakeslee}, {Blanton}, {Bochanski}, {Boroski}, {Brewington}, {Brinchmann}, {Brinkmann}, {Brunner}, {Budav{\'a}ri}, {Carey}, {Carliles}, {Carr}, {Castander}, {Cinabro}, {Connolly}, {Csabai}, {Cunha}, {Czarapata}, {Davenport}, {de Haas}, {Dilday}, {Doi}, {Eisenstein}, {Evans}, {Evans}, {Fan}, {Friedman}, {Frieman}, {Fukugita}, {G{\"a}nsicke}, {Gates}, {Gillespie}, {Gilmore}, {Gonzalez}, {Gonzalez}, {Grebel}, {Gunn}, {Gy{\"o}ry}, {Hall}, {Harding}, {Harris}, {Harvanek}, {Hawley}, {Hayes}, {Heckman}, {Hendry}, {Hennessy}, {Hindsley}, {Hoblitt}, {Hogan}, {Hogg}, {Holtzman}, {Hyde}, {Ichikawa}, {Ichikawa}, {Im}, {Ivezi{\'c}}, {Jester}, {Jiang}, {Johnson}, {Jorgensen}, {Juri{\'c}}, {Kent}, {Kessler}, {Kleinman}, {Knapp},
  {Konishi}, {Kron}, {Krzesinski}, {Kuropatkin}, {Lampeitl}, {Lebedeva}, {Lee}, {Lee}, {French Leger}, {L{\'e}pine}, {Li}, {Lima}, {Lin}, {Long}, {Loomis}, {Loveday}, {Lupton}, {Magnier}, {Malanushenko}, {Malanushenko}, {Mandelbaum}, {Margon}, {Marriner}, {Mart{\'\i}nez-Delgado}, {Matsubara}, {McGehee}, {McKay}, {Meiksin}, {Morrison}, {Mullally}, {Munn}, {Murphy}, {Nash}, {Nebot}, {Neilsen}, {Newberg}, {Newman}, {Nichol}, {Nicinski}, {Nieto-Santisteban}, {Nitta}, {Okamura}, {Oravetz}, {Ostriker}, {Owen}, {Padmanabhan}, {Pan}, {Park}, {Pauls}, {Peoples}, {Percival}, {Pier}, {Pope}, {Pourbaix}, {Price}, {Purger}, {Quinn}, {Raddick}, {Re Fiorentin}, {Richards}, {Richmond}, {Riess}, {Rix}, {Rockosi}, {Sako}, {Schlegel}, {Schneider}, {Scholz}, {Schreiber}, {Schwope}, {Seljak}, {Sesar}, {Sheldon}, {Shimasaku}, {Sibley}, {Simmons}, {Sivarani}, {Allyn Smith}, {Smith}, {Smol{\v{c}}i{\'c}}, {Snedden}, {Stebbins}, {Steinmetz}, {Stoughton}, {Strauss}, {SubbaRao}, {Suto}, {Szalay}, {Szapudi}, {Szkody}, {Tanaka},
  {Tegmark}, {Teodoro}, {Thakar}, {Tremonti}, {Tucker}, {Uomoto}, {Vanden Berk}, {Vandenberg}, {Vidrih}, {Vogeley}, {Voges}, {Vogt}, {Wadadekar}, {Watters}, {Weinberg}, {West}, {White}, {Wilhite}, {Wonders}, {Yanny}, {Yocum}, {York}, {Zehavi}, {Zibetti}, \& {Zucker}}]{SDSSDR7}
{Abazajian}, K.~N., {Adelman-McCarthy}, J.~K., {Ag{\"u}eros}, M.~A., {et~al.} 2009, \apjs, 182, 543, \dodoi{10.1088/0067-0049/182/2/543}

\bibitem[{{Agnello} {et~al.}(2018){Agnello}, {Lin}, {Kuropatkin}, {Buckley-Geer}, {Anguita}, {Schechter}, {Morishita}, {Motta}, {Rojas}, {Treu}, {Amara}, {Auger}, {Courbin}, {Fassnacht}, {Frieman}, {More}, {Marshall}, {McMahon}, {Meylan}, {Suyu}, {Glazebrook}, {Morgan}, {Nord}, {Abbott}, {Abdalla}, {Annis}, {Bechtol}, {Benoit-L{\'e}vy}, {Bertin}, {Bernstein}, {Brooks}, {Burke}, {Rosell}, {Carretero}, {Cunha}, {D'Andrea}, {da Costa}, {Desai}, {Drlica-Wagner}, {Eifler}, {Flaugher}, {Garc{\'\i}a-Bellido}, {Gaztanaga}, {Gerdes}, {Gruen}, {Gruendl}, {Gschwend}, {Gutierrez}, {Honscheid}, {James}, {Kuehn}, {Lahav}, {Lima}, {Maia}, {March}, {Menanteau}, {Miquel}, {Ogando}, {Plazas}, {Sanchez}, {Scarpine}, {Schindler}, {Schubnell}, {Sevilla-Noarbe}, {Smith}, {Soares-Santos}, {Sobreira}, {Suchyta}, {Swanson}, {Tarle}, {Tucker}, \& {Wechsler}}]{2018MNRAS.479.4345A}
{Agnello}, A., {Lin}, H., {Kuropatkin}, N., {et~al.} 2018, \mnras, 479, 4345, \dodoi{10.1093/mnras/sty1419}

\bibitem[{{Argo} {et~al.}(2003){Argo}, {Jackson}, {Browne}, {York}, {McKean}, {Biggs}, {Blandford}, {de Bruyn}, {Chae}, {Fassnacht}, {Koopmans}, {Myers}, {Norbury}, {Pearson}, {Phillips}, {Readhead}, {Rusin}, \& {Wilkinson}}]{2003MNRAS.338..957A}
{Argo}, M.~K., {Jackson}, N.~J., {Browne}, I.~W.~A., {et~al.} 2003, \mnras, 338, 957, \dodoi{10.1046/j.1365-8711.2003.06138.x}

\bibitem[{{Astropy Collaboration} {et~al.}(2013){Astropy Collaboration}, {Robitaille}, {Tollerud}, {Greenfield}, {Droettboom}, {Bray}, {Aldcroft}, {Davis}, {Ginsburg}, {Price-Whelan}, {Kerzendorf}, {Conley}, {Crighton}, {Barbary}, {Muna}, {Ferguson}, {Grollier}, {Parikh}, {Nair}, {Unther}, {Deil}, {Woillez}, {Conseil}, {Kramer}, {Turner}, {Singer}, {Fox}, {Weaver}, {Zabalza}, {Edwards}, {Azalee Bostroem}, {Burke}, {Casey}, {Crawford}, {Dencheva}, {Ely}, {Jenness}, {Labrie}, {Lim}, {Pierfederici}, {Pontzen}, {Ptak}, {Refsdal}, {Servillat}, \& {Streicher}}]{2013A&A...558A..33A}
{Astropy Collaboration}, {Robitaille}, T.~P., {Tollerud}, E.~J., {et~al.} 2013, \aap, 558, A33, \dodoi{10.1051/0004-6361/201322068}

\bibitem[{{Astropy Collaboration} {et~al.}(2018){Astropy Collaboration}, {Price-Whelan}, {Sip{\H{o}}cz}, {G{\"u}nther}, {Lim}, {Crawford}, {Conseil}, {Shupe}, {Craig}, {Dencheva}, {Ginsburg}, {VanderPlas}, {Bradley}, {P{\'e}rez-Su{\'a}rez}, {de Val-Borro}, {Aldcroft}, {Cruz}, {Robitaille}, {Tollerud}, {Ardelean}, {Babej}, {Bach}, {Bachetti}, {Bakanov}, {Bamford}, {Barentsen}, {Barmby}, {Baumbach}, {Berry}, {Biscani}, {Boquien}, {Bostroem}, {Bouma}, {Brammer}, {Bray}, {Breytenbach}, {Buddelmeijer}, {Burke}, {Calderone}, {Cano Rodr{\'\i}guez}, {Cara}, {Cardoso}, {Cheedella}, {Copin}, {Corrales}, {Crichton}, {D'Avella}, {Deil}, {Depagne}, {Dietrich}, {Donath}, {Droettboom}, {Earl}, {Erben}, {Fabbro}, {Ferreira}, {Finethy}, {Fox}, {Garrison}, {Gibbons}, {Goldstein}, {Gommers}, {Greco}, {Greenfield}, {Groener}, {Grollier}, {Hagen}, {Hirst}, {Homeier}, {Horton}, {Hosseinzadeh}, {Hu}, {Hunkeler}, {Ivezi{\'c}}, {Jain}, {Jenness}, {Kanarek}, {Kendrew}, {Kern}, {Kerzendorf}, {Khvalko}, {King}, {Kirkby}, {Kulkarni},
  {Kumar}, {Lee}, {Lenz}, {Littlefair}, {Ma}, {Macleod}, {Mastropietro}, {McCully}, {Montagnac}, {Morris}, {Mueller}, {Mumford}, {Muna}, {Murphy}, {Nelson}, {Nguyen}, {Ninan}, {N{\"o}the}, {Ogaz}, {Oh}, {Parejko}, {Parley}, {Pascual}, {Patil}, {Patil}, {Plunkett}, {Prochaska}, {Rastogi}, {Reddy Janga}, {Sabater}, {Sakurikar}, {Seifert}, {Sherbert}, {Sherwood-Taylor}, {Shih}, {Sick}, {Silbiger}, {Singanamalla}, {Singer}, {Sladen}, {Sooley}, {Sornarajah}, {Streicher}, {Teuben}, {Thomas}, {Tremblay}, {Turner}, {Terr{\'o}n}, {van Kerkwijk}, {de la Vega}, {Watkins}, {Weaver}, {Whitmore}, {Woillez}, {Zabalza}, \& {Astropy Contributors}}]{2018AJ....156..123A}
{Astropy Collaboration}, {Price-Whelan}, A.~M., {Sip{\H{o}}cz}, B.~M., {et~al.} 2018, \aj, 156, 123, \dodoi{10.3847/1538-3881/aabc4f}

\bibitem[{{Bade} {et~al.}(1997){Bade}, {Siebert}, {Lopez}, {Voges}, \& {Reimers}}]{1997AaA...317L..13B}
{Bade}, N., {Siebert}, J., {Lopez}, S., {Voges}, W., \& {Reimers}, D. 1997, \aap, 317, L13

\bibitem[{{Bechtol} {et~al.}(2022){Bechtol}, {Birrer}, {Cyr-Racine}, {Schutz}, {Adhikari}, {Amin}, {Banerjee}, {Bird}, {Blinov}, {Boddy}, {Boehm}, {Bundy}, {Buschmann}, {Chakrabarti}, {Curtin}, {Dai}, {Drlica-Wagner}, {Dvorkin}, {Erickcek}, {Gilman}, {Heeba}, {Kim}, {Ir{\v{s}}i{\v{c}}}, {Leauthaud}, {Lovell}, {Luki{\'c}}, {Mao}, {Mau}, {Mitridate}, {Mocz}, {Mu{\~n}oz}, {Nadler}, {Peter}, {Price-Whelan}, {Robertson}, {Sabti}, {Sehgal}, {Shipp}, {Simon}, {Singh}, {Van Tilburg}, {Wechsler}, {Widmark}, \& {Yu}}]{bechtol22}
{Bechtol}, K., {Birrer}, S., {Cyr-Racine}, F.-Y., {et~al.} 2022, arXiv e-prints, arXiv:2203.07354, \dodoi{10.48550/arXiv.2203.07354}

\bibitem[{{Becker} {et~al.}(1995){Becker}, {White}, \& {Helfand}}]{FIRST}
{Becker}, R.~H., {White}, R.~L., \& {Helfand}, D.~J. 1995, \apj, 450, 559, \dodoi{10.1086/176166}

\bibitem[{{Biggs} {et~al.}(2003){Biggs}, {Rusin}, {Browne}, {de Bruyn}, {Jackson}, {Koopmans}, {McKean}, {Myers}, {Blandford}, {Chae}, {Fassnacht}, {Norbury}, {Pearson}, {Phillips}, {Readhead}, \& {Wilkinson}}]{2003MNRAS.338.1084B}
{Biggs}, A.~D., {Rusin}, D., {Browne}, I.~W.~A., {et~al.} 2003, \mnras, 338, 1084, \dodoi{10.1046/j.1365-8711.2003.06159.x}

\bibitem[{{Birrer} \& {Amara}(2018)}]{lenstronomy}
{Birrer}, S., \& {Amara}, A. 2018, Physics of the Dark Universe, 22, 189, \dodoi{10.1016/j.dark.2018.11.002}

\bibitem[{{Birrer} {et~al.}(2021){Birrer}, {Shajib}, {Gilman}, {Galan}, {Aalbers}, {Millon}, {Morgan}, {Pagano}, {Park}, {Teodori}, {Tessore}, {Ueland}, {Van de Vyvere}, {Wagner-Carena}, {Wempe}, {Yang}, {Ding}, {Schmidt}, {Sluse}, {Zhang}, \& {Amara}}]{lenstronomy2}
{Birrer}, S., {Shajib}, A., {Gilman}, D., {et~al.} 2021, The Journal of Open Source Software, 6, 3283, \dodoi{10.21105/joss.03283}

\bibitem[{{Bock} {et~al.}(1999){Bock}, {Large}, \& {Sadler}}]{Bock1999}
{Bock}, D.~C.~J., {Large}, M.~I., \& {Sadler}, E.~M. 1999, \aj, 117, 1578, \dodoi{10.1086/300786}

\bibitem[{{Boyce} {et~al.}(2007){Boyce}, {Myers}, {Browne}, {Stroman}, \& {Jackson}}]{2007MNRAS.381L..55B}
{Boyce}, E.~R., {Myers}, S.~T., {Browne}, I.~W.~A., {Stroman}, W.~J., \& {Jackson}, N.~J. 2007, \mnras, 381, L55, \dodoi{10.1111/j.1745-3933.2007.00365.x}

\bibitem[{{Braun} {et~al.}(2019){Braun}, {Bonaldi}, {Bourke}, {Keane}, \& {Wagg}}]{skaperformance}
{Braun}, R., {Bonaldi}, A., {Bourke}, T., {Keane}, E., \& {Wagg}, J. 2019, arXiv e-prints, arXiv:1912.12699, \dodoi{10.48550/arXiv.1912.12699}

\bibitem[{{Browne} {et~al.}(2003){Browne}, {Wilkinson}, {Jackson}, {Myers}, {Fassnacht}, {Koopmans}, {Marlow}, {Norbury}, {Rusin}, {Sykes}, {Biggs}, {Blandford}, {de Bruyn}, {Chae}, {Helbig}, {King}, {McKean}, {Pearson}, {Phillips}, {Readhead}, {Xanthopoulos}, \& {York}}]{2003MNRAS.341...13B}
{Browne}, I.~W.~A., {Wilkinson}, P.~N., {Jackson}, N.~J.~F., {et~al.} 2003, \mnras, 341, 13, \dodoi{10.1046/j.1365-8711.2003.06257.x}

\bibitem[{{Bruzewski} {et~al.}(2021){Bruzewski}, {Schinzel}, {Taylor}, \& {Petrov}}]{Bruzewski2021}
{Bruzewski}, S., {Schinzel}, F.~K., {Taylor}, G.~B., \& {Petrov}, L. 2021, \apj, 914, 42, \dodoi{10.3847/1538-4357/abf73b}

\bibitem[{{Carilli} {et~al.}(2015){Carilli}, {McKinnon}, {Ott}, {Beasley}, {Isella}, {Murphy}, {Leroy}, {Casey}, {Moullet}, {Lacy}, {Hodge}, {Bower}, {Demorest}, {Hull}, {Hughes}, {di Francesco}, {Narayanan}, {Kent}, {Clark}, \& {Butler}}]{Carilli2015}
{Carilli}, C.~L., {McKinnon}, M., {Ott}, J., {et~al.} 2015, {Next Generation Very Large Array Memo No. 5. Science Working Groups Project Overview}, \url{https://library.nrao.edu/public/memos/ngvla/NGVLA_05.pdf}

\bibitem[{{CASA Team} {et~al.}(2022){CASA Team}, {Bean}, {Bhatnagar}, {Castro}, {Donovan Meyer}, {Emonts}, {Garcia}, {Garwood}, {Golap}, {Gonzalez Villalba}, {Harris}, {Hayashi}, {Hoskins}, {Hsieh}, {Jagannathan}, {Kawasaki}, {Keimpema}, {Kettenis}, {Lopez}, {Marvil}, {Masters}, {McNichols}, {Mehringer}, {Miel}, {Moellenbrock}, {Montesino}, {Nakazato}, {Ott}, {Petry}, {Pokorny}, {Raba}, {Rau}, {Schiebel}, {Schweighart}, {Sekhar}, {Shimada}, {Small}, {Steeb}, {Sugimoto}, {Suoranta}, {Tsutsumi}, {van Bemmel}, {Verkouter}, {Wells}, {Xiong}, {Szomoru}, {Griffith}, {Glendenning}, \& {Kern}}]{casa}
{CASA Team}, {Bean}, B., {Bhatnagar}, S., {et~al.} 2022, \pasp, 134, 114501, \dodoi{10.1088/1538-3873/ac9642}

\bibitem[{{Chambers} {et~al.}(2016){Chambers}, {Magnier}, {Metcalfe}, {Flewelling}, {Huber}, {Waters}, {Denneau}, {Draper}, {Farrow}, {Finkbeiner}, {Holmberg}, {Koppenhoefer}, {Price}, {Rest}, {Saglia}, {Schlafly}, {Smartt}, {Sweeney}, {Wainscoat}, {Burgett}, {Chastel}, {Grav}, {Heasley}, {Hodapp}, {Jedicke}, {Kaiser}, {Kudritzki}, {Luppino}, {Lupton}, {Monet}, {Morgan}, {Onaka}, {Shiao}, {Stubbs}, {Tonry}, {White}, {Ba{\~n}ados}, {Bell}, {Bender}, {Bernard}, {Boegner}, {Boffi}, {Botticella}, {Calamida}, {Casertano}, {Chen}, {Chen}, {Cole}, {Deacon}, {Frenk}, {Fitzsimmons}, {Gezari}, {Gibbs}, {Goessl}, {Goggia}, {Gourgue}, {Goldman}, {Grant}, {Grebel}, {Hambly}, {Hasinger}, {Heavens}, {Heckman}, {Henderson}, {Henning}, {Holman}, {Hopp}, {Ip}, {Isani}, {Jackson}, {Keyes}, {Koekemoer}, {Kotak}, {Le}, {Liska}, {Long}, {Lucey}, {Liu}, {Martin}, {Masci}, {McLean}, {Mindel}, {Misra}, {Morganson}, {Murphy}, {Obaika}, {Narayan}, {Nieto-Santisteban}, {Norberg}, {Peacock}, {Pier}, {Postman}, {Primak}, {Rae}, {Rai},
  {Riess}, {Riffeser}, {Rix}, {R{\"o}ser}, {Russel}, {Rutz}, {Schilbach}, {Schultz}, {Scolnic}, {Strolger}, {Szalay}, {Seitz}, {Small}, {Smith}, {Soderblom}, {Taylor}, {Thomson}, {Taylor}, {Thakar}, {Thiel}, {Thilker}, {Unger}, {Urata}, {Valenti}, {Wagner}, {Walder}, {Walter}, {Watters}, {Werner}, {Wood-Vasey}, \& {Wyse}}]{Chambers2016}
{Chambers}, K.~C., {Magnier}, E.~A., {Metcalfe}, N., {et~al.} 2016, arXiv e-prints, arXiv:1612.05560, \dodoi{10.48550/arXiv.1612.05560}

\bibitem[{{Collett}(2015)}]{collett15}
{Collett}, T.~E. 2015, \apj, 811, 20, \dodoi{10.1088/0004-637X/811/1/20}

\bibitem[{{Condon}(1992)}]{condon92}
{Condon}, J.~J. 1992, \araa, 30, 575, \dodoi{10.1146/annurev.aa.30.090192.003043}

\bibitem[{{Condon} {et~al.}(1998){Condon}, {Cotton}, {Greisen}, {Yin}, {Perley}, {Taylor}, \& {Broderick}}]{Condon1998}
{Condon}, J.~J., {Cotton}, W.~D., {Greisen}, E.~W., {et~al.} 1998, \aj, 115, 1693, \dodoi{10.1086/300337}

\bibitem[{{Conway} {et~al.}(1990){Conway}, {Cornwell}, \& {Wilkinson}}]{mfs}
{Conway}, J.~E., {Cornwell}, T.~J., \& {Wilkinson}, P.~N. 1990, \mnras, 246, 490

\bibitem[{{Dark Energy Survey Collaboration} {et~al.}(2016){Dark Energy Survey Collaboration}, {Abbott}, {Abdalla}, {Aleksi{\'c}}, {Allam}, {Amara}, {Bacon}, {Balbinot}, {Banerji}, {Bechtol}, {Benoit-L{\'e}vy}, {Bernstein}, {Bertin}, {Blazek}, {Bonnett}, {Bridle}, {Brooks}, {Brunner}, {Buckley-Geer}, {Burke}, {Caminha}, {Capozzi}, {Carlsen}, {Carnero-Rosell}, {Carollo}, {Carrasco-Kind}, {Carretero}, {Castander}, {Clerkin}, {Collett}, {Conselice}, {Crocce}, {Cunha}, {D'Andrea}, {da Costa}, {Davis}, {Desai}, {Diehl}, {Dietrich}, {Dodelson}, {Doel}, {Drlica-Wagner}, {Estrada}, {Etherington}, {Evrard}, {Fabbri}, {Finley}, {Flaugher}, {Foley}, {Fosalba}, {Frieman}, {Garc{\'\i}a-Bellido}, {Gaztanaga}, {Gerdes}, {Giannantonio}, {Goldstein}, {Gruen}, {Gruendl}, {Guarnieri}, {Gutierrez}, {Hartley}, {Honscheid}, {Jain}, {James}, {Jeltema}, {Jouvel}, {Kessler}, {King}, {Kirk}, {Kron}, {Kuehn}, {Kuropatkin}, {Lahav}, {Li}, {Lima}, {Lin}, {Maia}, {Makler}, {Manera}, {Maraston}, {Marshall}, {Martini}, {McMahon},
  {Melchior}, {Merson}, {Miller}, {Miquel}, {Mohr}, {Morice-Atkinson}, {Naidoo}, {Neilsen}, {Nichol}, {Nord}, {Ogando}, {Ostrovski}, {Palmese}, {Papadopoulos}, {Peiris}, {Peoples}, {Percival}, {Plazas}, {Reed}, {Refregier}, {Romer}, {Roodman}, {Ross}, {Rozo}, {Rykoff}, {Sadeh}, {Sako}, {S{\'a}nchez}, {Sanchez}, {Santiago}, {Scarpine}, {Schubnell}, {Sevilla-Noarbe}, {Sheldon}, {Smith}, {Smith}, {Soares-Santos}, {Sobreira}, {Soumagnac}, {Suchyta}, {Sullivan}, {Swanson}, {Tarle}, {Thaler}, {Thomas}, {Thomas}, {Tucker}, {Vieira}, {Vikram}, {Walker}, {Wechsler}, {Weller}, {Wester}, {Whiteway}, {Wilcox}, {Yanny}, {Zhang}, \& {Zuntz}}]{DES2016}
{Dark Energy Survey Collaboration}, {Abbott}, T., {Abdalla}, F.~B., {et~al.} 2016, \mnras, 460, 1270, \dodoi{10.1093/mnras/stw641}

\bibitem[{{Dawes} {et~al.}(2023){Dawes}, {Storfer}, {Huang}, {Aldering}, {Cikota}, {Dey}, \& {Schlegel}}]{dawes23}
{Dawes}, C., {Storfer}, C., {Huang}, X., {et~al.} 2023, \apjs, 269, 61, \dodoi{10.3847/1538-4365/ad015a}

\bibitem[{{Deane} {et~al.}(2013){Deane}, {Rawlings}, {Garrett}, {Heywood}, {Jarvis}, {Kl{\"o}ckner}, {Marshall}, \& {McKean}}]{2013MNRAS.434.3322D}
{Deane}, R.~P., {Rawlings}, S., {Garrett}, M.~A., {et~al.} 2013, \mnras, 434, 3322, \dodoi{10.1093/mnras/stt1241}

\bibitem[{{Delchambre} {et~al.}(2019{\natexlab{a}}){Delchambre}, {Krone-Martins}, {Wertz}, {Ducourant}, {Galluccio}, {Kl{\"u}ter}, {Mignard}, {Teixeira}, {Djorgovski}, {Stern}, {Graham}, {Surdej}, {Bastian}, {Wambsganss}, {Le Campion}, \& {Slezak}}]{1817}
{Delchambre}, L., {Krone-Martins}, A., {Wertz}, O., {et~al.} 2019{\natexlab{a}}, \aap, 622, A165, \dodoi{10.1051/0004-6361/201833802}

\bibitem[{{Delchambre} {et~al.}(2019{\natexlab{b}}){Delchambre}, {Krone-Martins}, {Wertz}, {Ducourant}, {Galluccio}, {Kl{\"u}ter}, {Mignard}, {Teixeira}, {Djorgovski}, {Stern}, {Graham}, {Surdej}, {Bastian}, {Wambsganss}, {Le Campion}, \& {Slezak}}]{2019AaA...622A.165D}
---. 2019{\natexlab{b}}, \aap, 622, A165, \dodoi{10.1051/0004-6361/201833802}

\bibitem[{{Dey} {et~al.}(2019){Dey}, {Schlegel}, {Lang}, {Blum}, {Burleigh}, {Fan}, {Findlay}, {Finkbeiner}, {Herrera}, {Juneau}, {Landriau}, {Levi}, {McGreer}, {Meisner}, {Myers}, {Moustakas}, {Nugent}, {Patej}, {Schlafly}, {Walker}, {Valdes}, {Weaver}, {Y{\`e}che}, {Zou}, {Zhou}, {Abareshi}, {Abbott}, {Abolfathi}, {Aguilera}, {Alam}, {Allen}, {Alvarez}, {Annis}, {Ansarinejad}, {Aubert}, {Beechert}, {Bell}, {BenZvi}, {Beutler}, {Bielby}, {Bolton}, {Brice{\~n}o}, {Buckley-Geer}, {Butler}, {Calamida}, {Carlberg}, {Carter}, {Casas}, {Castander}, {Choi}, {Comparat}, {Cukanovaite}, {Delubac}, {DeVries}, {Dey}, {Dhungana}, {Dickinson}, {Ding}, {Donaldson}, {Duan}, {Duckworth}, {Eftekharzadeh}, {Eisenstein}, {Etourneau}, {Fagrelius}, {Farihi}, {Fitzpatrick}, {Font-Ribera}, {Fulmer}, {G{\"a}nsicke}, {Gaztanaga}, {George}, {Gerdes}, {Gontcho}, {Gorgoni}, {Green}, {Guy}, {Harmer}, {Hernandez}, {Honscheid}, {Huang}, {James}, {Jannuzi}, {Jiang}, {Joyce}, {Karcher}, {Karkar}, {Kehoe}, {Kneib}, {Kueter-Young}, {Lan},
  {Lauer}, {Le Guillou}, {Le Van Suu}, {Lee}, {Lesser}, {Perreault Levasseur}, {Li}, {Mann}, {Marshall}, {Mart{\'\i}nez-V{\'a}zquez}, {Martini}, {du Mas des Bourboux}, {McManus}, {Meier}, {M{\'e}nard}, {Metcalfe}, {Mu{\~n}oz-Guti{\'e}rrez}, {Najita}, {Napier}, {Narayan}, {Newman}, {Nie}, {Nord}, {Norman}, {Olsen}, {Paat}, {Palanque-Delabrouille}, {Peng}, {Poppett}, {Poremba}, {Prakash}, {Rabinowitz}, {Raichoor}, {Rezaie}, {Robertson}, {Roe}, {Ross}, {Ross}, {Rudnick}, {Safonova}, {Saha}, {S{\'a}nchez}, {Savary}, {Schweiker}, {Scott}, {Seo}, {Shan}, {Silva}, {Slepian}, {Soto}, {Sprayberry}, {Staten}, {Stillman}, {Stupak}, {Summers}, {Sien Tie}, {Tirado}, {Vargas-Maga{\~n}a}, {Vivas}, {Wechsler}, {Williams}, {Yang}, {Yang}, {Yapici}, {Zaritsky}, {Zenteno}, {Zhang}, {Zhang}, {Zhou}, \& {Zhou}}]{Dey2019}
{Dey}, A., {Schlegel}, D.~J., {Lang}, D., {et~al.} 2019, \aj, 157, 168, \dodoi{10.3847/1538-3881/ab089d}

\bibitem[{{Dobie} {et~al.}(2024){Dobie}, {Sluse}, {Deller}, {Murphy}, {Krone-Martins}, {Stern}, {Wang}, {Wang}, {B{\oe}hm}, {Djorgovski}, {Galluccio}, {Delchambre}, {Connor}, {den Brok}, {Do Vale Cunha}, {Ducourant}, {Graham}, {Jalan}, {Klioner}, {Kl{\"u}ter}, {Mignard}, {Negi}, {Petit}, {Scarano}, {Slezak}, {Surdej}, {Teixeira}, {Walton}, \& {Wambsganss}}]{dobie23}
{Dobie}, D., {Sluse}, D., {Deller}, A., {et~al.} 2024, \mnras, 528, 5880, \dodoi{10.1093/mnras/stad4002}

\bibitem[{{Falco} {et~al.}(1996){Falco}, {Lehar}, {Perley}, {Wambsganss}, \& {Gorenstein}}]{1996AJ....112..897F}
{Falco}, E.~E., {Lehar}, J., {Perley}, R.~A., {Wambsganss}, J., \& {Gorenstein}, M.~V. 1996, \aj, 112, 897, \dodoi{10.1086/118062}

\bibitem[{{Fassnacht} {et~al.}(1999){Fassnacht}, {Blandford}, {Cohen}, {Matthews}, {Pearson}, {Readhead}, {Womble}, {Myers}, {Browne}, {Jackson}, {Marlow}, {Wilkinson}, {Koopmans}, {de Bruyn}, {Schilizzi}, {Bremer}, \& {Miley}}]{1999AJ....117..658F}
{Fassnacht}, C.~D., {Blandford}, R.~D., {Cohen}, J.~G., {et~al.} 1999, \aj, 117, 658, \dodoi{10.1086/300724}

\bibitem[{{Gaia Collaboration} {et~al.}(2016){Gaia Collaboration}, {Prusti}, {de Bruijne}, {Brown}, {Vallenari}, {Babusiaux}, {Bailer-Jones}, {Bastian}, {Biermann}, {Evans}, {Eyer}, {Jansen}, {Jordi}, {Klioner}, {Lammers}, {Lindegren}, {Luri}, {Mignard}, {Milligan}, {Panem}, {Poinsignon}, {Pourbaix}, {Randich}, {Sarri}, {Sartoretti}, {Siddiqui}, {Soubiran}, {Valette}, {van Leeuwen}, {Walton}, {Aerts}, {Arenou}, {Cropper}, {Drimmel}, {H{\o}g}, {Katz}, {Lattanzi}, {O'Mullane}, {Grebel}, {Holland}, {Huc}, {Passot}, {Bramante}, {Cacciari}, {Casta{\~n}eda}, {Chaoul}, {Cheek}, {De Angeli}, {Fabricius}, {Guerra}, {Hern{\'a}ndez}, {Jean-Antoine-Piccolo}, {Masana}, {Messineo}, {Mowlavi}, {Nienartowicz}, {Ord{\'o}{\~n}ez-Blanco}, {Panuzzo}, {Portell}, {Richards}, {Riello}, {Seabroke}, {Tanga}, {Th{\'e}venin}, {Torra}, {Els}, {Gracia-Abril}, {Comoretto}, {Garcia-Reinaldos}, {Lock}, {Mercier}, {Altmann}, {Andrae}, {Astraatmadja}, {Bellas-Velidis}, {Benson}, {Berthier}, {Blomme}, {Busso}, {Carry}, {Cellino}, {Clementini},
  {Cowell}, {Creevey}, {Cuypers}, {Davidson}, {De Ridder}, {de Torres}, {Delchambre}, {Dell'Oro}, {Ducourant}, {Fr{\'e}mat}, {Garc{\'\i}a-Torres}, {Gosset}, {Halbwachs}, {Hambly}, {Harrison}, {Hauser}, {Hestroffer}, {Hodgkin}, {Huckle}, {Hutton}, {Jasniewicz}, {Jordan}, {Kontizas}, {Korn}, {Lanzafame}, {Manteiga}, {Moitinho}, {Muinonen}, {Osinde}, {Pancino}, {Pauwels}, {Petit}, {Recio-Blanco}, {Robin}, {Sarro}, {Siopis}, {Smith}, {Smith}, {Sozzetti}, {Thuillot}, {van Reeven}, {Viala}, {Abbas}, {Abreu Aramburu}, {Accart}, {Aguado}, {Allan}, {Allasia}, {Altavilla}, {{\'A}lvarez}, {Alves}, {Anderson}, {Andrei}, {Anglada Varela}, {Antiche}, {Antoja}, {Ant{\'o}n}, {Arcay}, {Atzei}, {Ayache}, {Bach}, {Baker}, {Balaguer-N{\'u}{\~n}ez}, {Barache}, {Barata}, {Barbier}, {Barblan}, {Baroni}, {Barrado y Navascu{\'e}s}, {Barros}, {Barstow}, {Becciani}, {Bellazzini}, {Bellei}, {Bello Garc{\'\i}a}, {Belokurov}, {Bendjoya}, {Berihuete}, {Bianchi}, {Bienaym{\'e}}, {Billebaud}, {Blagorodnova}, {Blanco-Cuaresma}, {Boch},
  {Bombrun}, {Borrachero}, {Bouquillon}, {Bourda}, {Bouy}, {Bragaglia}, {Breddels}, {Brouillet}, {Br{\"u}semeister}, {Bucciarelli}, {Budnik}, {Burgess}, {Burgon}, {Burlacu}, {Busonero}, {Buzzi}, {Caffau}, {Cambras}, {Campbell}, {Cancelliere}, {Cantat-Gaudin}, {Carlucci}, {Carrasco}, {Castellani}, {Charlot}, {Charnas}, {Charvet}, {Chassat}, {Chiavassa}, {Clotet}, {Cocozza}, {Collins}, {Collins}, {Costigan}, {Crifo}, {Cross}, {Crosta}, {Crowley}, {Dafonte}, {Damerdji}, {Dapergolas}, {David}, {David}, {De Cat}, {de Felice}, {de Laverny}, {De Luise}, {De March}, {de Martino}, {de Souza}, {Debosscher}, {del Pozo}, {Delbo}, {Delgado}, {Delgado}, {di Marco}, {Di Matteo}, {Diakite}, {Distefano}, {Dolding}, {Dos Anjos}, {Drazinos}, {Dur{\'a}n}, {Dzigan}, {Ecale}, {Edvardsson}, {Enke}, {Erdmann}, {Escolar}, {Espina}, {Evans}, {Eynard Bontemps}, {Fabre}, {Fabrizio}, {Faigler}, {Falc{\~a}o}, {Farr{\`a}s Casas}, {Faye}, {Federici}, {Fedorets}, {Fern{\'a}ndez-Hern{\'a}ndez}, {Fernique}, {Fienga}, {Figueras}, {Filippi},
  {Findeisen}, {Fonti}, {Fouesneau}, {Fraile}, {Fraser}, {Fuchs}, {Furnell}, {Gai}, {Galleti}, {Galluccio}, {Garabato}, {Garc{\'\i}a-Sedano}, {Gar{\'e}}, {Garofalo}, {Garralda}, {Gavras}, {Gerssen}, {Geyer}, {Gilmore}, {Girona}, {Giuffrida}, {Gomes}, {Gonz{\'a}lez-Marcos}, {Gonz{\'a}lez-N{\'u}{\~n}ez}, {Gonz{\'a}lez-Vidal}, {Granvik}, {Guerrier}, {Guillout}, {Guiraud}, {G{\'u}rpide}, {Guti{\'e}rrez-S{\'a}nchez}, {Guy}, {Haigron}, {Hatzidimitriou}, {Haywood}, {Heiter}, {Helmi}, {Hobbs}, {Hofmann}, {Holl}, {Holland}, {Hunt}, {Hypki}, {Icardi}, {Irwin}, {Jevardat de Fombelle}, {Jofr{\'e}}, {Jonker}, {Jorissen}, {Julbe}, {Karampelas}, {Kochoska}, {Kohley}, {Kolenberg}, {Kontizas}, {Koposov}, {Kordopatis}, {Koubsky}, {Kowalczyk}, {Krone-Martins}, {Kudryashova}, {Kull}, {Bachchan}, {Lacoste-Seris}, {Lanza}, {Lavigne}, {Le Poncin-Lafitte}, {Lebreton}, {Lebzelter}, {Leccia}, {Leclerc}, {Lecoeur-Taibi}, {Lemaitre}, {Lenhardt}, {Leroux}, {Liao}, {Licata}, {Lindstr{\o}m}, {Lister}, {Livanou}, {Lobel}, {L{\"o}ffler},
  {L{\'o}pez}, {Lopez-Lozano}, {Lorenz}, {Loureiro}, {MacDonald}, {Magalh{\~a}es Fernandes}, {Managau}, {Mann}, {Mantelet}, {Marchal}, {Marchant}, {Marconi}, {Marie}, {Marinoni}, {Marrese}, {Marschalk{\'o}}, {Marshall}, {Mart{\'\i}n-Fleitas}, {Martino}, {Mary}, {Matijevi{\v{c}}}, {Mazeh}, {McMillan}, {Messina}, {Mestre}, {Michalik}, {Millar}, {Miranda}, {Molina}, {Molinaro}, {Molinaro}, {Moln{\'a}r}, {Moniez}, {Montegriffo}, {Monteiro}, {Mor}, {Mora}, {Morbidelli}, {Morel}, {Morgenthaler}, {Morley}, {Morris}, {Mulone}, {Muraveva}, {Musella}, {Narbonne}, {Nelemans}, {Nicastro}, {Noval}, {Ord{\'e}novic}, {Ordieres-Mer{\'e}}, {Osborne}, {Pagani}, {Pagano}, {Pailler}, {Palacin}, {Palaversa}, {Parsons}, {Paulsen}, {Pecoraro}, {Pedrosa}, {Pentik{\"a}inen}, {Pereira}, {Pichon}, {Piersimoni}, {Pineau}, {Plachy}, {Plum}, {Poujoulet}, {Pr{\v{s}}a}, {Pulone}, {Ragaini}, {Rago}, {Rambaux}, {Ramos-Lerate}, {Ranalli}, {Rauw}, {Read}, {Regibo}, {Renk}, {Reyl{\'e}}, {Ribeiro}, {Rimoldini}, {Ripepi}, {Riva}, {Rixon},
  {Roelens}, {Romero-G{\'o}mez}, {Rowell}, {Royer}, {Rudolph}, {Ruiz-Dern}, {Sadowski}, {Sagrist{\`a} Sell{\'e}s}, {Sahlmann}, {Salgado}, {Salguero}, {Sarasso}, {Savietto}, {Schnorhk}, {Schultheis}, {Sciacca}, {Segol}, {Segovia}, {Segransan}, {Serpell}, {Shih}, {Smareglia}, {Smart}, {Smith}, {Solano}, {Solitro}, {Sordo}, {Soria Nieto}, {Souchay}, {Spagna}, {Spoto}, {Stampa}, {Steele}, {Steidelm{\"u}ller}, {Stephenson}, {Stoev}, {Suess}, {S{\"u}veges}, {Surdej}, {Szabados}, {Szegedi-Elek}, {Tapiador}, {Taris}, {Tauran}, {Taylor}, {Teixeira}, {Terrett}, {Tingley}, {Trager}, {Turon}, {Ulla}, {Utrilla}, {Valentini}, {van Elteren}, {Van Hemelryck}, {van Leeuwen}, {Varadi}, {Vecchiato}, {Veljanoski}, {Via}, {Vicente}, {Vogt}, {Voss}, {Votruba}, {Voutsinas}, {Walmsley}, {Weiler}, {Weingrill}, {Werner}, {Wevers}, {Whitehead}, {Wyrzykowski}, {Yoldas}, {{\v{Z}}erjal}, {Zucker}, {Zurbach}, {Zwitter}, {Alecu}, {Allen}, {Allende Prieto}, {Amorim}, {Anglada-Escud{\'e}}, {Arsenijevic}, {Azaz}, {Balm}, {Beck}, {Bernstein},
  {Bigot}, {Bijaoui}, {Blasco}, {Bonfigli}, {Bono}, {Boudreault}, {Bressan}, {Brown}, {Brunet}, {Bunclark}, {Buonanno}, {Butkevich}, {Carret}, {Carrion}, {Chemin}, {Ch{\'e}reau}, {Corcione}, {Darmigny}, {de Boer}, {de Teodoro}, {de Zeeuw}, {Delle Luche}, {Domingues}, {Dubath}, {Fodor}, {Fr{\'e}zouls}, {Fries}, {Fustes}, {Fyfe}, {Gallardo}, {Gallegos}, {Gardiol}, {Gebran}, {Gomboc}, {G{\'o}mez}, {Grux}, {Gueguen}, {Heyrovsky}, {Hoar}, {Iannicola}, {Isasi Parache}, {Janotto}, {Joliet}, {Jonckheere}, {Keil}, {Kim}, {Klagyivik}, {Klar}, {Knude}, {Kochukhov}, {Kolka}, {Kos}, {Kutka}, {Lainey}, {LeBouquin}, {Liu}, {Loreggia}, {Makarov}, {Marseille}, {Martayan}, {Martinez-Rubi}, {Massart}, {Meynadier}, {Mignot}, {Munari}, {Nguyen}, {Nordlander}, {Ocvirk}, {O'Flaherty}, {Olias Sanz}, {Ortiz}, {Osorio}, {Oszkiewicz}, {Ouzounis}, {Palmer}, {Park}, {Pasquato}, {Peltzer}, {Peralta}, {P{\'e}turaud}, {Pieniluoma}, {Pigozzi}, {Poels}, {Prat}, {Prod'homme}, {Raison}, {Rebordao}, {Risquez}, {Rocca-Volmerange}, {Rosen},
  {Ruiz-Fuertes}, {Russo}, {Sembay}, {Serraller Vizcaino}, {Short}, {Siebert}, {Silva}, {Sinachopoulos}, {Slezak}, {Soffel}, {Sosnowska}, {Strai{\v{z}}ys}, {ter Linden}, {Terrell}, {Theil}, {Tiede}, {Troisi}, {Tsalmantza}, {Tur}, {Vaccari}, {Vachier}, {Valles}, {Van Hamme}, {Veltz}, {Virtanen}, {Wallut}, {Wichmann}, {Wilkinson}, {Ziaeepour}, \& {Zschocke}}]{Gaia2016}
{Gaia Collaboration}, {Prusti}, T., {de Bruijne}, J.~H.~J., {et~al.} 2016, \aap, 595, A1, \dodoi{10.1051/0004-6361/201629272}

\bibitem[{{Gaia Collaboration} {et~al.}(2023){Gaia Collaboration}, {Vallenari}, {Brown}, {Prusti}, {de Bruijne}, {Arenou}, {Babusiaux}, {Biermann}, {Creevey}, {Ducourant}, {Evans}, {Eyer}, {Guerra}, {Hutton}, {Jordi}, {Klioner}, {Lammers}, {Lindegren}, {Luri}, {Mignard}, {Panem}, {Pourbaix}, {Randich}, {Sartoretti}, {Soubiran}, {Tanga}, {Walton}, {Bailer-Jones}, {Bastian}, {Drimmel}, {Jansen}, {Katz}, {Lattanzi}, {van Leeuwen}, {Bakker}, {Cacciari}, {Casta{\~n}eda}, {De Angeli}, {Fabricius}, {Fouesneau}, {Fr{\'e}mat}, {Galluccio}, {Guerrier}, {Heiter}, {Masana}, {Messineo}, {Mowlavi}, {Nicolas}, {Nienartowicz}, {Pailler}, {Panuzzo}, {Riclet}, {Roux}, {Seabroke}, {Sordo}, {Th{\'e}venin}, {Gracia-Abril}, {Portell}, {Teyssier}, {Altmann}, {Andrae}, {Audard}, {Bellas-Velidis}, {Benson}, {Berthier}, {Blomme}, {Burgess}, {Busonero}, {Busso}, {C{\'a}novas}, {Carry}, {Cellino}, {Cheek}, {Clementini}, {Damerdji}, {Davidson}, {de Teodoro}, {Nu{\~n}ez Campos}, {Delchambre}, {Dell'Oro}, {Esquej},
  {Fern{\'a}ndez-Hern{\'a}ndez}, {Fraile}, {Garabato}, {Garc{\'\i}a-Lario}, {Gosset}, {Haigron}, {Halbwachs}, {Hambly}, {Harrison}, {Hern{\'a}ndez}, {Hestroffer}, {Hodgkin}, {Holl}, {Jan{\ss}en}, {Jevardat de Fombelle}, {Jordan}, {Krone-Martins}, {Lanzafame}, {L{\"o}ffler}, {Marchal}, {Marrese}, {Moitinho}, {Muinonen}, {Osborne}, {Pancino}, {Pauwels}, {Recio-Blanco}, {Reyl{\'e}}, {Riello}, {Rimoldini}, {Roegiers}, {Rybizki}, {Sarro}, {Siopis}, {Smith}, {Sozzetti}, {Utrilla}, {van Leeuwen}, {Abbas}, {{\'A}brah{\'a}m}, {Abreu Aramburu}, {Aerts}, {Aguado}, {Ajaj}, {Aldea-Montero}, {Altavilla}, {{\'A}lvarez}, {Alves}, {Anders}, {Anderson}, {Anglada Varela}, {Antoja}, {Baines}, {Baker}, {Balaguer-N{\'u}{\~n}ez}, {Balbinot}, {Balog}, {Barache}, {Barbato}, {Barros}, {Barstow}, {Bartolom{\'e}}, {Bassilana}, {Bauchet}, {Becciani}, {Bellazzini}, {Berihuete}, {Bernet}, {Bertone}, {Bianchi}, {Binnenfeld}, {Blanco-Cuaresma}, {Blazere}, {Boch}, {Bombrun}, {Bossini}, {Bouquillon}, {Bragaglia}, {Bramante}, {Breedt},
  {Bressan}, {Brouillet}, {Brugaletta}, {Bucciarelli}, {Burlacu}, {Butkevich}, {Buzzi}, {Caffau}, {Cancelliere}, {Cantat-Gaudin}, {Carballo}, {Carlucci}, {Carnerero}, {Carrasco}, {Casamiquela}, {Castellani}, {Castro-Ginard}, {Chaoul}, {Charlot}, {Chemin}, {Chiaramida}, {Chiavassa}, {Chornay}, {Comoretto}, {Contursi}, {Cooper}, {Cornez}, {Cowell}, {Crifo}, {Cropper}, {Crosta}, {Crowley}, {Dafonte}, {Dapergolas}, {David}, {David}, {de Laverny}, {De Luise}, {De March}, {De Ridder}, {de Souza}, {de Torres}, {del Peloso}, {del Pozo}, {Delbo}, {Delgado}, {Delisle}, {Demouchy}, {Dharmawardena}, {Di Matteo}, {Diakite}, {Diener}, {Distefano}, {Dolding}, {Edvardsson}, {Enke}, {Fabre}, {Fabrizio}, {Faigler}, {Fedorets}, {Fernique}, {Fienga}, {Figueras}, {Fournier}, {Fouron}, {Fragkoudi}, {Gai}, {Garcia-Gutierrez}, {Garcia-Reinaldos}, {Garc{\'\i}a-Torres}, {Garofalo}, {Gavel}, {Gavras}, {Gerlach}, {Geyer}, {Giacobbe}, {Gilmore}, {Girona}, {Giuffrida}, {Gomel}, {Gomez}, {Gonz{\'a}lez-N{\'u}{\~n}ez},
  {Gonz{\'a}lez-Santamar{\'\i}a}, {Gonz{\'a}lez-Vidal}, {Granvik}, {Guillout}, {Guiraud}, {Guti{\'e}rrez-S{\'a}nchez}, {Guy}, {Hatzidimitriou}, {Hauser}, {Haywood}, {Helmer}, {Helmi}, {Sarmiento}, {Hidalgo}, {Hilger}, {H{\l}adczuk}, {Hobbs}, {Holland}, {Huckle}, {Jardine}, {Jasniewicz}, {Jean-Antoine Piccolo}, {Jim{\'e}nez-Arranz}, {Jorissen}, {Juaristi Campillo}, {Julbe}, {Karbevska}, {Kervella}, {Khanna}, {Kontizas}, {Kordopatis}, {Korn}, {K{\'o}sp{\'a}l}, {Kostrzewa-Rutkowska}, {Kruszy{\'n}ska}, {Kun}, {Laizeau}, {Lambert}, {Lanza}, {Lasne}, {Le Campion}, {Lebreton}, {Lebzelter}, {Leccia}, {Leclerc}, {Lecoeur-Taibi}, {Liao}, {Licata}, {Lindstr{\o}m}, {Lister}, {Livanou}, {Lobel}, {Lorca}, {Loup}, {Madrero Pardo}, {Magdaleno Romeo}, {Managau}, {Mann}, {Manteiga}, {Marchant}, {Marconi}, {Marcos}, {Marcos Santos}, {Mar{\'\i}n Pina}, {Marinoni}, {Marocco}, {Marshall}, {Martin Polo}, {Mart{\'\i}n-Fleitas}, {Marton}, {Mary}, {Masip}, {Massari}, {Mastrobuono-Battisti}, {Mazeh}, {McMillan}, {Messina}, {Michalik},
  {Millar}, {Mints}, {Molina}, {Molinaro}, {Moln{\'a}r}, {Monari}, {Mongui{\'o}}, {Montegriffo}, {Montero}, {Mor}, {Mora}, {Morbidelli}, {Morel}, {Morris}, {Muraveva}, {Murphy}, {Musella}, {Nagy}, {Noval}, {Oca{\~n}a}, {Ogden}, {Ordenovic}, {Osinde}, {Pagani}, {Pagano}, {Palaversa}, {Palicio}, {Pallas-Quintela}, {Panahi}, {Payne-Wardenaar}, {Pe{\~n}alosa Esteller}, {Penttil{\"a}}, {Pichon}, {Piersimoni}, {Pineau}, {Plachy}, {Plum}, {Poggio}, {Pr{\v{s}}a}, {Pulone}, {Racero}, {Ragaini}, {Rainer}, {Raiteri}, {Rambaux}, {Ramos}, {Ramos-Lerate}, {Re Fiorentin}, {Regibo}, {Richards}, {Rios Diaz}, {Ripepi}, {Riva}, {Rix}, {Rixon}, {Robichon}, {Robin}, {Robin}, {Roelens}, {Rogues}, {Rohrbasser}, {Romero-G{\'o}mez}, {Rowell}, {Royer}, {Ruz Mieres}, {Rybicki}, {Sadowski}, {S{\'a}ez N{\'u}{\~n}ez}, {Sagrist{\`a} Sell{\'e}s}, {Sahlmann}, {Salguero}, {Samaras}, {Sanchez Gimenez}, {Sanna}, {Santove{\~n}a}, {Sarasso}, {Schultheis}, {Sciacca}, {Segol}, {Segovia}, {S{\'e}gransan}, {Semeux}, {Shahaf}, {Siddiqui}, {Siebert},
  {Siltala}, {Silvelo}, {Slezak}, {Slezak}, {Smart}, {Snaith}, {Solano}, {Solitro}, {Souami}, {Souchay}, {Spagna}, {Spina}, {Spoto}, {Steele}, {Steidelm{\"u}ller}, {Stephenson}, {S{\"u}veges}, {Surdej}, {Szabados}, {Szegedi-Elek}, {Taris}, {Taylor}, {Teixeira}, {Tolomei}, {Tonello}, {Torra}, {Torra}, {Torralba Elipe}, {Trabucchi}, {Tsounis}, {Turon}, {Ulla}, {Unger}, {Vaillant}, {van Dillen}, {van Reeven}, {Vanel}, {Vecchiato}, {Viala}, {Vicente}, {Voutsinas}, {Weiler}, {Wevers}, {Wyrzykowski}, {Yoldas}, {Yvard}, {Zhao}, {Zorec}, {Zucker}, \& {Zwitter}}]{Gaiadr3}
{Gaia Collaboration}, {Vallenari}, A., {Brown}, A.~G.~A., {et~al.} 2023, \aap, 674, A1, \dodoi{10.1051/0004-6361/202243940}

\bibitem[{{Galvin} {et~al.}(2020){Galvin}, {Huynh}, {Norris}, {Wang}, {Hopkins}, {Polsterer}, {Ralph}, {O'Brien}, \& {Heald}}]{skacatalog}
{Galvin}, T.~J., {Huynh}, M.~T., {Norris}, R.~P., {et~al.} 2020, \mnras, 497, 2730, \dodoi{10.1093/mnras/staa1890}

\bibitem[{{Gordon} {et~al.}(2021){Gordon}, {Boyce}, {O'Dea}, {Rudnick}, {Andernach}, {Vantyghem}, {Baum}, {Bui}, {Dionyssiou}, {Safi-Harb}, \& {Sander}}]{Gordon2021}
{Gordon}, Y.~A., {Boyce}, M.~M., {O'Dea}, C.~P., {et~al.} 2021, \apjs, 255, 30, \dodoi{10.3847/1538-4365/ac05c0}

\bibitem[{{Graham} \& {Liu}(1995)}]{1995ApJ...449L..29G}
{Graham}, J.~R., \& {Liu}, M.~C. 1995, \apjl, 449, L29, \dodoi{10.1086/309629}

\bibitem[{{Gross} {et~al.}(2023){Gross}, {Chen}, {Foord}, {Liu}, {Shen}, {Oguri}, {Goulding}, {Hwang}, {Zakamska}, {Ma}, \& {Nolan}}]{2023ApJ...956..117G}
{Gross}, A.~C., {Chen}, Y.-C., {Foord}, A., {et~al.} 2023, \apj, 956, 117, \dodoi{10.3847/1538-4357/acf469}

\bibitem[{{Haarsma} {et~al.}(2005){Haarsma}, {Winn}, {Falco}, {Kochanek}, {Ammar}, {Boersma}, {Fogwell}, {Muxlow}, {McLeod}, \& {Leh{\'a}r}}]{2005AJ....130.1977H}
{Haarsma}, D.~B., {Winn}, J.~N., {Falco}, E.~E., {et~al.} 2005, \aj, 130, 1977, \dodoi{10.1086/466513}

\bibitem[{{Hagen} {et~al.}(1999){Hagen}, {Engels}, \& {Reimers}}]{hamburg}
{Hagen}, H.~J., {Engels}, D., \& {Reimers}, D. 1999, \aaps, 134, 483, \dodoi{10.1051/aas:1999442}

\bibitem[{{Hartley} {et~al.}(2021){Hartley}, {Jackson}, {Badole}, {McKean}, {Sluse}, \& {Vives-Arias}}]{2021MNRAS.508.4625H}
{Hartley}, P., {Jackson}, N., {Badole}, S., {et~al.} 2021, \mnras, 508, 4625, \dodoi{10.1093/mnras/stab2758}

\bibitem[{{He} {et~al.}(2023){He}, {Li}, {Cao}, {Li}, {Zou}, \& {Dye}}]{he23}
{He}, Z., {Li}, N., {Cao}, X., {et~al.} 2023, \aap, 672, A123, \dodoi{10.1051/0004-6361/202245484}

\bibitem[{{Helfand} {et~al.}(2015){Helfand}, {White}, \& {Becker}}]{FIRSTfinal}
{Helfand}, D.~J., {White}, R.~L., \& {Becker}, R.~H. 2015, \apj, 801, 26, \dodoi{10.1088/0004-637X/801/1/26}

\bibitem[{{Hennawi} {et~al.}(2006){Hennawi}, {Strauss}, {Oguri}, {Inada}, {Richards}, {Pindor}, {Schneider}, {Becker}, {Gregg}, {Hall}, {Johnston}, {Fan}, {Burles}, {Schlegel}, {Gunn}, {Lupton}, {Bahcall}, {Brunner}, \& {Brinkmann}}]{2006AJ....131....1H}
{Hennawi}, J.~F., {Strauss}, M.~A., {Oguri}, M., {et~al.} 2006, \aj, 131, 1, \dodoi{10.1086/498235}

\bibitem[{{Hewitt} {et~al.}(1992){Hewitt}, {Turner}, {Lawrence}, {Schneider}, \& {Brody}}]{1992AJ....104..968H}
{Hewitt}, J.~N., {Turner}, E.~L., {Lawrence}, C.~R., {Schneider}, D.~P., \& {Brody}, J.~P. 1992, \aj, 104, 968, \dodoi{10.1086/116290}

\bibitem[{{Hewitt} {et~al.}(1988){Hewitt}, {Turner}, {Schneider}, {Burke}, \& {Langston}}]{1988Natur.333..537H}
{Hewitt}, J.~N., {Turner}, E.~L., {Schneider}, D.~P., {Burke}, B.~F., \& {Langston}, G.~I. 1988, \nat, 333, 537, \dodoi{10.1038/333537a0}

\bibitem[{{H{\"o}gbom}(1974)}]{hogbom}
{H{\"o}gbom}, J.~A. 1974, \aaps, 15, 417

\bibitem[{{Huang} {et~al.}(2020){Huang}, {Storfer}, {Ravi}, {Pilon}, {Domingo}, {Schlegel}, {Bailey}, {Dey}, {Gupta}, {Herrera}, {Juneau}, {Landriau}, {Lang}, {Meisner}, {Moustakas}, {Myers}, {Schlafly}, {Valdes}, {Weaver}, {Yang}, \& {Y{\`e}che}}]{Huang2020}
{Huang}, X., {Storfer}, C., {Ravi}, V., {et~al.} 2020, \apj, 894, 78, \dodoi{10.3847/1538-4357/ab7ffb}

\bibitem[{{Huchra} {et~al.}(1985){Huchra}, {Gorenstein}, {Kent}, {Shapiro}, {Smith}, {Horine}, \& {Perley}}]{1985AJ.....90..691H}
{Huchra}, J., {Gorenstein}, M., {Kent}, S., {et~al.} 1985, \aj, 90, 691, \dodoi{10.1086/113777}

\bibitem[{{Ibata} {et~al.}(1999){Ibata}, {Lewis}, {Irwin}, {Leh{\'a}r}, \& {Totten}}]{1999AJ....118.1922I}
{Ibata}, R.~A., {Lewis}, G.~F., {Irwin}, M.~J., {Leh{\'a}r}, J., \& {Totten}, E.~J. 1999, \aj, 118, 1922, \dodoi{10.1086/301111}

\bibitem[{{Inada} {et~al.}(2003{\natexlab{a}}){Inada}, {Oguri}, {Pindor}, {Hennawi}, {Chiu}, {Zheng}, {Ichikawa}, {Gregg}, {Becker}, {Suto}, {Strauss}, {Turner}, {Keeton}, {Annis}, {Castander}, {Eisenstein}, {Frieman}, {Fukugita}, {Gunn}, {Johnston}, {Kent}, {Nichol}, {Richards}, {Rix}, {Sheldon}, {Bahcall}, {Brinkmann}, {Ivezi{\'c}}, {Lamb}, {McKay}, {Schneider}, \& {York}}]{2003Natur.426..810I}
{Inada}, N., {Oguri}, M., {Pindor}, B., {et~al.} 2003{\natexlab{a}}, \nat, 426, 810, \dodoi{10.1038/nature02153}

\bibitem[{{Inada} {et~al.}(2003{\natexlab{b}}){Inada}, {Becker}, {Burles}, {Castander}, {Eisenstein}, {Hall}, {Johnston}, {Pindor}, {Richards}, {Schechter}, {Sekiguchi}, {White}, {Brinkmann}, {Frieman}, {Kleinman}, {Krzesi{\'n}ski}, {Long}, {Neilsen}, {Newman}, {Nitta}, {Schneider}, {Snedden}, \& {York}}]{2003AJ....126..666I}
{Inada}, N., {Becker}, R.~H., {Burles}, S., {et~al.} 2003{\natexlab{b}}, \aj, 126, 666, \dodoi{10.1086/375906}

\bibitem[{{Intema} {et~al.}(2017){Intema}, {Jagannathan}, {Mooley}, \& {Frail}}]{Intema2017}
{Intema}, H.~T., {Jagannathan}, P., {Mooley}, K.~P., \& {Frail}, D.~A. 2017, \aap, 598, A78, \dodoi{10.1051/0004-6361/201628536}

\bibitem[{{Irwin} {et~al.}(1998){Irwin}, {Ibata}, {Lewis}, \& {Totten}}]{1998ApJ...505..529I}
{Irwin}, M.~J., {Ibata}, R.~A., {Lewis}, G.~F., \& {Totten}, E.~J. 1998, \apj, 505, 529, \dodoi{10.1086/306213}

\bibitem[{{Ivezi{\'c}} {et~al.}(2019){Ivezi{\'c}}, {Kahn}, {Tyson}, {Abel}, {Acosta}, {Allsman}, {Alonso}, {AlSayyad}, {Anderson}, {Andrew}, {Angel}, {Angeli}, {Ansari}, {Antilogus}, {Araujo}, {Armstrong}, {Arndt}, {Astier}, {Aubourg}, {Auza}, {Axelrod}, {Bard}, {Barr}, {Barrau}, {Bartlett}, {Bauer}, {Bauman}, {Baumont}, {Bechtol}, {Bechtol}, {Becker}, {Becla}, {Beldica}, {Bellavia}, {Bianco}, {Biswas}, {Blanc}, {Blazek}, {Blandford}, {Bloom}, {Bogart}, {Bond}, {Booth}, {Borgland}, {Borne}, {Bosch}, {Boutigny}, {Brackett}, {Bradshaw}, {Brandt}, {Brown}, {Bullock}, {Burchat}, {Burke}, {Cagnoli}, {Calabrese}, {Callahan}, {Callen}, {Carlin}, {Carlson}, {Chandrasekharan}, {Charles-Emerson}, {Chesley}, {Cheu}, {Chiang}, {Chiang}, {Chirino}, {Chow}, {Ciardi}, {Claver}, {Cohen-Tanugi}, {Cockrum}, {Coles}, {Connolly}, {Cook}, {Cooray}, {Covey}, {Cribbs}, {Cui}, {Cutri}, {Daly}, {Daniel}, {Daruich}, {Daubard}, {Daues}, {Dawson}, {Delgado}, {Dellapenna}, {de Peyster}, {de Val-Borro}, {Digel}, {Doherty}, {Dubois},
  {Dubois-Felsmann}, {Durech}, {Economou}, {Eifler}, {Eracleous}, {Emmons}, {Fausti Neto}, {Ferguson}, {Figueroa}, {Fisher-Levine}, {Focke}, {Foss}, {Frank}, {Freemon}, {Gangler}, {Gawiser}, {Geary}, {Gee}, {Geha}, {Gessner}, {Gibson}, {Gilmore}, {Glanzman}, {Glick}, {Goldina}, {Goldstein}, {Goodenow}, {Graham}, {Gressler}, {Gris}, {Guy}, {Guyonnet}, {Haller}, {Harris}, {Hascall}, {Haupt}, {Hernandez}, {Herrmann}, {Hileman}, {Hoblitt}, {Hodgson}, {Hogan}, {Howard}, {Huang}, {Huffer}, {Ingraham}, {Innes}, {Jacoby}, {Jain}, {Jammes}, {Jee}, {Jenness}, {Jernigan}, {Jevremovi{\'c}}, {Johns}, {Johnson}, {Johnson}, {Jones}, {Juramy-Gilles}, {Juri{\'c}}, {Kalirai}, {Kallivayalil}, {Kalmbach}, {Kantor}, {Karst}, {Kasliwal}, {Kelly}, {Kessler}, {Kinnison}, {Kirkby}, {Knox}, {Kotov}, {Krabbendam}, {Krughoff}, {Kub{\'a}nek}, {Kuczewski}, {Kulkarni}, {Ku}, {Kurita}, {Lage}, {Lambert}, {Lange}, {Langton}, {Le Guillou}, {Levine}, {Liang}, {Lim}, {Lintott}, {Long}, {Lopez}, {Lotz}, {Lupton}, {Lust}, {MacArthur}, {Mahabal},
  {Mandelbaum}, {Markiewicz}, {Marsh}, {Marshall}, {Marshall}, {May}, {McKercher}, {McQueen}, {Meyers}, {Migliore}, {Miller}, {Mills}, {Miraval}, {Moeyens}, {Moolekamp}, {Monet}, {Moniez}, {Monkewitz}, {Montgomery}, {Morrison}, {Mueller}, {Muller}, {Mu{\~n}oz Arancibia}, {Neill}, {Newbry}, {Nief}, {Nomerotski}, {Nordby}, {O'Connor}, {Oliver}, {Olivier}, {Olsen}, {O'Mullane}, {Ortiz}, {Osier}, {Owen}, {Pain}, {Palecek}, {Parejko}, {Parsons}, {Pease}, {Peterson}, {Peterson}, {Petravick}, {Libby Petrick}, {Petry}, {Pierfederici}, {Pietrowicz}, {Pike}, {Pinto}, {Plante}, {Plate}, {Plutchak}, {Price}, {Prouza}, {Radeka}, {Rajagopal}, {Rasmussen}, {Regnault}, {Reil}, {Reiss}, {Reuter}, {Ridgway}, {Riot}, {Ritz}, {Robinson}, {Roby}, {Roodman}, {Rosing}, {Roucelle}, {Rumore}, {Russo}, {Saha}, {Sassolas}, {Schalk}, {Schellart}, {Schindler}, {Schmidt}, {Schneider}, {Schneider}, {Schoening}, {Schumacher}, {Schwamb}, {Sebag}, {Selvy}, {Sembroski}, {Seppala}, {Serio}, {Serrano}, {Shaw}, {Shipsey}, {Sick}, {Silvestri},
  {Slater}, {Smith}, {Smith}, {Sobhani}, {Soldahl}, {Storrie-Lombardi}, {Stover}, {Strauss}, {Street}, {Stubbs}, {Sullivan}, {Sweeney}, {Swinbank}, {Szalay}, {Takacs}, {Tether}, {Thaler}, {Thayer}, {Thomas}, {Thornton}, {Thukral}, {Tice}, {Trilling}, {Turri}, {Van Berg}, {Vanden Berk}, {Vetter}, {Virieux}, {Vucina}, {Wahl}, {Walkowicz}, {Walsh}, {Walter}, {Wang}, {Wang}, {Warner}, {Wiecha}, {Willman}, {Winters}, {Wittman}, {Wolff}, {Wood-Vasey}, {Wu}, {Xin}, {Yoachim}, \& {Zhan}}]{Ivezic2019}
{Ivezi{\'c}}, {\v{Z}}., {Kahn}, S.~M., {Tyson}, J.~A., {et~al.} 2019, \apj, 873, 111, \dodoi{10.3847/1538-4357/ab042c}

\bibitem[{{Jackson}(2011)}]{2011ApJ...739L..28J}
{Jackson}, N. 2011, \apjl, 739, L28, \dodoi{10.1088/2041-8205/739/1/L28}

\bibitem[{{Jackson} {et~al.}(2024){Jackson}, {Badole}, {Dugdale}, {Stacey}, {Hartley}, \& {McKean}}]{Jackson2024}
{Jackson}, N., {Badole}, S., {Dugdale}, T., {et~al.} 2024, \mnras, \dodoi{10.1093/mnras/stae916}

\bibitem[{{Jackson} \& {Browne}(2007)}]{JB07}
{Jackson}, N., \& {Browne}, I.~W.~A. 2007, \mnras, 374, 168, \dodoi{10.1111/j.1365-2966.2006.11126.x}

\bibitem[{{Jackson} {et~al.}(2015){Jackson}, {Tagore}, {Roberts}, {Sluse}, {Stacey}, {Vives-Arias}, {Wucknitz}, \& {Volino}}]{2015MNRAS.454..287J}
{Jackson}, N., {Tagore}, A.~S., {Roberts}, C., {et~al.} 2015, \mnras, 454, 287, \dodoi{10.1093/mnras/stv1982}

\bibitem[{{Jackson} {et~al.}(1995){Jackson}, {de Bruyn}, {Myers}, {Bremer}, {Miley}, {Schilizzi}, {Browne}, {Nair}, {Wilkinson}, {Blandford}, {Pearson}, \& {Readhead}}]{1995MNRAS.274L..25J}
{Jackson}, N., {de Bruyn}, A.~G., {Myers}, S., {et~al.} 1995, \mnras, 274, L25, \dodoi{10.1093/mnras/274.1.L25}

\bibitem[{{Jackson} {et~al.}(1998){Jackson}, {Nair}, {Browne}, {Wilkinson}, {Muxlow}, {de Bruyn}, {Koopmans}, {Bremer}, {Snellen}, {Miley}, {Schilizzi}, {Myers}, {Fassnacht}, {Womble}, {Readhead}, {Blandford}, \& {Pearson}}]{1998MNRAS.296..483J}
{Jackson}, N., {Nair}, S., {Browne}, I.~W.~A., {et~al.} 1998, \mnras, 296, 483, \dodoi{10.1046/j.1365-8711.1998.01304.x}

\bibitem[{{Jacobs} {et~al.}(2019){Jacobs}, {Collett}, {Glazebrook}, {Buckley-Geer}, {Diehl}, {Lin}, {McCarthy}, {Qin}, {Odden}, {Caso Escudero}, {Dial}, {Yung}, {Gaitsch}, {Pellico}, {Lindgren}, {Abbott}, {Annis}, {Avila}, {Brooks}, {Burke}, {Carnero Rosell}, {Carrasco Kind}, {Carretero}, {da Costa}, {De Vicente}, {Fosalba}, {Frieman}, {Garc{\'\i}a-Bellido}, {Gaztanaga}, {Goldstein}, {Gruen}, {Gruendl}, {Gschwend}, {Hollowood}, {Honscheid}, {Hoyle}, {James}, {Krause}, {Kuropatkin}, {Lahav}, {Lima}, {Maia}, {Marshall}, {Miquel}, {Plazas}, {Roodman}, {Sanchez}, {Scarpine}, {Serrano}, {Sevilla-Noarbe}, {Smith}, {Sobreira}, {Suchyta}, {Swanson}, {Tarle}, {Vikram}, {Walker}, {Zhang}, \& {DES Collaboration}}]{Jacobs2019}
{Jacobs}, C., {Collett}, T., {Glazebrook}, K., {et~al.} 2019, \apjs, 243, 17, \dodoi{10.3847/1538-4365/ab26b6}

\bibitem[{{King} {et~al.}(1999){King}, {Browne}, {Marlow}, {Patnaik}, \& {Wilkinson}}]{1999MNRAS.307..225K}
{King}, L.~J., {Browne}, I. W.~A., {Marlow}, D.~R., {Patnaik}, A.~R., \& {Wilkinson}, P.~N. 1999, \mnras, 307, 225, \dodoi{10.1046/j.1365-8711.1999.02328.x}

\bibitem[{{King} {et~al.}(1997){King}, {Browne}, {Muxlow}, {Narasimha}, {Patnaik}, {Porcas}, \& {Wilkinson}}]{1997MNRAS.289..450K}
{King}, L.~J., {Browne}, I.~W.~A., {Muxlow}, T.~W.~B., {et~al.} 1997, \mnras, 289, 450, \dodoi{10.1093/mnras/289.2.450}

\bibitem[{{Koopmans} {et~al.}(1999){Koopmans}, {de Bruyn}, {Marlow}, {Jackson}, {Blandford}, {Browne}, {Fassnacht}, {Myers}, {Pearson}, {Readhead}, {Wilkinson}, \& {Womble}}]{1999MNRAS.303..727K}
{Koopmans}, L.~V.~E., {de Bruyn}, A.~G., {Marlow}, D.~R., {et~al.} 1999, \mnras, 303, 727, \dodoi{10.1046/j.1365-8711.1999.02342.x}

\bibitem[{{Kormann} {et~al.}(1994){Kormann}, {Schneider}, \& {Bartelmann}}]{SIE}
{Kormann}, R., {Schneider}, P., \& {Bartelmann}, M. 1994, \aap, 284, 285

\bibitem[{{Krone-Martins} {et~al.}(2018){Krone-Martins}, {Delchambre}, {Wertz}, {Ducourant}, {Mignard}, {Teixeira}, {Kl{\"u}ter}, {Le Campion}, {Galluccio}, {Surdej}, {Bastian}, {Wambsganss}, {Graham}, {Djorgovski}, \& {Slezak}}]{2018AaA...616L..11K}
{Krone-Martins}, A., {Delchambre}, L., {Wertz}, O., {et~al.} 2018, \aap, 616, L11, \dodoi{10.1051/0004-6361/201833337}

\bibitem[{{Krone-Martins} {et~al.}(2019){Krone-Martins}, {Graham}, {Stern}, {Djorgovski}, {Delchambre}, {Ducourant}, {Teixeira}, {Drake}, {Scarano}, {Surdej}, {Galluccio}, {Jalan}, {Wertz}, {Kl{\"u}ter}, {Mignard}, {Spindola-Duarte}, {Dobie}, {Slezak}, {Sluse}, {Murphy}, {Boehm}, {Nierenberg}, {Bastian}, {Wambsganss}, \& {LeCampion}}]{2019arXiv191208977K}
{Krone-Martins}, A., {Graham}, M.~J., {Stern}, D., {et~al.} 2019, arXiv e-prints, arXiv:1912.08977, \dodoi{10.48550/arXiv.1912.08977}

\bibitem[{{Lacy} {et~al.}(2002){Lacy}, {Gregg}, {Becker}, {White}, {Glikman}, {Helfand}, \& {Winn}}]{2002AJ....123.2925L}
{Lacy}, M., {Gregg}, M., {Becker}, R.~H., {et~al.} 2002, \aj, 123, 2925, \dodoi{10.1086/340568}

\bibitem[{{Lacy} {et~al.}(2020){Lacy}, {Baum}, {Chandler}, {Chatterjee}, {Clarke}, {Deustua}, {English}, {Farnes}, {Gaensler}, {Gugliucci}, {Hallinan}, {Kent}, {Kimball}, {Law}, {Lazio}, {Marvil}, {Mao}, {Medlin}, {Mooley}, {Murphy}, {Myers}, {Osten}, {Richards}, {Rosolowsky}, {Rudnick}, {Schinzel}, {Sivakoff}, {Sjouwerman}, {Taylor}, {White}, {Wrobel}, {Andernach}, {Beasley}, {Berger}, {Bhatnager}, {Birkinshaw}, {Bower}, {Brandt}, {Brown}, {Burke-Spolaor}, {Butler}, {Comerford}, {Demorest}, {Fu}, {Giacintucci}, {Golap}, {G{\"u}th}, {Hales}, {Hiriart}, {Hodge}, {Horesh}, {Ivezi{\'c}}, {Jarvis}, {Kamble}, {Kassim}, {Liu}, {Loinard}, {Lyons}, {Masters}, {Mezcua}, {Moellenbrock}, {Mroczkowski}, {Nyland}, {O'Dea}, {O'Sullivan}, {Peters}, {Radford}, {Rao}, {Robnett}, {Salcido}, {Shen}, {Sobotka}, {Witz}, {Vaccari}, {van Weeren}, {Vargas}, {Williams}, \& {Yoon}}]{VLASS}
{Lacy}, M., {Baum}, S.~A., {Chandler}, C.~J., {et~al.} 2020, \pasp, 132, 035001, \dodoi{10.1088/1538-3873/ab63eb}

\bibitem[{{Lang}(2014)}]{unWISE}
{Lang}, D. 2014, \aj, 147, 108, \dodoi{10.1088/0004-6256/147/5/108}

\bibitem[{{Langston} {et~al.}(1989){Langston}, {Schneider}, {Conner}, {Carilli}, {Lehar}, {Burke}, {Turner}, {Gunn}, {Hewitt}, \& {Schmidt}}]{1989AJ.....97.1283L}
{Langston}, G.~I., {Schneider}, D.~P., {Conner}, S., {et~al.} 1989, \aj, 97, 1283, \dodoi{10.1086/115071}

\bibitem[{{Laureijs} {et~al.}(2011){Laureijs}, {Amiaux}, {Arduini}, {Augu{\`e}res}, {Brinchmann}, {Cole}, {Cropper}, {Dabin}, {Duvet}, {Ealet}, {Garilli}, {Gondoin}, {Guzzo}, {Hoar}, {Hoekstra}, {Holmes}, {Kitching}, {Maciaszek}, {Mellier}, {Pasian}, {Percival}, {Rhodes}, {Saavedra Criado}, {Sauvage}, {Scaramella}, {Valenziano}, {Warren}, {Bender}, {Castander}, {Cimatti}, {Le F{\`e}vre}, {Kurki-Suonio}, {Levi}, {Lilje}, {Meylan}, {Nichol}, {Pedersen}, {Popa}, {Rebolo Lopez}, {Rix}, {Rottgering}, {Zeilinger}, {Grupp}, {Hudelot}, {Massey}, {Meneghetti}, {Miller}, {Paltani}, {Paulin-Henriksson}, {Pires}, {Saxton}, {Schrabback}, {Seidel}, {Walsh}, {Aghanim}, {Amendola}, {Bartlett}, {Baccigalupi}, {Beaulieu}, {Benabed}, {Cuby}, {Elbaz}, {Fosalba}, {Gavazzi}, {Helmi}, {Hook}, {Irwin}, {Kneib}, {Kunz}, {Mannucci}, {Moscardini}, {Tao}, {Teyssier}, {Weller}, {Zamorani}, {Zapatero Osorio}, {Boulade}, {Foumond}, {Di Giorgio}, {Guttridge}, {James}, {Kemp}, {Martignac}, {Spencer}, {Walton}, {Bl{\"u}mchen}, {Bonoli},
  {Bortoletto}, {Cerna}, {Corcione}, {Fabron}, {Jahnke}, {Ligori}, {Madrid}, {Martin}, {Morgante}, {Pamplona}, {Prieto}, {Riva}, {Toledo}, {Trifoglio}, {Zerbi}, {Abdalla}, {Douspis}, {Grenet}, {Borgani}, {Bouwens}, {Courbin}, {Delouis}, {Dubath}, {Fontana}, {Frailis}, {Grazian}, {Koppenh{\"o}fer}, {Mansutti}, {Melchior}, {Mignoli}, {Mohr}, {Neissner}, {Noddle}, {Poncet}, {Scodeggio}, {Serrano}, {Shane}, {Starck}, {Surace}, {Taylor}, {Verdoes-Kleijn}, {Vuerli}, {Williams}, {Zacchei}, {Altieri}, {Escudero Sanz}, {Kohley}, {Oosterbroek}, {Astier}, {Bacon}, {Bardelli}, {Baugh}, {Bellagamba}, {Benoist}, {Bianchi}, {Biviano}, {Branchini}, {Carbone}, {Cardone}, {Clements}, {Colombi}, {Conselice}, {Cresci}, {Deacon}, {Dunlop}, {Fedeli}, {Fontanot}, {Franzetti}, {Giocoli}, {Garcia-Bellido}, {Gow}, {Heavens}, {Hewett}, {Heymans}, {Holland}, {Huang}, {Ilbert}, {Joachimi}, {Jennins}, {Kerins}, {Kiessling}, {Kirk}, {Kotak}, {Krause}, {Lahav}, {van Leeuwen}, {Lesgourgues}, {Lombardi}, {Magliocchetti}, {Maguire},
  {Majerotto}, {Maoli}, {Marulli}, {Maurogordato}, {McCracken}, {McLure}, {Melchiorri}, {Merson}, {Moresco}, {Nonino}, {Norberg}, {Peacock}, {Pello}, {Penny}, {Pettorino}, {Di Porto}, {Pozzetti}, {Quercellini}, {Radovich}, {Rassat}, {Roche}, {Ronayette}, {Rossetti}, {Sartoris}, {Schneider}, {Semboloni}, {Serjeant}, {Simpson}, {Skordis}, {Smadja}, {Smartt}, {Spano}, {Spiro}, {Sullivan}, {Tilquin}, {Trotta}, {Verde}, {Wang}, {Williger}, {Zhao}, {Zoubian}, \& {Zucca}}]{Laureijs2011}
{Laureijs}, R., {Amiaux}, J., {Arduini}, S., {et~al.} 2011, arXiv e-prints, arXiv:1110.3193, \dodoi{10.48550/arXiv.1110.3193}

\bibitem[{{Lawrence} {et~al.}(1986){Lawrence}, {Bennett}, {Hewitt}, {Langston}, {Klotz}, {Burke}, \& {Turner}}]{MGVLA}
{Lawrence}, C.~R., {Bennett}, C.~L., {Hewitt}, J.~N., {et~al.} 1986, \apjs, 61, 105, \dodoi{10.1086/191109}

\bibitem[{{Lawrence} {et~al.}(1984){Lawrence}, {Schneider}, {Schmidt}, {Bennett}, {Hewitt}, {Burke}, {Turner}, \& {Gunn}}]{1984Sci...223...46L}
{Lawrence}, C.~R., {Schneider}, D.~P., {Schmidt}, M., {et~al.} 1984, Science, 223, 46, \dodoi{10.1126/science.223.4631.46}

\bibitem[{{Leh{\'a}r} {et~al.}(2001){Leh{\'a}r}, {Buchalter}, {McMahon}, {Kochanek}, \& {Muxlow}}]{2001ApJ...547...60L}
{Leh{\'a}r}, J., {Buchalter}, A., {McMahon}, R.~G., {Kochanek}, C.~S., \& {Muxlow}, T.~W.~B. 2001, \apj, 547, 60, \dodoi{10.1086/318367}

\bibitem[{{Lehar} {et~al.}(1993){Lehar}, {Langston}, {Silber}, {Lawrence}, \& {Burke}}]{1993AJ....105..847L}
{Lehar}, J., {Langston}, G.~I., {Silber}, A., {Lawrence}, C.~R., \& {Burke}, B.~F. 1993, \aj, 105, 847, \dodoi{10.1086/116476}

\bibitem[{{Lehar} {et~al.}(1997){Lehar}, {Burke}, {Conner}, {Falco}, {Fletcher}, {Irwin}, {McMahon}, {Muslow}, \& {Schechter}}]{1997AJ....114...48L}
{Lehar}, J., {Burke}, B.~F., {Conner}, S.~R., {et~al.} 1997, \aj, 114, 48, \dodoi{10.1086/118451}

\bibitem[{{Lemon} {et~al.}(2020){Lemon}, {Auger}, {McMahon}, {Anguita}, {Apostolovski}, {Chen}, {Fassnacht}, {Melo}, {Motta}, {Shajib}, {Treu}, {Agnello}, {Buckley-Geer}, {Schechter}, {Birrer}, {Collett}, {Courbin}, {Rusu}, {Abbott}, {Allam}, {Annis}, {Avila}, {Bertin}, {Brooks}, {Burke}, {Carnero Rosell}, {Carrasco Kind}, {Carretero}, {Costanzi}, {da Costa}, {De Vicente}, {Desai}, {Eifler}, {Flaugher}, {Frieman}, {Garc{\'\i}a-Bellido}, {Gaztanaga}, {Gerdes}, {Gruen}, {Gruendl}, {Gschwend}, {Gutierrez}, {Honscheid}, {James}, {Kim}, {Krause}, {Kuehn}, {Kuropatkin}, {Lahav}, {Lima}, {Lin}, {Maia}, {March}, {Marshall}, {Menanteau}, {Miquel}, {Palmese}, {Paz-Chinch{\'o}n}, {Plazas}, {Roodman}, {Sanchez}, {Schubnell}, {Serrano}, {Smith}, {Soares-Santos}, {Suchyta}, {Tarle}, \& {Walker}}]{2020MNRAS.494.3491L}
{Lemon}, C., {Auger}, M.~W., {McMahon}, R., {et~al.} 2020, \mnras, 494, 3491, \dodoi{10.1093/mnras/staa652}

\bibitem[{{Lemon} {et~al.}(2023){Lemon}, {Anguita}, {Auger-Williams}, {Courbin}, {Galan}, {McMahon}, {Neira}, {Oguri}, {Schechter}, {Shajib}, {Treu}, {Agnello}, \& {Spiniello}}]{lemon2023}
{Lemon}, C., {Anguita}, T., {Auger-Williams}, M.~W., {et~al.} 2023, \mnras, 520, 3305, \dodoi{10.1093/mnras/stac3721}

\bibitem[{{Lemon} {et~al.}(2024){Lemon}, {Courbin}, {More}, {Schechter}, {Ca{\~n}ameras}, {Delchambre}, {Leung}, {Shu}, {Spiniello}, {Hezaveh}, {Kl{\"u}ter}, \& {McMahon}}]{Lemon2024}
{Lemon}, C., {Courbin}, F., {More}, A., {et~al.} 2024, \ssr, 220, 23, \dodoi{10.1007/s11214-024-01042-9}

\bibitem[{{Lemon} {et~al.}(2019){Lemon}, {Auger}, \& {McMahon}}]{Lemon19}
{Lemon}, C.~A., {Auger}, M.~W., \& {McMahon}, R.~G. 2019, \mnras, 483, 4242, \dodoi{10.1093/mnras/sty3366}

\bibitem[{{Lemon} {et~al.}(2017){Lemon}, {Auger}, {McMahon}, \& {Koposov}}]{Lemon2017}
{Lemon}, C.~A., {Auger}, M.~W., {McMahon}, R.~G., \& {Koposov}, S.~E. 2017, \mnras, 472, 5023, \dodoi{10.1093/mnras/stx2094}

\bibitem[{{Lemon} {et~al.}(2018){Lemon}, {Auger}, {McMahon}, \& {Ostrovski}}]{Lemon2018}
{Lemon}, C.~A., {Auger}, M.~W., {McMahon}, R.~G., \& {Ostrovski}, F. 2018, \mnras, 479, 5060, \dodoi{10.1093/mnras/sty911}

\bibitem[{{Magain} {et~al.}(1988){Magain}, {Surdej}, {Swings}, {Borgeest}, \& {Kayser}}]{1988Natur.334..325M}
{Magain}, P., {Surdej}, J., {Swings}, J.~P., {Borgeest}, U., \& {Kayser}, R. 1988, \nat, 334, 325, \dodoi{10.1038/334325a0}

\bibitem[{{Mangat} {et~al.}(2021){Mangat}, {McKean}, {Brilenkov}, {Hartley}, {Stacey}, {Vegetti}, \& {Wen}}]{2021MNRAS.508L..64M}
{Mangat}, C.~S., {McKean}, J.~P., {Brilenkov}, R., {et~al.} 2021, \mnras, 508, L64, \dodoi{10.1093/mnrasl/slab106}

\bibitem[{{Mao} {et~al.}(2017){Mao}, {Carilli}, {Gaensler}, {Wucknitz}, {Keeton}, {Basu}, {Beck}, {Kronberg}, \& {Zweibel}}]{mao17}
{Mao}, S.~A., {Carilli}, C., {Gaensler}, B.~M., {et~al.} 2017, Nature Astronomy, 1, 621, \dodoi{10.1038/s41550-017-0218-x}

\bibitem[{{Marlow} {et~al.}(1999){Marlow}, {Myers}, {Rusin}, {Jackson}, {Browne}, {Wilkinson}, {Muxlow}, {Fassnacht}, {Lubin}, {Kundi{\'c}}, {Blandford}, {Pearson}, {Readhead}, {Koopmans}, \& {de Bruyn}}]{1999AJ....118..654M}
{Marlow}, D.~R., {Myers}, S.~T., {Rusin}, D., {et~al.} 1999, \aj, 118, 654, \dodoi{10.1086/300987}

\bibitem[{{Marlow} {et~al.}(2001){Marlow}, {Rusin}, {Norbury}, {Jackson}, {Browne}, {Wilkinson}, {Fassnacht}, {Myers}, {Koopmans}, {Blandford}, {Pearson}, {Readhead}, \& {de Bruyn}}]{2001AJ....121..619M}
{Marlow}, D.~R., {Rusin}, D., {Norbury}, M., {et~al.} 2001, \aj, 121, 619, \dodoi{10.1086/318735}

\bibitem[{McKean(2023)}]{mckeancomm}
McKean, J. 2023, personal communication

\bibitem[{{McKean} {et~al.}(2015){McKean}, {Jackson}, {Vegetti}, {Rybak}, {Serjeant}, {Koopmans}, {Metcalf}, {Fassnacht}, {Marshall}, \& {Pandey-Pommier}}]{McKean2015}
{McKean}, J., {Jackson}, N., {Vegetti}, S., {et~al.} 2015, in Advancing Astrophysics with the Square Kilometre Array (AASKA14), 84, \dodoi{10.22323/1.215.0084}

\bibitem[{{Metcalf} \& {Madau}(2001)}]{Metcalf2001}
{Metcalf}, R.~B., \& {Madau}, P. 2001, \apj, 563, 9, \dodoi{10.1086/323695}

\bibitem[{{Morabito} {et~al.}(2022){Morabito}, {Jackson}, {Mooney}, {Sweijen}, {Badole}, {Kukreti}, {Venkattu}, {Groeneveld}, {Kappes}, {Bonnassieux}, {Drabent}, {Iacobelli}, {Croston}, {Best}, {Bondi}, {Callingham}, {Conway}, {Deller}, {Hardcastle}, {McKean}, {Miley}, {Moldon}, {R{\"o}ttgering}, {Tasse}, {Shimwell}, {van Weeren}, {Anderson}, {Asgekar}, {Avruch}, {van Bemmel}, {Bentum}, {Bonafede}, {Brouw}, {Butcher}, {Ciardi}, {Corstanje}, {Coolen}, {Damstra}, {de Gasperin}, {Duscha}, {Eisl{\"o}ffel}, {Engels}, {Falcke}, {Garrett}, {Griessmeier}, {Gunst}, {van Haarlem}, {Hoeft}, {van der Horst}, {J{\"u}tte}, {Kadler}, {Koopmans}, {Krankowski}, {Mann}, {Nelles}, {Oonk}, {Orru}, {Paas}, {Pandey}, {Pizzo}, {Pandey-Pommier}, {Reich}, {Rothkaehl}, {Ruiter}, {Schwarz}, {Shulevski}, {Soida}, {Tagger}, {Vocks}, {Wijers}, {Wijnholds}, {Wucknitz}, {Zarka}, \& {Zucca}}]{Morabito2022}
{Morabito}, L.~K., {Jackson}, N.~J., {Mooney}, S., {et~al.} 2022, \aap, 658, A1, \dodoi{10.1051/0004-6361/202140649}

\bibitem[{{Myers} {et~al.}(1995){Myers}, {Fassnacht}, {Djorgovski}, {Blandford}, {Matthews}, {Neugebauer}, {Pearson}, {Readhead}, {Smith}, {Thompson}, {Womble}, {Browne}, {Wilkinson}, {Nair}, {Jackson}, {Snellen}, {Miley}, {de Bruyn}, \& {Schilizzi}}]{1995ApJ...447L...5M}
{Myers}, S.~T., {Fassnacht}, C.~D., {Djorgovski}, S.~G., {et~al.} 1995, \apjl, 447, L5, \dodoi{10.1086/309556}

\bibitem[{{Myers} {et~al.}(1999){Myers}, {Rusin}, {Fassnacht}, {Blandford}, {Pearson}, {Readhead}, {Jackson}, {Browne}, {Marlow}, {Wilkinson}, {Koopmans}, \& {de Bruyn}}]{1999AJ....117.2565M}
{Myers}, S.~T., {Rusin}, D., {Fassnacht}, C.~D., {et~al.} 1999, \aj, 117, 2565, \dodoi{10.1086/300875}

\bibitem[{{Myers} {et~al.}(2003){Myers}, {Jackson}, {Browne}, {de Bruyn}, {Pearson}, {Readhead}, {Wilkinson}, {Biggs}, {Blandford}, {Fassnacht}, {Koopmans}, {Marlow}, {McKean}, {Norbury}, {Phillips}, {Rusin}, {Shepherd}, \& {Sykes}}]{Myers2003}
{Myers}, S.~T., {Jackson}, N.~J., {Browne}, I.~W.~A., {et~al.} 2003, \mnras, 341, 1, \dodoi{10.1046/j.1365-8711.2003.06256.x}

\bibitem[{{Nyland} {et~al.}(2023){Nyland}, {Alexander}, {Andernach}, {Callingham}, {Cigan}, {Clarke}, {Dong}, {Gordon}, {Kent}, {Lacy}, {Law}, {Myers}, {Ott}, {Peters}, {Polisensky}, {Sivakoff}, {Tremblay}, {Ward}, {Birmingham}, {Patil}, {Petric}, {Kooi}, \& {The VLASS SSG}}]{Nyland2023}
{Nyland}, K., {Alexander}, K., {Andernach}, H., {et~al.} 2023, {VLASS Epoch 4 Science Case}, \url{https://science.nrao.edu/vlass/library/vlass-epoch-4-science-case}

\bibitem[{{O'Dea} \& {Saikia}(2021)}]{ODea2021}
{O'Dea}, C.~P., \& {Saikia}, D.~J. 2021, \aapr, 29, 3, \dodoi{10.1007/s00159-021-00131-w}

\bibitem[{{Patnaik} {et~al.}(1993){Patnaik}, {Browne}, {King}, {Muxlow}, {Walsh}, \& {Wilkinson}}]{1993MNRAS.261..435P}
{Patnaik}, A.~R., {Browne}, I.~W.~A., {King}, L.~J., {et~al.} 1993, \mnras, 261, 435, \dodoi{10.1093/mnras/261.2.435}

\bibitem[{{Patnaik} {et~al.}(1992){Patnaik}, {Browne}, {Walsh}, {Chaffee}, \& {Foltz}}]{1992MNRAS.259P...1P}
{Patnaik}, A.~R., {Browne}, I.~W.~A., {Walsh}, D., {Chaffee}, F.~H., \& {Foltz}, C.~B. 1992, \mnras, 259, 1P, \dodoi{10.1093/mnras/259.1.1P}

\bibitem[{{Phillips} {et~al.}(2000){Phillips}, {Norbury}, {Koopmans}, {Browne}, {Jackson}, {Wilkinson}, {Biggs}, {Blandford}, {de Bruyn}, {Fassnacht}, {Helbig}, {Mao}, {Marlow}, {Myers}, {Pearson}, {Readhead}, {Rusin}, \& {Xanthopoulos}}]{2000MNRAS.319L...7P}
{Phillips}, P.~M., {Norbury}, M.~A., {Koopmans}, L.~V.~E., {et~al.} 2000, \mnras, 319, L7, \dodoi{10.1046/j.1365-8711.2000.04033.x}

\bibitem[{{Powell} {et~al.}(2023){Powell}, {Vegetti}, {McKean}, {White}, {Ferreira}, {May}, \& {Spingola}}]{powell23}
{Powell}, D.~M., {Vegetti}, S., {McKean}, J.~P., {et~al.} 2023, \mnras, 524, L84, \dodoi{10.1093/mnrasl/slad074}

\bibitem[{{Pramesh Rao} \& {Subrahmanyan}(1988)}]{1988MNRAS.231..229P}
{Pramesh Rao}, A., \& {Subrahmanyan}, R. 1988, \mnras, 231, 229, \dodoi{10.1093/mnras/231.2.229}

\bibitem[{{Readhead} \& {Wilkinson}(1978)}]{selfcal}
{Readhead}, A.~C.~S., \& {Wilkinson}, P.~N. 1978, \apj, 223, 25, \dodoi{10.1086/156232}

\bibitem[{{Reimers} {et~al.}(2002){Reimers}, {Hagen}, {Baade}, {Lopez}, \& {Tytler}}]{2002AaA...382L..26R}
{Reimers}, D., {Hagen}, H.~J., {Baade}, R., {Lopez}, S., \& {Tytler}, D. 2002, \aap, 382, L26, \dodoi{10.1051/0004-6361:20011798}

\bibitem[{{Riechers} {et~al.}(2008){Riechers}, {Walter}, {Brewer}, {Carilli}, {Lewis}, {Bertoldi}, \& {Cox}}]{2008ApJ...686..851R}
{Riechers}, D.~A., {Walter}, F., {Brewer}, B.~J., {et~al.} 2008, \apj, 686, 851, \dodoi{10.1086/591434}

\bibitem[{{Rusin} {et~al.}(2001){Rusin}, {Marlow}, {Norbury}, {Browne}, {Jackson}, {Wilkinson}, {Fassnacht}, {Myers}, {Koopmans}, {Blandford}, {Pearson}, {Readhead}, \& {de Bruyn}}]{2001AJ....122..591R}
{Rusin}, D., {Marlow}, D.~R., {Norbury}, M., {et~al.} 2001, \aj, 122, 591, \dodoi{10.1086/321156}

\bibitem[{{Schechter} {et~al.}(1998){Schechter}, {Gregg}, {Becker}, {Helfand}, \& {White}}]{1998AJ....115.1371S}
{Schechter}, P.~L., {Gregg}, M.~D., {Becker}, R.~H., {Helfand}, D.~J., \& {White}, R.~L. 1998, \aj, 115, 1371, \dodoi{10.1086/300294}

\bibitem[{{Schechter} {et~al.}(2017){Schechter}, {Morgan}, {Chehade}, {Metcalfe}, {Shanks}, \& {McDonald}}]{2329disc}
{Schechter}, P.~L., {Morgan}, N.~D., {Chehade}, B., {et~al.} 2017, \aj, 153, 219, \dodoi{10.3847/1538-3881/aa6899}

\bibitem[{{Shajib} {et~al.}(2021){Shajib}, {Molina}, {Agnello}, {Williams}, {Birrer}, {Treu}, {Fassnacht}, {Morishita}, {Abramson}, {Schechter}, \& {Wisotzki}}]{shajib}
{Shajib}, A.~J., {Molina}, E., {Agnello}, A., {et~al.} 2021, \mnras, 503, 1557, \dodoi{10.1093/mnras/stab532}

\bibitem[{{Sluse} {et~al.}(2003){Sluse}, {Surdej}, {Claeskens}, {Hutsem{\'e}kers}, {Jean}, {Courbin}, {Nakos}, {Billeres}, \& {Khmil}}]{2003AaA...406L..43S}
{Sluse}, D., {Surdej}, J., {Claeskens}, J.~F., {et~al.} 2003, \aap, 406, L43, \dodoi{10.1051/0004-6361:20030904}

\bibitem[{{Spergel} {et~al.}(2015){Spergel}, {Gehrels}, {Baltay}, {Bennett}, {Breckinridge}, {Donahue}, {Dressler}, {Gaudi}, {Greene}, {Guyon}, {Hirata}, {Kalirai}, {Kasdin}, {Macintosh}, {Moos}, {Perlmutter}, {Postman}, {Rauscher}, {Rhodes}, {Wang}, {Weinberg}, {Benford}, {Hudson}, {Jeong}, {Mellier}, {Traub}, {Yamada}, {Capak}, {Colbert}, {Masters}, {Penny}, {Savransky}, {Stern}, {Zimmerman}, {Barry}, {Bartusek}, {Carpenter}, {Cheng}, {Content}, {Dekens}, {Demers}, {Grady}, {Jackson}, {Kuan}, {Kruk}, {Melton}, {Nemati}, {Parvin}, {Poberezhskiy}, {Peddie}, {Ruffa}, {Wallace}, {Whipple}, {Wollack}, \& {Zhao}}]{Spergel2015}
{Spergel}, D., {Gehrels}, N., {Baltay}, C., {et~al.} 2015, arXiv e-prints, arXiv:1503.03757, \dodoi{10.48550/arXiv.1503.03757}

\bibitem[{{Spingola} {et~al.}(2019{\natexlab{a}}){Spingola}, {McKean}, {Lee}, {Deller}, \& {Moldon}}]{2019MNRAS.483.2125S}
{Spingola}, C., {McKean}, J.~P., {Lee}, M., {Deller}, A., \& {Moldon}, J. 2019{\natexlab{a}}, \mnras, 483, 2125, \dodoi{10.1093/mnras/sty3189}

\bibitem[{{Spingola} {et~al.}(2019{\natexlab{b}}){Spingola}, {McKean}, {Massari}, \& {Koopmans}}]{spignola2019}
{Spingola}, C., {McKean}, J.~P., {Massari}, D., \& {Koopmans}, L.~V.~E. 2019{\natexlab{b}}, \aap, 630, A108, \dodoi{10.1051/0004-6361/201935427}

\bibitem[{{Stern} {et~al.}(2021){Stern}, {Djorgovski}, {Krone-Martins}, {Sluse}, {Delchambre}, {Ducourant}, {Teixeira}, {Surdej}, {Boehm}, {den Brok}, {Dobie}, {Drake}, {Galluccio}, {Graham}, {Jalan}, {Kl{\"u}ter}, {Le Campion}, {Mahabal}, {Mignard}, {Murphy}, {Nierenberg}, {Scarano}, {Simon}, {Slezak}, {Spindola-Duarte}, \& {Wambsganss}}]{2021ApJ...921...42S}
{Stern}, D., {Djorgovski}, S.~G., {Krone-Martins}, A., {et~al.} 2021, \apj, 921, 42, \dodoi{10.3847/1538-4357/ac0f04}

\bibitem[{{Sykes} {et~al.}(1998){Sykes}, {Browne}, {Jackson}, {Marlow}, {Nair}, {Wilkinson}, {Blandford}, {Cohen}, {Fassnacht}, {Hogg}, {Pearson}, {Readhead}, {Womble}, {Myers}, {De Bruyn}, {Bremer}, {Miley}, \& {Schilizzi}}]{1998MNRAS.301..310S}
{Sykes}, C.~M., {Browne}, I.~W.~A., {Jackson}, N.~J., {et~al.} 1998, \mnras, 301, 310, \dodoi{10.1046/j.1365-8711.1998.02081.x}

\bibitem[{{Vegetti} {et~al.}(2023){Vegetti}, {Birrer}, {Despali}, {Fassnacht}, {Gilman}, {Hezaveh}, {Perreault Levasseur}, {McKean}, {Powell}, {O'Riordan}, \& {Vernardos}}]{vegetti23}
{Vegetti}, S., {Birrer}, S., {Despali}, G., {et~al.} 2023, arXiv e-prints, arXiv:2306.11781, \dodoi{10.48550/arXiv.2306.11781}

\bibitem[{{Walsh} {et~al.}(1979){Walsh}, {Carswell}, \& {Weymann}}]{walsh79}
{Walsh}, D., {Carswell}, R.~F., \& {Weymann}, R.~J. 1979, \nat, 279, 381, \dodoi{10.1038/279381a0}

\bibitem[{{Weymann} {et~al.}(1980){Weymann}, {Latham}, {Angel}, {Green}, {Liebert}, {Turnshek}, {Turnshek}, \& {Tyson}}]{1980Natur.285..641W}
{Weymann}, R.~J., {Latham}, D., {Angel}, J. R.~P., {et~al.} 1980, \nat, 285, 641, \dodoi{10.1038/285641a0}

\bibitem[{{Winn} {et~al.}(2001){Winn}, {Hewitt}, {Patnaik}, {Schechter}, {Schommer}, {L{\'o}pez}, {Maza}, \& {Wachter}}]{2001AJ....121.1223W}
{Winn}, J.~N., {Hewitt}, J.~N., {Patnaik}, A.~R., {et~al.} 2001, \aj, 121, 1223, \dodoi{10.1086/319403}

\bibitem[{{Winn} {et~al.}(2002{\natexlab{a}}){Winn}, {Lovell}, {Chen}, {Fletcher}, {Hewitt}, {Patnaik}, \& {Schechter}}]{2002ApJ...564..143W}
{Winn}, J.~N., {Lovell}, J. E.~J., {Chen}, H.-W., {et~al.} 2002{\natexlab{a}}, \apj, 564, 143, \dodoi{10.1086/324144}

\bibitem[{{Winn} {et~al.}(2000){Winn}, {Hewitt}, {Schechter}, {Dressler}, {Falco}, {Impey}, {Kochanek}, {Leh{\'a}r}, {Lovell}, {McLeod}, {Morgan}, {Mu{\~n}oz}, {Rix}, \& {Ruiz}}]{2000AJ....120.2868W}
{Winn}, J.~N., {Hewitt}, J.~N., {Schechter}, P.~L., {et~al.} 2000, \aj, 120, 2868, \dodoi{10.1086/316874}

\bibitem[{{Winn} {et~al.}(2002{\natexlab{b}}){Winn}, {Morgan}, {Hewitt}, {Kochanek}, {Lovell}, {Patnaik}, {Pindor}, {Schechter}, \& {Schommer}}]{2002AJ....123...10W}
{Winn}, J.~N., {Morgan}, N.~D., {Hewitt}, J.~N., {et~al.} 2002{\natexlab{b}}, \aj, 123, 10, \dodoi{10.1086/338094}

\bibitem[{{Wisotzki} {et~al.}(2002){Wisotzki}, {Schechter}, {Bradt}, {Heinm{\"u}ller}, \& {Reimers}}]{2002AaA...395...17W}
{Wisotzki}, L., {Schechter}, P.~L., {Bradt}, H.~V., {Heinm{\"u}ller}, J., \& {Reimers}, D. 2002, \aap, 395, 17, \dodoi{10.1051/0004-6361:20021213}

\bibitem[{{Wright} {et~al.}(2010){Wright}, {Eisenhardt}, {Mainzer}, {Ressler}, {Cutri}, {Jarrett}, {Kirkpatrick}, {Padgett}, {McMillan}, {Skrutskie}, {Stanford}, {Cohen}, {Walker}, {Mather}, {Leisawitz}, {Gautier}, {McLean}, {Benford}, {Lonsdale}, {Blain}, {Mendez}, {Irace}, {Duval}, {Liu}, {Royer}, {Heinrichsen}, {Howard}, {Shannon}, {Kendall}, {Walsh}, {Larsen}, {Cardon}, {Schick}, {Schwalm}, {Abid}, {Fabinsky}, {Naes}, \& {Tsai}}]{WISE}
{Wright}, E.~L., {Eisenhardt}, P. R.~M., {Mainzer}, A.~K., {et~al.} 2010, \aj, 140, 1868, \dodoi{10.1088/0004-6256/140/6/1868}

\bibitem[{Wucknitz(2009)}]{Wucknitz:2009xu}
Wucknitz, O. 2009, PoS, IX EVN Symposium, 102, \dodoi{10.22323/1.072.0102}

\bibitem[{{Xanthopoulos} {et~al.}(1998){Xanthopoulos}, {Browne}, {King}, {Koopmans}, {Jackson}, {Marlow}, {Patnaik}, {Porcas}, \& {Wilkinson}}]{1998MNRAS.300..649X}
{Xanthopoulos}, E., {Browne}, I.~W.~A., {King}, L.~J., {et~al.} 1998, \mnras, 300, 649, \dodoi{10.1046/j.1365-8711.1998.01804.x}

\bibitem[{{York} {et~al.}(2000){York}, {Adelman}, {Anderson}, {Anderson}, {Annis}, {Bahcall}, {Bakken}, {Barkhouser}, {Bastian}, {Berman}, {Boroski}, {Bracker}, {Briegel}, {Briggs}, {Brinkmann}, {Brunner}, {Burles}, {Carey}, {Carr}, {Castander}, {Chen}, {Colestock}, {Connolly}, {Crocker}, {Csabai}, {Czarapata}, {Davis}, {Doi}, {Dombeck}, {Eisenstein}, {Ellman}, {Elms}, {Evans}, {Fan}, {Federwitz}, {Fiscelli}, {Friedman}, {Frieman}, {Fukugita}, {Gillespie}, {Gunn}, {Gurbani}, {de Haas}, {Haldeman}, {Harris}, {Hayes}, {Heckman}, {Hennessy}, {Hindsley}, {Holm}, {Holmgren}, {Huang}, {Hull}, {Husby}, {Ichikawa}, {Ichikawa}, {Ivezi{\'c}}, {Kent}, {Kim}, {Kinney}, {Klaene}, {Kleinman}, {Kleinman}, {Knapp}, {Korienek}, {Kron}, {Kunszt}, {Lamb}, {Lee}, {Leger}, {Limmongkol}, {Lindenmeyer}, {Long}, {Loomis}, {Loveday}, {Lucinio}, {Lupton}, {MacKinnon}, {Mannery}, {Mantsch}, {Margon}, {McGehee}, {McKay}, {Meiksin}, {Merelli}, {Monet}, {Munn}, {Narayanan}, {Nash}, {Neilsen}, {Neswold}, {Newberg}, {Nichol}, {Nicinski},
  {Nonino}, {Okada}, {Okamura}, {Ostriker}, {Owen}, {Pauls}, {Peoples}, {Peterson}, {Petravick}, {Pier}, {Pope}, {Pordes}, {Prosapio}, {Rechenmacher}, {Quinn}, {Richards}, {Richmond}, {Rivetta}, {Rockosi}, {Ruthmansdorfer}, {Sandford}, {Schlegel}, {Schneider}, {Sekiguchi}, {Sergey}, {Shimasaku}, {Siegmund}, {Smee}, {Smith}, {Snedden}, {Stone}, {Stoughton}, {Strauss}, {Stubbs}, {SubbaRao}, {Szalay}, {Szapudi}, {Szokoly}, {Thakar}, {Tremonti}, {Tucker}, {Uomoto}, {Vanden Berk}, {Vogeley}, {Waddell}, {Wang}, {Watanabe}, {Weinberg}, {Yanny}, {Yasuda}, \& {SDSS Collaboration}}]{SDSS}
{York}, D.~G., {Adelman}, J., {Anderson}, John~E., J., {et~al.} 2000, \aj, 120, 1579, \dodoi{10.1086/301513}

\bibitem[{{Yue} {et~al.}(2022){Yue}, {Fan}, {Yang}, \& {Wang}}]{Yue2022}
{Yue}, M., {Fan}, X., {Yang}, J., \& {Wang}, F. 2022, \aj, 163, 139, \dodoi{10.3847/1538-3881/ac4cb0}

\bibitem[{{Zaborowski} {et~al.}(2023){Zaborowski}, {Drlica-Wagner}, {Ashmead}, {Wu}, {Morgan}, {Bom}, {Shajib}, {Birrer}, {Cerny}, {Buckley-Geer}, {Mutlu-Pakdil}, {Ferguson}, {Glazebrook}, {Lozano}, {Gordon}, {Martinez}, {Manwadkar}, {O'Donnell}, {Poh}, {Riley}, {Sakowska}, {Santana-Silva}, {Santiago}, {Sluse}, {Tan}, {Tollerud}, {Verma}, {Carballo-Bello}, {Choi}, {James}, {Kuropatkin}, {Mart{\'\i}nez-V{\'a}zquez}, {Nidever}, {Castellon}, {No{\"e}l}, {Olsen}, {Pace}, {Mau}, {Yanny}, {Zenteno}, {Abbott}, {Aguena}, {Alves}, {Andrade-Oliveira}, {Bocquet}, {Brooks}, {Burke}, {Carnero Rosell}, {Carrasco Kind}, {Carretero}, {Castander}, {Conselice}, {Costanzi}, {Pereira}, {de Vicente}, {Desai}, {Dietrich}, {Doel}, {Everett}, {Ferrero}, {Flaugher}, {Friedel}, {Frieman}, {Garc{\'\i}a-Bellido}, {Gruen}, {Gruendl}, {Gutierrez}, {Hinton}, {Hollowood}, {Honscheid}, {Kuehn}, {Lin}, {Marshall}, {Melchior}, {Mena-Fern{\'a}ndez}, {Menanteau}, {Miquel}, {Palmese}, {Paz-Chinch{\'o}n}, {Pieres}, {Malag{\'o}n}, {Prat},
  {Rodriguez-Monroy}, {Romer}, {Sanchez}, {Scarpine}, {Sevilla-Noarbe}, {Smith}, {Suchyta}, {To}, {Weaverdyck}, {Delve Collaboration}, \& {Des Collaboration}}]{zaborowski23}
{Zaborowski}, E.~A., {Drlica-Wagner}, A., {Ashmead}, F., {et~al.} 2023, \apj, 954, 68, \dodoi{10.3847/1538-4357/ace4ba}

\bibitem[{{Zhang} {et~al.}(2023){Zhang}, {Zhang}, {Nightingale}, {Zou}, {Cao}, {Tsai}, {Yang}, {Shi}, {Wang}, {Xu}, {Lin}, {Zhou}, \& {Li}}]{2023MNRAS.524.3671Z}
{Zhang}, L., {Zhang}, Z.-Y., {Nightingale}, J.~W., {et~al.} 2023, \mnras, 524, 3671, \dodoi{10.1093/mnras/stad2069}

\bibitem[{{Zhou} {et~al.}(2020){Zhou}, {Newman}, {Dawson}, {Eisenstein}, {Brooks}, {Dey}, {Dey}, {Duan}, {Eftekharzadeh}, {Gazta{\~n}aga}, {Kehoe}, {Landriau}, {Levi}, {Licquia}, {Meisner}, {Moustakas}, {Myers}, {Palanque-Delabrouille}, {Poppett}, {Prada}, {Raichoor}, {Schlegel}, {Schubnell}, {Staten}, {Tarl{\'e}}, \& {Y{\`e}che}}]{Zhou2020}
{Zhou}, R., {Newman}, J.~A., {Dawson}, K.~S., {et~al.} 2020, RNAAS, 4, 181, \dodoi{10.3847/2515-5172/abc0f4}

\end{thebibliography}
\bibliographystyle{aasjournal}



\end{document}